\documentclass[12pt, draftclsnofoot, onecolumn]{IEEEtran}

\IEEEoverridecommandlockouts

\usepackage{cite}
\usepackage{amsmath,amssymb,amsfonts,bbold}
\usepackage{algorithmic}
\usepackage{graphicx}
\usepackage[dvipsnames]{xcolor}
\usepackage{tikz}
\usepackage{tikz-3dplot}

\usepackage{hyperref}
\usetikzlibrary{arrows}
\usetikzlibrary{arrows.meta}
\usetikzlibrary{automata,positioning}   
\usepackage{lipsum}
\usetikzlibrary{decorations.markings}
\usetikzlibrary{patterns}
\usetikzlibrary{fit}

\usepackage{textcomp}
\usepackage{xcolor}

\ifCLASSOPTIONcompsoc
    \usepackage[caption=false, font=normalsize, labelfont=sf, textfont=sf]{subfig}
\else
\usepackage[caption=false, font=footnotesize]{subfig}
\fi

\def\BibTeX{{\rm B\kern-.05em{\sc i\kern-.025em b}\kern-.08em
    T\kern-.1667em\lower.7ex\hbox{E}\kern-.125emX}}
\usepackage{mathtools}

\usetikzlibrary{math}
\usepackage{amsthm}
\usepackage{amssymb}
\usepackage[utf8]{inputenc}
\usepackage{algorithm2e}
\usepackage{amsmath}

\definecolor{darkcerulean}{rgb}{0.0, 0.0, 0.55}
\definecolor{darkcerulean}{rgb}{0.03, 0.27, 0.49}
\definecolor{darkgreen}{rgb}{0.0, 0.2, 0.13}
\definecolor{applegreen}{rgb}{0.55, 0.71, 0.0}
\definecolor{ashgrey}{rgb}{0.7, 0.75, 0.71}
\definecolor{aurometalsaurus}{rgb}{0.43, 0.5, 0.5}
\definecolor{darkgray}{rgb}{0.66, 0.66, 0.66}

\begin{document}
\bstctlcite{IEEEexample:BSTcontrol}
\tikzset{->-/.style={decoration={
			markings,
			mark=at position #1 with {\arrow{>}}},postaction={decorate}}}

\title{Device-Agnostic Millimeter Wave Beam Selection using Machine Learning

\thanks{This work is supported by the Danish Council for Independent Research, grant no. DFF 8022-00371B.}
}

\author{Sajad~Rezaie,~João~Morais,~Ahmed~Alkhateeb,~and~Carles~Navarro~Manch\'on
\thanks{S.~Rezaie~and~C.~Navarro~Manch\'on are with the Department of Electronic Systems, Aalborg University, Aalborg, Denmark. (Email: sre@es.aau.dk, cnm@es.aau.dk)} 

\thanks{J.~Morais~and~A.~Alkhateeb are with the School of Electrical, Computer and Energy Engineering, Arizona State University, Tempe, AZ, USA. (Email: joao@asu.edu, alkhateeb@asu.edu)}}


\maketitle

\begin{abstract}
Most research in the area of machine learning-based user beam selection considers a structure where the model proposes appropriate user beams. However, this design requires a specific model for each user-device beam codebook, where a model learned for a device with a particular codebook can not be reused for another device with a different codebook. Moreover, this design requires training and test samples for each antenna placement configuration/codebook. This paper proposes a device-agnostic beam selection framework that leverages context information to propose appropriate user beams using a generic model and a post processing unit. The generic neural network predicts the potential angles of arrival, and the post processing unit maps these directions to beams based on the specific device's codebook. The proposed beam selection framework works well for user devices with antenna configuration/codebook unseen in the training dataset. Also, the proposed generic network has the option to be trained with a dataset mixed of samples with different antenna configurations/codebooks, which significantly eases the burden of effective model training.
\end{abstract}


\section{Introduction} \label{Intro}
\IEEEPARstart{I}{nitial} beam alignment (BA) in multiple-input multiple-output (MIMO) systems operating at millimeter wave (mmWave) band is critical for establishing a reliable communication link. Using predefined beams in codebook-based beamforming is a popular approach to simplify the BA process. An exhaustive search over all possible beam pairs at the transceivers finds the optimal beam configuration, but results in undesirable overhead and latency for the network. Although hierarchical beam search imposes a limited overhead to the network, it is prone to errors due to the low signal-to-noise ratio (SNR) at the first search stages with wide beams \cite{giordani_toward_2020}. Machine learning (ML)-based methods are proposed to overcome these challenges by leveraging the knowledge available in data captured in a specific environment. In \cite{sohrabi_deep_2021-1}, an ML-based procedure is proposed to estimate the dominant path's angle of arrival (AoA) with an adaptive compressed sensing approach. Also, the authors of \cite{anton-haro_learning_2019} use the ML capabilities in a beam selection framework to recommend the best beam for the base station (BS) in a multi-user SIMO scenario. However, these methods assume a single path channel for the mmWave link, therefore the proposed frameworks may not work well in multipath channels.

Another research track has focused on context information (CI)-based beam selection methods leveraging additional information like user position, camera images, and sub-6 GHz channel state information. The extra information provides the opportunity to reduce the beam search space, which subsequently leads to reducing the latency and increasing the accuracy of the BA process \cite{va_inverse_2018, alrabeiah_millimeter_2020, alrabeiah_deep_2020}. As it is not a straightforward task to use CI in the beam selection process, deep learning (DL) capabilities are often used to boost the performance of CI-based methods \cite{xu_3d_2020, dias_position_2019, gao_fusionnet_2021}. The acquisition of CI may apply additional latency or overhead to the system, which constitutes the main challenge for the CI-based approach in some applications.

\begin{figure}[!t]
	\centering
    \scalebox{1}{\includegraphics[trim={0cm 0cm 0cm 0cm},clip, width=0.95\textwidth]{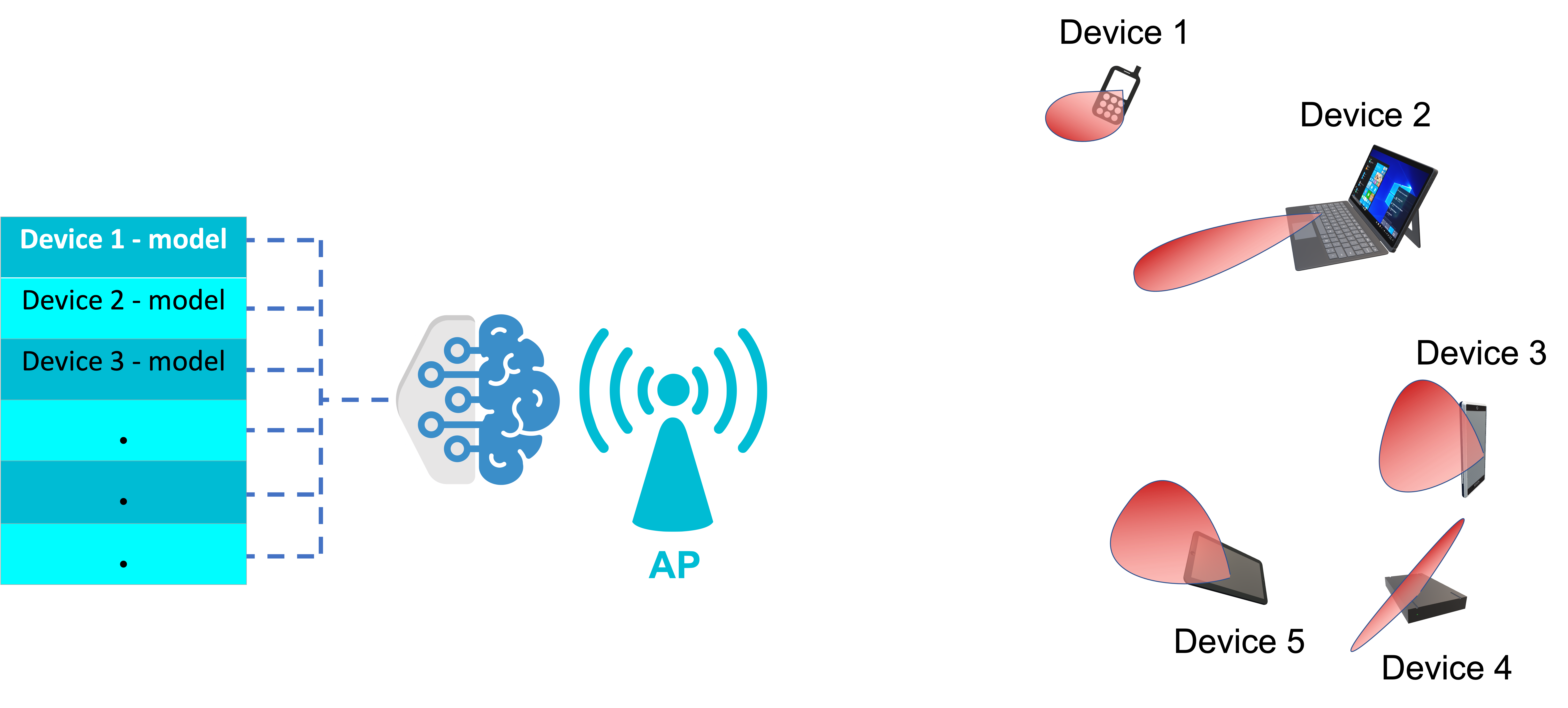}}
    \caption{An access point communicates with different user terminals, such as smartphones, tablets, and laptops from different vendors. The user terminals have different beam codebooks because of different antenna configurations and hardware, which requires the network to maintain a specific ML model for each type of user device.}
    \label{Fig:WirelessDevices}
\end{figure}

However, in both context-free and context-aware ML-based beam selection methods, most research assumes a network structure specific to the user antenna configuration/beam-codebook to recommend the best user beam from the codebook. In this structure, the ML model estimates the optimality probability for each beam from the codebook. However, as shown in Fig.~\ref{Fig:WirelessDevices}, an access point may serve several user terminals (UTs) from different vendors, which may have different antenna configurations/codebooks. Thus, different ML networks are needed for different device codebook configurations, making the ML-based beam selection less attractive. Another challenge is collecting training datasets for different antenna placement designs/beam-codebook configurations \cite{va_position-aided_2017, anton-haro_learning_2019, dias_position_2019}. Here, UT terminal indicates the type of device, its antenna configuration, and its beam codebook. Thus, using UT-specific ML models for each possible terminal implies the collection of UT-specific training datasets, along with the associated training computational and memory challenges of having to train, store, and manage a multiplicity of models.

Transfer learning (TL) provides the option to reuse the information learned by a network with a large training dataset and a given antenna configuration to reach an acceptable performance for a new antenna configuration with a smaller training dataset \cite{rezaie_deep_2021}. However, directly reusing the learned network without additional samples is not feasible with this technique. Moreover, the TL method in \cite{rezaie_deep_2021} does not solve the issue of training a network with samples from different antenna configurations/codebooks.

One possible avenue to overcome the need for UT-specific beam selection NNs is to design models that rank beamforming directions rather than specific codebook beams. To discretize the angular direction space in such an approach, Fibonacci grids \cite{swinbank_fibonacci_2006} can employed. A Fibonacci grid (FG) is a set of points in which the points are placed evenly on the sphere. FGs have already been applied to solve other problems in mmWave beam management. For instance, an orientation-assisted beam alignment method using particle filters is proposed in \cite{ali_orientation-assisted_2021}, where a FG is used for initialization of particles. To design codebooks that account for coupling and the impacts of the user hand grip, the authors in \cite{mo_beam_2019,alammouri_hand_2019} use the FG as a way to pick uniformly distributed sampling points on the unit sphere. Also, a similarity metric/measure using the FG is used to match beams in two different antenna configurations \cite{mo_sub-chain_2021}. 

\subsection{Contributions}
As a solution to the above mentioned challenges with device-specific structure for ML-based beam alignment methods, this paper proposes a device-agnostic ML-based beam selection framework. The proposed framework includes two main parts: a generic neural network (NN) that recommends directions instead of UT beams, and a post processing unit that maps the proposed directions to the UT beams in the codebook. The main contributions of this paper are:
\begin{enumerate}
    \item We use the FG to discretize the AoAs at the UT side. Using this technique, we replace the device-specific beam selection task with the discretized direction selection as a device-agnostic task. The generic NN provides the optimality probability for each point in the Fibonacci grid.
    \item\label{PostP} The generic NN is followed by a post processing unit implementing a deterministic mapping function. The proposed unit calculates the optimality probability of UT beams using the optimality probability of the FG points.
    \item We consider a multi-labeling training procedure for the generic NN. By reversing the mapping process in \ref{PostP}), the optimal UT beam corresponds to a subset of the Fibonacci grid, and we mark those points as labels in the generic NN training procedure.
    \item To illustrate the proposed device-agnostic framework in the ML-based beam selection methods, we implement the device-agnostic framework in the two CI-based beam alignment methods respectively proposed in \cite{rezaie_location-_2022} and \cite{alrabeiah_deep_2020}: (a) Beam pair selection for user handsets leveraging the user location and orientation information, and (b) mmWave beam pair selection using the uplink sub-6 GHz channel information.
    \item We use accurate 3-dimensional (3D) ray-tracing tools to generate the training and test samples for the two CI-based studies. The numerical evaluations show that the proposed device-agnostic framework works well even for devices with beam codebooks that were unseen in the training samples. In addition, the generic NN can be trained with a dataset measured with different devices.
    \item We extended the work presented in \cite{alrabeiah_deep_2020} to support beamforming at the user side. The results show that the sub-6 GHz channel as CI can be used for mmWave UT beamforming in both line-of-sight (LOS) and non-line-of-sight (NLOS) conditions. 
\end{enumerate}


\subsection{Organization and Notations}
The rest of the paper is organized as follows. First, the overall view of the beam selection problem, the components of the proposed device-agnostic framework, and the labeling vector for the generic NN are described in Section~\ref{Sec:DevAgn}. In Section~\ref{Sec:LocOri}, we explain the system model of the location and orientation-aware beam selection method and show how the network structure can be updated based on the proposed device-agnostic design. Also, the numerical results of this work are shown in this section. The system model of the beam selection method leveraging the sub-6 GHz channel is explained in Section~\ref{Sec:Sub6}. This section includes the results before and after integrating the device-agnostic framework in this study. Finally, we summarise the conclusions of the study in Section~\ref{Sec:Conc}.

The notations used in this paper are as follows. $\mathbb{R}$ and $\mathbb{C}$ denote the fields of real and complex numbers, respectively. We denote by $\mathcal{A}$ a finite set and by $|\mathcal{A}|$ its cardinality. In addition, we use $a$, $\boldsymbol{a}$, and $\boldsymbol{A}$ to denote a scalar, a vector, and a matrix, respectively. $a_i$ is the $i$th entry of the column vector $\boldsymbol{a}$, and $A_{i,j}$ represents the entry in the $i$th row and $j$th column of the matrix $\boldsymbol{A}$. Notations $(\cdot)^T$ and $(\cdot)^H$ denote, respectively, transposition and complex transposition of vectors and matrices. Also, $\otimes$ is the Kronecker product. $\underset{i, j}{\mathrm{arg\hspace{2pt}max} \hspace{0.5pt} A_{i, j}}$ denotes the row and column index of the maximum entry of matrix $\boldsymbol{A}$ as a tuple. $\boldsymbol{0}_t$ and $\boldsymbol{I}_p$ denote a zero column vector of length $t$ and the $p \times p$ identity matrix, respectively. $\mathbb{P} (A)$ denotes the probability of the event $A$. 

\section{Device-agnostic Beam Selection framework}\label{Sec:DevAgn}
A brief background on the codebook-based beam selection procedure from a general perspective is introduced in Section~\ref{SubSec:CodeBS}. Then, the widely used device-specific DL-based beam selection is explained in Section~\ref{SubSec:Device-Specific}. Our proposed device-agnostic beam selection framework is introduced in Section~\ref{SubSec:Device-Agnostic}, including the generic NN and post processing unit as the main blocks of the framework. Lastly, Section~\ref{SubSec:DataCollection} describes the dataset collection and labeling procedure for the device-specific and device-agnostic frameworks.

\subsection{Codebook-based Beam Selection}\label{SubSec:CodeBS}
Consider $\boldsymbol{H}$ as the mmWave downlink channel between an access point (AP) and UT placed in an environment, and the AP and UT respectively use $\boldsymbol{u}$ and $\boldsymbol{v}$ beamforming vectors for communication. The received signal at the UT may be written as
\begin{equation}\label{Y_Rec_Sig_Gen}
    y = \sqrt{P_\mathrm{AP}} \boldsymbol{v}^H \boldsymbol{H} \boldsymbol{u} s + \boldsymbol{v}^H \boldsymbol{n}
\end{equation}
where $s$ and $P_\mathrm{AP}$ are the unit power symbol $s$ and the AP transmission power, respectively. In addition, $\boldsymbol{n}$ is a zero-mean complex Gaussian noise with variance $\sigma_n^2$. 

Consider the UT has a codebook with $N_\mathrm{UT}$ beams, and the set $\mathcal{B}_\mathrm{UT}$ includes the indices of all beams in the UT codebook. In this section, for the sake of simplicity of explanation, we focus only on the selection of the UT beamforming vector $\boldsymbol{v}$. We assume that the AP beamforming vector $\boldsymbol{u}$ has already been chosen by other means. However, Sections~\ref{Sec:LocOri} and~\ref{Sec:Sub6} show that the proposed technique can be applied to joint beamforming applications where both the AP and UT are capable of beamforming. In the beam selection approach, we target predicting a beam candidate list, $\mathcal{S} \subset \mathcal{B}_\mathrm{UT}$, for sensing the environment, which leads to a smaller beam search space. The optimal choice of $\mathcal{S}$ depends on the propagation properties of the environment and the UT properties like its location and orientation. As it is proven in \cite{va_inverse_2018}, the optimal UT beam candidate list for a given UT condition includes the beam indices that have the highest probabilities of optimality. The probability of optimality for the $i$th beam may be expressed as $P_{j}^{B} = \mathbb{P} \left[ j = j^\star \right]$, where $j^\star$ is the optimal UT beam for communication providing the highest received signal strength (RSS).

\subsection{Device-Specific framework}\label{SubSec:Device-Specific}
In prior work, the CI-based beam selection was mainly performed using a common device-specific NN that takes some assistance information (such as LIDAR, camera images, user position, etc.) as input and provides the optimality probability for each beam in the codebook, $P_{j}^{B}, j = 1, 2, \cdots, N_\mathrm{UT}$, as shown in Fig.~\ref{Fig:SpcGenNN}\subref{1_a}. 
The device-specific NN structure includes a series of dense and/or convolutional layers followed by an output layer with $N_\mathrm{UT}$ neurons. The weights of the network are trained for a specific device with a specific codebook. Thus, several sets of trained weights need to be stored to serve different UTs, which causes challenges with memory and management of models.

\subsection{Device-Agnostic framework}\label{SubSec:Device-Agnostic}
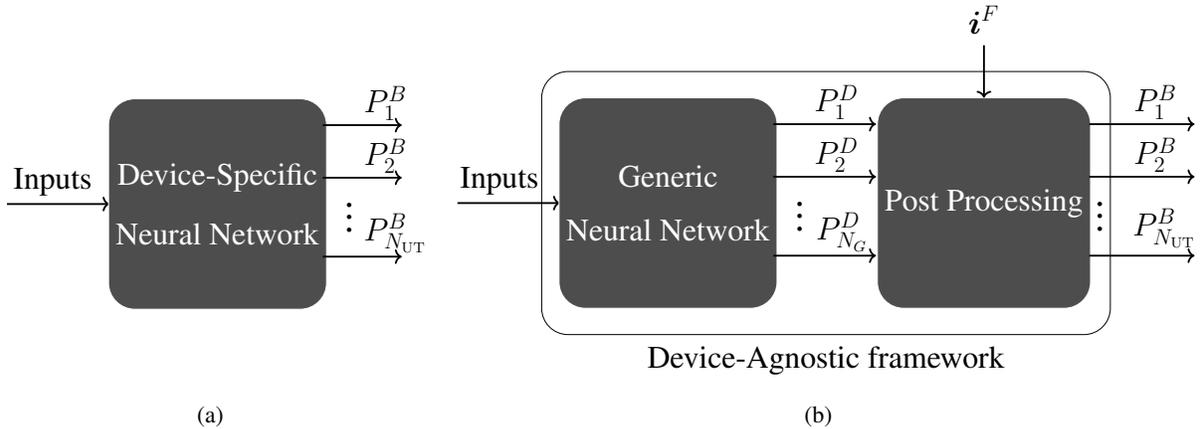
\begin{figure}[t]
    \definecolor{darkgray}{rgb}{0.3, 0.3, 0.3}
    \centering
    \subfloat[\label{1_a}]{
	\scalebox{0.70}{
		\begin{tikzpicture}[cross/.style={path picture={ 
				\draw[red, line width=3pt]
				(path picture bounding box.south east) -- (path picture bounding box.north west) (path picture bounding box.south west) -- (path picture bounding box.north east);
		}}]
		\draw[->, line width=1pt] (0, 0) -- (1.95, 0) node [font=\Large, above left = 0pt and 5pt] {\Large Inputs};
		
		\node at (4,0) [draw,white,minimum size=4cm,rounded corners=5mm,fill=darkgray] (v100) {\shortstack{\Large Device-Specific \\[0.5cm] \Large Neural Network}};
		
		\path (6.5,0.5) -- (6.5,-1) node [black, font=\Huge, midway, sloped] {\textbf{$...$}};

		\draw[->, line width=1pt] (6,-1) -- (7.5,-1) node [font=\Large, above right = -2pt and -25pt] {$P_{N_\mathrm{UT}}^{B}$};
		\draw[->, line width=1pt] (6,0.5) -- (7.5,0.5) node [font=\Large, above right = -2pt and -25pt] {$P_{2}^{B}$};
		\draw[->, line width=1pt] (6,1.5) -- (7.5,1.5) node [font=\Large, above right = -2pt and -25 pt] {$P_{1}^{B}$};
		
		\node at (5,-3.25) (v2) {\shortstack{}};
		
		\end{tikzpicture}}}
	\subfloat[\label{1_b}]{
	\scalebox{0.70}{
		\begin{tikzpicture}[cross/.style={path picture={ 
				\draw[red, line width=3pt]
				(path picture bounding box.south east) -- (path picture bounding box.north west) (path picture bounding box.south west) -- (path picture bounding box.north east);
		}}]
		\draw[->, line width=1pt] (0, 0) -- (1.95, 0) node [font=\Large, above left = 0pt and 7pt] {\Large Inputs};
		
		\node at (4,0) [draw,white,minimum size=4cm,rounded corners=5mm,fill=darkgray] (v100) {\shortstack{\Large Generic \\[0.5cm] \Large Neural Network}};
		
		\path (6.5,0.5) -- (6.5,-1) node [black, font=\Huge, midway, sloped] {\textbf{$...$}};

		\draw[->, line width=1pt] (6,-1) -- (7.95,-1) node [font=\Large, above left = -2pt and 0pt] {$P_{N_{G}}^{D}$};
		\draw[->, line width=1pt] (6,0.5) -- (7.95,0.5) node [font=\Large, above left = -2pt and 5pt] {$P_{2}^{D}$};
		\draw[->, line width=1pt] (6,1.5) -- (7.95,1.5) node [font=\Large, above left = -2pt and 5pt] {$P_{1}^{D}$};
		
		\node at (10,0) [draw,white,minimum size=4cm,rounded corners=5mm,fill=darkgray] (v100) {\shortstack{\Large Post Processing}};
		
		\draw[->, line width=1pt] (12,-1) -- (14,-1) node [font=\Large, above left = 0pt and -5pt] {$P_{N_\mathrm{UT}}^{B}$};
		\draw[->, line width=1pt] (12,0.5) -- (14,0.5) node [font=\Large, above left = -2pt and 5pt] {$P_{2}^{B}$};
		\draw[->, line width=1pt] (12,1.5) -- (14,1.5) node [font=\Large, above left = -2pt and 5pt] {$P_{1}^{B}$};
		\path (12.2,0.5) -- (12.2,-1) node [black, font=\Huge, midway, sloped] {\textbf{$...$}};
		
		\draw[->, line width=1pt] (10,3) -- (10,2) node [sloped, font=\Large, above = 30pt] {$\boldsymbol{i}^{F}$};
		
		\node at (7,0) [rectangle,draw,minimum width=10.8cm, minimum height=5cm,rounded corners=5mm] (v1) {};
		\node at (7,-3) (v2) {\shortstack{\Large Device-Agnostic framework}};
		
		\end{tikzpicture}}}	
	\caption{Neural network-based beam selection using (a) a device-specific and (b) the proposed device-agnostic structures. In both cases, the inputs to the networks may include data relevant for beam management, such as location information, orientation information, out-of-band channel state information, RGB camera images, RADAR, LIDAR, etc.}
	\label{Fig:SpcGenNN}
\end{figure}

In this study, we propose a device-agnostic framework that can solve the above mentioned problem and provides performance as good as the device-specific design. As shown in Fig.~\ref{Fig:SpcGenNN}\subref{1_b}, the proposed device-agnostic framework includes two main parts, the generic NN and a post processing unit.
\subsubsection{Generic NN}\label{SubSec:GenericNN}
\begin{figure}[t]
	\centering
    \scalebox{1}{\includegraphics[trim={4.5cm 1.5cm 5.5cm 2cm},clip, width=0.5\textwidth]{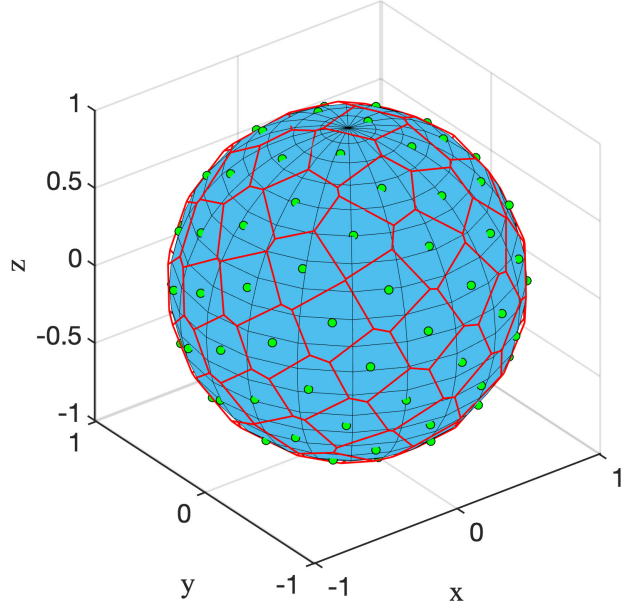}}
    \caption{Fibonacci grid with $100$ points. Cells corresponding to each Fibonacci point cover equal areas on the sphere surface.}
    \label{Fig:FibGrid}
\end{figure}

As each beam of the UT codebook covers parts of the unit sphere, it is sufficient to know the AoA in a discretized set of directions. We use the spherical Fibonacci grid to obtain an equal-area grid of points on the sphere \cite{swinbank_fibonacci_2006}. Set $\mathcal{A}=\{(\phi_k, \theta_k), k=1,..., n_{Fib}\}$ containing $n_{Fib}$ elements, where each element is a tuple (azimuth, elevation) describing potential directions of transmission/reception. We consider that set $\mathcal{I} = \{1, 2, \cdots, n_{Fib}\}$ includes the indexes of elements in set $\mathcal{A}$. Fig.~\ref{Fig:FibGrid} shows the Fibonacci grid with $n_{Fib} = 100$ points on the unit sphere.

The generic network predicts the optimality probabilities of the UT AoA in the UT local coordinate system (LCS), i.e.,
\begin{equation}
    P_{k}^{D} = \mathbb{P} \left[ k = k^\star \right]
\end{equation}
where $k^\star$ denotes the index of element of $\mathcal{A}$ containing the optimal direction of transmission/reception in the UT LCS.

\subsubsection{Post Processing}\label{SubSec:PostProcessing}
In the post processing unit, the probabilities of optimal direction, $P_{k}^{D}$, are combined to construct the optimality probabilities of UT beams, $P_{j}$. Thus, the generic probabilities as outcomes of the generic NN are combined based on the UT-specific codebook mapping information to generate the UT-specific beam probabilities. This operation needs information about the beam index providing the highest gain for each Fibonacci point, $\boldsymbol{i}^{F} \in \mathbb{N}^{n_{Fib}\times1}$, which depends solely on the device codebook and antenna geometry. The $k$th entry of $\boldsymbol{i}^{F}$ indicates the index of the beam which has highest gain in the direction of the $k$th element of $\mathcal{A}$. The mapping is required to be shared only one time by the UT, and we assume the AP have knowledge of such mapping. We define set $\mathcal{I}^{j}, j=1, \cdots, N_\mathrm{UT}$, a partition of set $\mathcal{I}$, including the index of Fibonacci points that the UT beam $i$ provides the highest gains in those directions, i.e., $\boldsymbol{i}^{F}_k = j \iff k \in \mathcal{I}^{j}$. $\mathcal{I}^{j}$ are non-overlapping sets whose union is equal to $\mathcal{I}$. Thus, the post processing unit uses the subsets $\mathcal{I}^{j}$ to map the FG optimality probabilities, $P_{k}^{D}$, onto UT beam optimality probabilities as
\begin{equation}
    P_{j}^{B} = \sum_{k \in \mathcal{I}^{j}} P_{k}^{D}.
\end{equation}
In the last step, the beam candidate list $\mathcal{S}$ is proposed by including the indices of beams that have the highest optimality likelihood, $P_{j}^{B}$.

\subsection{Dataset Generation and Labeling}\label{SubSec:DataCollection}
To construct the training and evaluation datasets, all the possible UT beam are sensed for each realization of the scenario. Considering beam $j^\star$ provides the highest RSS, the labeling vector of the device-specific network is 
\begin{equation} \label{eq:LabelSpec}
    L^{S}_{j} = 
    \begin{cases}
    	1, & \text{if $j = j^\star$,} \\
    	0, & \text{otherwise}.
    \end{cases}
\end{equation}
However, for training the generic NN, the Fibonacci points in the beam region of the optimal UT beam are marked, i.e., 
\begin{equation} \label{eq:LabelGen}
    L^{G}_{k} = 
    \begin{cases}
    	1, & \text{if $k \in \mathcal{I}^{j^\star}$,} \\
    	0, & \text{otherwise}.
    \end{cases}
\end{equation}
We use the mapping function between the Fibonacci points and the UT beams, $\mathcal{I}^{j^\star}$, to mark the right neurons at the output layer of the generic network. Thus, we mark all the directions where the UT beam $j^\star$ provides the highest gain in the labeling vector. Each dataset sample includes the inputs to the network, the RSS measurements for all the UT beams, and the corresponding labeling vector.

\section{Location- and Orientation-aware Beam Selection}\label{Sec:LocOri}
The study in this section extends the beam selection method proposed in \cite{rezaie_location-_2022} by integrating the described device-agnostic framework. This work shows how the device-specific structure can be replaced with the device-agnostic framework without (noticeable) losses in performance.

\subsection{System and Channel Model}\label{Sec:SysModel}
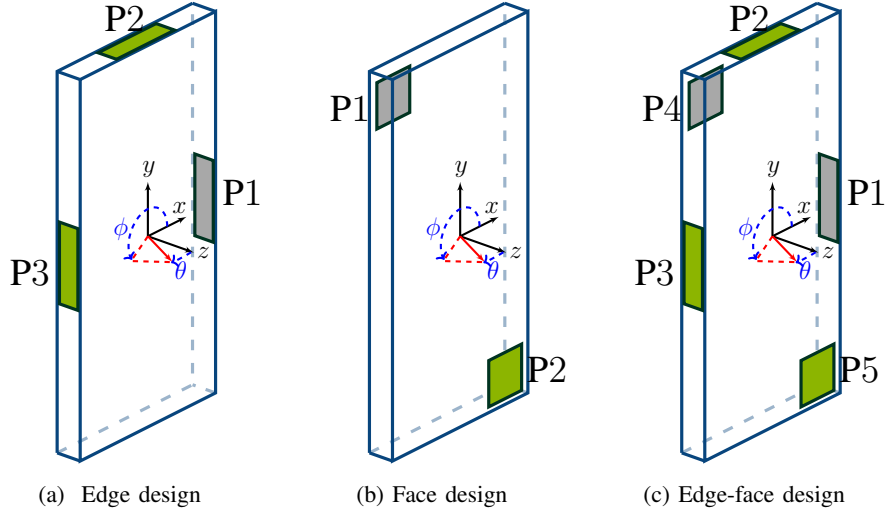
\begin{figure}[t]
	\centering
	\subfloat[ Edge design\label{1a}]{%
       \tdplotsetmaincoords{-25}{0}
	\scalebox{0.4}{
	\begin{tikzpicture}[tdplot_main_coords]
    
	\pgfmathsetmacro{\cubex}{7}
	\pgfmathsetmacro{\cubey}{14}
	\pgfmathsetmacro{\cubez}{-1}
	
	\pgfmathsetmacro{\patchl}{3}
	\pgfmathsetmacro{\patchw}{0.8}
	\pgfmathsetmacro{\patchs}{1.7}
	
	
	\begin{scope}[rotate around y=50]
	
	\draw[color=darkcerulean, line width = 3pt] (0,0,0) -- ++(\cubex,0,0) -- ++(0,\cubey,0) -- ++(-\cubex,0,0) -- cycle;
	\draw[color=darkcerulean, line width = 3pt] (0,0,\cubez) -- ++(0,\cubey,0) -- ++(\cubex,0,0);
	\draw[color=darkcerulean, line width = 3pt, dash pattern=on 10pt off 12pt, opacity=0.4] (0,0,\cubez) -- ++(\cubex,0,0) -- ++(0,\cubey,0);
	\draw[color=darkcerulean, line width = 3pt] (0,0,0) -- ++(0,0,\cubez);
	\draw[color=darkcerulean, line width = 3pt] (0,\cubey,0) -- ++(0,0,\cubez);
	\draw[color=darkcerulean, line width = 3pt] (\cubex,\cubey,0) -- ++(0,0,\cubez);
	\draw[color=darkcerulean, line width = 3pt, dash pattern=on 10pt off 10pt, opacity=0.4] (\cubex,0,0) -- ++(0,0,\cubez);

    \draw[color=darkgreen, fill=applegreen, line width = 3pt] (0,\cubey/2 - \patchl/2,-0.1) -- ++(0,\patchl,0) -- ++(0,0,-\patchw) -- ++(0,-\patchl,0) -- cycle;
    \node[scale=3] at (-2, \cubey/2+0.5,\cubez/2) {P$3$};
    
    \draw[color=darkgreen, fill=applegreen, line width = 3pt] (\cubex/2 - \patchl/2,\cubey,-0.1) -- ++(\patchl,0,0) -- ++(0,0,-\patchw) -- ++(-\patchl,0,0) -- cycle;
    \node[scale=3] at (\cubex/2, \cubey+0.7,\cubez) {P$2$};
    
    \draw[color=darkgreen, fill=darkgray, line width = 3pt] (\cubex,\cubey/2 - \patchl/2,-0.1) -- ++(0,\patchl,0) -- ++(0,0,-\patchw) -- ++(0,-\patchl,0) -- cycle;
    \node[scale=3] at (\cubex+2, \cubey/2-0.5,\cubez/2) {P$1$};
    
    \draw[>=latex',  ->, line width  = 2pt] (\cubex/2,\cubey/2,0) -- ++ (2,0,0) node[above left =-6pt and -10pt, scale=2]{$x$};  
    \draw[>=latex',  ->, line width  = 2pt] (\cubex/2,\cubey/2,0) -- ++ (0,2,0) node[above right=-5pt and -15pt, scale=2]{$y$};  
    \draw[>=latex',  ->, line width  = 2pt] (\cubex/2,\cubey/2,0) -- ++ (0,0,2) node[above right=-15pt and -7pt, scale=2]{$z$};  
    
    \draw[>=latex',  ->, line width = 2pt, red] (\cubex/2,\cubey/2,0) -- ++ (-0.87,-0.1,1.8) node{};  
    \draw[dash pattern=on 5pt off 5pt, line width  = 2pt, red] (\cubex/2,\cubey/2,0) -- ++ (-0.87,-0.6,0) --++ (0,0.5,1.8);  
    
    \tdplotdrawarc[color=blue, ->, line width = 2pt, dash pattern=on 5pt off 5pt]{(\cubex/2,\cubey/2,0)}{1}{0}{215}{above left =-35pt and 3pt,color=blue, scale=2}{$\phi$}
    
    \tdplotsetthetaplanecoords{215}
    \tdplotdrawarc[tdplot_rotated_coords, color=blue, ->, line width = 2pt, dash pattern=on 5pt off 5pt]{(-12.9,-4.5,0)}{2}{0}{24}{below = -7pt, scale=2}{$\theta$}
    
    \end{scope}
    
	\end{tikzpicture} }}
  \subfloat[Face design\label{1b}]{%
       \tdplotsetmaincoords{-25}{0}
	\scalebox{0.4}{
	\begin{tikzpicture}[tdplot_main_coords]
    
	\pgfmathsetmacro{\cubex}{7}
	\pgfmathsetmacro{\cubey}{14}
	\pgfmathsetmacro{\cubez}{-1}
	
	\pgfmathsetmacro{\patchl}{3}
	\pgfmathsetmacro{\patchw}{0.8}
	\pgfmathsetmacro{\patchs}{1.7}
	
	
	\begin{scope}[rotate around y=50]
	
	\draw[color=darkgreen, fill=darkgray, line width = 3pt] (-2,13.2,-\cubez) -- ++(0,\patchs,0) -- ++(\patchs,0,0) -- ++(0,-\patchs,0) -- cycle;
	\node[scale=3] at (-1, \cubey-1,\cubez) {P$1$};
	
	\draw[color=darkcerulean, line width = 3pt] (0,0,0) -- ++(\cubex,0,0) -- ++(0,\cubey,0) -- ++(-\cubex,0,0) -- cycle;
	\draw[color=darkcerulean, line width = 3pt] (0,0,\cubez) -- ++(0,\cubey,0) -- ++(\cubex,0,0);
	\draw[color=darkcerulean, line width = 3pt, dash pattern=on 10pt off 12pt, opacity=0.4] (0,0,\cubez) -- ++(\cubex,0,0) -- ++(0,\cubey,0);
	\draw[color=darkcerulean, line width = 3pt] (0,0,0) -- ++(0,0,\cubez);
	\draw[color=darkcerulean, line width = 3pt] (0,\cubey,0) -- ++(0,0,\cubez);
	\draw[color=darkcerulean, line width = 3pt] (\cubex,\cubey,0) -- ++(0,0,\cubez);
	\draw[color=darkcerulean, line width = 3pt, dash pattern=on 10pt off 10pt, opacity=0.4] (\cubex,0,0) -- ++(0,0,\cubez);
    
    \draw[color=darkgreen, fill=applegreen, line width = 3pt] (\cubex-2.5
    ,0.5,0.4) -- ++(0,\patchs,0) -- ++(\patchs,0,0) -- ++(0,-\patchs,0) -- cycle;
    \node[scale=3] at (\cubex+1,0.5) {P$2$};
    
    \draw[>=latex',  ->, line width  = 2pt] (\cubex/2,\cubey/2,0) -- ++ (2,0,0) node[above left =-6pt and -10pt, scale=2]{$x$};  
    \draw[>=latex',  ->, line width  = 2pt] (\cubex/2,\cubey/2,0) -- ++ (0,2,0) node[above right=-5pt and -15pt, scale=2]{$y$};  
    \draw[>=latex',  ->, line width  = 2pt] (\cubex/2,\cubey/2,0) -- ++ (0,0,2) node[above right=-15pt and -7pt, scale=2]{$z$};  
    
    \draw[>=latex',  ->, line width = 2pt, red] (\cubex/2,\cubey/2,0) -- ++ (-0.87,-0.1,1.8) node{};  
    \draw[dash pattern=on 5pt off 5pt, line width  = 2pt, red] (\cubex/2,\cubey/2,0) -- ++ (-0.87,-0.6,0) --++ (0,0.5,1.8);  
    
    \tdplotdrawarc[color=blue, ->, line width = 2pt, dash pattern=on 5pt off 5pt]{(\cubex/2,\cubey/2,0)}{1}{0}{215}{above left =-35pt and 3pt,color=blue, scale=2}{$\phi$}
    
    \tdplotsetthetaplanecoords{215}
    \tdplotdrawarc[tdplot_rotated_coords, color=blue, ->, line width = 2pt, dash pattern=on 5pt off 5pt]{(-12.9,-4.5,0)}{2}{0}{24}{below = -7pt, scale=2}{$\theta$}
    
    \end{scope}
	\end{tikzpicture} }}
  \subfloat[Edge-face design\label{1c}]{%
       \tdplotsetmaincoords{-25}{0}
	\scalebox{0.4}{
	\begin{tikzpicture}[tdplot_main_coords]
    
	\pgfmathsetmacro{\cubex}{7}
	\pgfmathsetmacro{\cubey}{14}
	\pgfmathsetmacro{\cubez}{-1}
	
	\pgfmathsetmacro{\patchl}{3}
	\pgfmathsetmacro{\patchw}{0.8}
	\pgfmathsetmacro{\patchs}{1.7}
	
	
	\begin{scope}[rotate around y=50]
	
	\draw[color=darkgreen, fill=darkgray, line width = 3pt] (-2,13.2,-\cubez) -- ++(0,\patchs,0) -- ++(\patchs,0,0) -- ++(0,-\patchs,0) -- cycle;
	\node[scale=3] at (-1, \cubey-1,\cubez) {P$4$};
	
	\draw[color=darkcerulean, line width = 3pt] (0,0,0) -- ++(\cubex,0,0) -- ++(0,\cubey,0) -- ++(-\cubex,0,0) -- cycle;
	\draw[color=darkcerulean, line width = 3pt] (0,0,\cubez) -- ++(0,\cubey,0) -- ++(\cubex,0,0);
	\draw[color=darkcerulean, line width = 3pt, dash pattern=on 10pt off 12pt, opacity=0.4] (0,0,\cubez) -- ++(\cubex,0,0) -- ++(0,\cubey,0);
	\draw[color=darkcerulean, line width = 3pt] (0,0,0) -- ++(0,0,\cubez);
	\draw[color=darkcerulean, line width = 3pt] (0,\cubey,0) -- ++(0,0,\cubez);
	\draw[color=darkcerulean, line width = 3pt] (\cubex,\cubey,0) -- ++(0,0,\cubez);
	\draw[color=darkcerulean, line width = 3pt, dash pattern=on 10pt off 10pt, opacity=0.4] (\cubex,0,0) -- ++(0,0,\cubez);

    \draw[color=darkgreen, fill=applegreen, line width = 3pt] (0,\cubey/2 - \patchl/2,-0.1) -- ++(0,\patchl,0) -- ++(0,0,-\patchw) -- ++(0,-\patchl,0) -- cycle;
    \node[scale=3] at (-2, \cubey/2+0.5,\cubez/2) {P$3$};
    
    \draw[color=darkgreen, fill=applegreen, line width = 3pt] (\cubex/2 - \patchl/2,\cubey,-0.1) -- ++(\patchl,0,0) -- ++(0,0,-\patchw) -- ++(-\patchl,0,0) -- cycle;
    \node[scale=3] at (\cubex/2, \cubey+0.7,\cubez) {P$2$};
    
    \draw[color=darkgreen, fill=darkgray, line width = 3pt] (\cubex,\cubey/2 - \patchl/2,-0.1) -- ++(0,\patchl,0) -- ++(0,0,-\patchw) -- ++(0,-\patchl,0) -- cycle;
    \node[scale=3] at (\cubex+2, \cubey/2-0.5,\cubez/2) {P$1$};
    
    \draw[color=darkgreen, fill=applegreen, line width = 3pt] (\cubex-2.5
    ,0.5,0.4) -- ++(0,\patchs,0) -- ++(\patchs,0,0) -- ++(0,-\patchs,0) -- cycle;
    \node[scale=3] at (\cubex+1,0.5) {P$5$};
    
    \draw[>=latex',  ->, line width  = 2pt] (\cubex/2,\cubey/2,0) -- ++ (2,0,0) node[above left =-6pt and -10pt, scale=2]{$x$};  
    \draw[>=latex',  ->, line width  = 2pt] (\cubex/2,\cubey/2,0) -- ++ (0,2,0) node[above right=-5pt and -15pt, scale=2]{$y$};  
    \draw[>=latex',  ->, line width  = 2pt] (\cubex/2,\cubey/2,0) -- ++ (0,0,2) node[above right=-15pt and -7pt, scale=2]{$z$};  
    
    \draw[>=latex',  ->, line width = 2pt, red] (\cubex/2,\cubey/2,0) -- ++ (-0.87,-0.1,1.8) node{};  
    \draw[dash pattern=on 5pt off 5pt, line width  = 2pt, red] (\cubex/2,\cubey/2,0) -- ++ (-0.87,-0.6,0) --++ (0,0.5,1.8);  
    
    \tdplotdrawarc[color=blue, ->, line width = 2pt, dash pattern=on 5pt off 5pt]{(\cubex/2,\cubey/2,0)}{1}{0}{215}{above left =-35pt and 3pt,color=blue, scale=2}{$\phi$}
    
    \tdplotsetthetaplanecoords{215}
    \tdplotdrawarc[tdplot_rotated_coords, color=blue, ->, line width = 2pt, dash pattern=on 5pt off 5pt]{(-12.9,-4.5,0)}{2}{0}{24}{below = -7pt, scale=2}{$\theta$}
    
    \end{scope}
	\end{tikzpicture} }}
	\caption{Antenna placement designs inspired from \cite{raghavan_antenna_2019} (a) edge design with $3$ ULAs on the device's edges (b) face design with $2$ UPAs on the device's face and back (c) edge-face design with $5$ panels.}
	\label{Fig:Designs}
\end{figure}

We consider a point to point communication between a fixed AP and a mobile UT in an indoor environment. The AP uses a uniform planar array (UPA) with array size of $\{N_\mathrm{AP_x}, N_\mathrm{AP_y}, N_\mathrm{AP_z}\}$. The UT is made of $N_{P}$ panels, where the $p$th panel has an uniform linear array (ULA) or UPA of size $\{N_\mathrm{UT_x}^{(p)}, N_\mathrm{UT_y}^{(p)}, N_\mathrm{UT_z}^{(p)}\}$. Thus, the AP, the $p$th UT panel, and the UT include $N_\mathrm{AP} = N_\mathrm{AP_x} N_\mathrm{AP_y} N_\mathrm{AP_z}$, $N_\mathrm{UT}^{(p)} = N_\mathrm{UT_x}^{(p)} N_\mathrm{UT_y}^{(p)} N_\mathrm{UT_z}^{(p)}$, and $N_\mathrm{UT} = \sum_{p=1}^{N_{P}} N_\mathrm{UT}^{(p)}$ antenna elements, respectively. Inspired by \cite{raghavan_antenna_2019}, we consider the edge (E), face (F), and edge-face (EF) antenna placement designs, respectively, with $3$, $2$, and $5$ panels at UTs, which are illustrated in Fig.~\ref{Fig:Designs}.

A global coordinate system (GCS) is considered for the environment and we define the position and orientation of the transceivers in the GCS \cite{ali_orientation-assisted_2021}. In addition, we consider an LCS for each transceiver. The LCSs are defined in a way that the $yz$ and $xy$ plane of the AP and UT LCSs are alligned parallel to the AP's UPA and UT's screen, respectively. $\boldsymbol{p}_\mathrm{AP} = (x_\mathrm{AP}, y_\mathrm{AP}, z_\mathrm{AP}) \in \mathbb{R}^3$ and $\boldsymbol{p}_\mathrm{UT} = (x_\mathrm{UT}, y_\mathrm{UT}, z_\mathrm{UT}) \in \mathbb{R}^3$ denote the positions of the AP and UT in the GCS. The AP and UT LCSs are rotated about $z$, $y$, and $x$ axes of the GCS by angles $\boldsymbol{\psi}_\mathrm{AP} = (\alpha_\mathrm{AP}, \beta_\mathrm{AP}, \gamma_\mathrm{AP})$ and $\boldsymbol{\psi}_\mathrm{UT} = (\alpha_\mathrm{UT}, \beta_\mathrm{UT}, \gamma_\mathrm{UT})$, respectively \cite{rezaie_deep_2022}.

\subsubsection{Channel Model}
We consider a narrow band channel between the AP and the $p$th UT panel, $\boldsymbol{H}^{(p)} \in \mathbb{C}^{N_\mathrm{UT}^{(p)} \times N_\mathrm{AP}}$. The channel considering the contribution of $L^{(p)}$ paths may be modeled as
\begin{equation}\label{Eq:Channel}
    \boldsymbol{H}^{(p)} = \sum_{l=0}^{L^{(p)}} \sqrt{\rho_l^{(p)}} \hspace{2pt} e^{j\vartheta_l^{(p)}} \hspace{2pt} \boldsymbol{a}_\mathrm{UT}^{(p)}(\phi^{(p)}_{l}, \theta^{(p)}_{l}) \hspace{2pt} \boldsymbol{a}_\mathrm{AP}^H(\psi^{(p)}_{l},\omega^{(p)}_{l})
\end{equation}
where $\rho_l^{(p)}$ and $\vartheta_l^{(p)}$ are, respectively, the power and phase of the $l$th path. The azimuth and elevation angle of departure (AoD) of the $l$th path in the AP LCS are $\phi^{(p)}_{l}$ and $\theta^{(p)}_{l}$, respectively. $\psi^{(p)}_{l}$ and $\omega^{(p)}_{l}$ denote, analogously, the azimuth and elevation AoA for the $l$th path in the UT LCS. $\boldsymbol{a}_\mathrm{UT}^{(p)}$ and $\boldsymbol{a}_\mathrm{AP}$ respectively stands for the antenna array response of the $p$th UT panel and the AP. Considering an antenna array of size $\{N_\mathrm{x}, N_\mathrm{y}, N_\mathrm{z}\}$, the antenna array response can be defined as
\begin{equation}
    \boldsymbol{a}(\phi, \theta) = \frac{1}{\sqrt{N_\mathrm{a}}} g_\mathrm{a}(\phi, \theta) \boldsymbol{a}_\mathrm{z}(\theta) \otimes \boldsymbol{a}_\mathrm{y}(\phi, \theta) \otimes \boldsymbol{a}_\mathrm{x}(\phi, \theta)
\end{equation}
where $N_\mathrm{a} = N_\mathrm{x} N_\mathrm{y} N_\mathrm{z}$. At azimuth and elevation angles $\phi$ and $\theta$, $g_\mathrm{a}(\phi, \theta)$ denotes the antenna gain of an antenna element\footnote{We assume all antennas have the same radiation pattern.}. Also, $\boldsymbol{a}_\mathrm{x} \in \mathbb{C}^{N_\mathrm{x} \times 1}$, $\boldsymbol{a}_\mathrm{y} \in \mathbb{C}^{N_\mathrm{y} \times 1}$, and $\boldsymbol{a}_\mathrm{z} \in \mathbb{C}^{N_\mathrm{z} \times 1}$ are defined as
\begin{equation}\label{eq:a_x}
     \boldsymbol{a}_\mathrm{x}(\phi, \theta) =  [1 , e^{j \pi sin(\theta) \hspace{2pt} cos(\phi) }, \dots 
     , e^{ j \pi (N_\mathrm{x} - 1) \hspace{2pt} sin(\theta) \hspace{2pt} cos(\phi) }]^T,
\end{equation}
\begin{equation}\label{eq:a_y}
     \boldsymbol{a}_\mathrm{y}(\phi, \theta) =  [1 , e^{j \pi sin(\theta) \hspace{2pt} sin(\phi) }, \dots 
     , e^{ j \pi (N_\mathrm{y} - 1) \hspace{2pt} sin(\theta) \hspace{2pt} sin(\phi) }]^T,
\end{equation}
\begin{equation}\label{eq:a_z}
     \boldsymbol{a}_\mathrm{z}(\theta) =  [1 , e^{j \pi cos(\theta)}, \dots 
     , e^{ j \pi (N_\mathrm{z} - 1) \hspace{2pt} cos(\theta) }]^T.
\end{equation}
Considering the AP transmits the unit power symbol $s$ with transmission power $P_\mathrm{AP}$, the received signal at the $p$th UT panel is
\begin{equation}\label{Y_Rec_Sig}
    y^{(p)}= \sqrt{P_\mathrm{AP}} {\boldsymbol{v}^{(p)}}^H \boldsymbol{H}^{(p)} \boldsymbol{u} s + {\boldsymbol{v}^{(p)}}^H \boldsymbol{n}^{(p)}
\end{equation}
where $\boldsymbol{u}$ and $\boldsymbol{v}^{(p)}$ respectively denote the beamforming vector at the AP and $p$th UT panel. Also, $\boldsymbol{n}^{(p)}\in \mathbb{C}^{N_\mathrm{UT}^{(p)}}$ denotes a complex Gaussian noise vector with zero mean and variance $\sigma_n^2$. 

\subsubsection{Codebook Definition}
We consider an analog phased array design for beamforming at the AP and each UT panel. Both the AP and UT have one RF chain, and the UT RF chain is connected to one of the panels.
For simplicity, multi-panel beamforming is not considered in this paper. We use discrete Fourier transform (DFT)-based analog codebook for each panel \cite{rezaie_location-_2020}. All the available beamforming vectors at the AP are included in the set $\boldsymbol{\mathcal{U}} = \{\boldsymbol{u}_1, \dots, \boldsymbol{u}_{N_\mathrm{AP}}\}$. $\boldsymbol{\mathcal{V}}^{p} = \{\boldsymbol{v}^{p}_{1}, \dots, \boldsymbol{v}^{p}_{{N^{(p)}_\mathrm{UT}}}\}$ includes all the beams for the $p$th UT panel. The beam codebook of the UT, $\boldsymbol{\mathcal{V}} = \{\boldsymbol{v}_1, \dots, \boldsymbol{v}_{N_\mathrm{UT}}\}$, can be expressed as union of all the panel codebooks. Using the beamforming vectors $\boldsymbol{u}_i$ and $\boldsymbol{v}_j$ at the AP and UT, the RSS at the UT is
\begin{equation}\label{Eq:RSS}
    R_{i, j} = \Big | \sqrt{P_\mathrm{AP}} \boldsymbol{v}_j^H \boldsymbol{H}^{(p_j)} \boldsymbol{u}_i s + \boldsymbol{v}_j^H \boldsymbol{n}  \Big | ^2
\end{equation}
where $p_j$ is the corresponding panel of combiner $\boldsymbol{v}_j$. We define set $\mathcal{B}$, including the indices of all beam pairs in the AP and UT codebooks.
\subsection{Deep Learning based Beam Selection}
The beam candidate list, $\mathcal{S}$, is the key component of the beam selection approach, which reduces the search space over all the possible combinations of AP and UT beams. As mentioned in Section~\ref{Sec:DevAgn}\cite{va_inverse_2018}, the optimal candidate list for a known UT position and orientation includes the beam pair indices that provides the highest likelihood of optimality. The optimality probability for the $(i, j)$th beam pair is $P_{i,j}^{B} = \mathbb{P} \left[ (i,j) = (i^\star, j^\star) \right]$, where $(i^\star, j^\star)$ is the optimal AP/UT beam pair for communication, i.e., 
\begin{equation} \label{eq:i_j_max}
    (i^\star, j^\star) = \underset{(i, j) \in\mathcal{B}}{\mathrm{arg\hspace{2pt}max}} \hspace{2pt} R_{i, j}.
\end{equation}

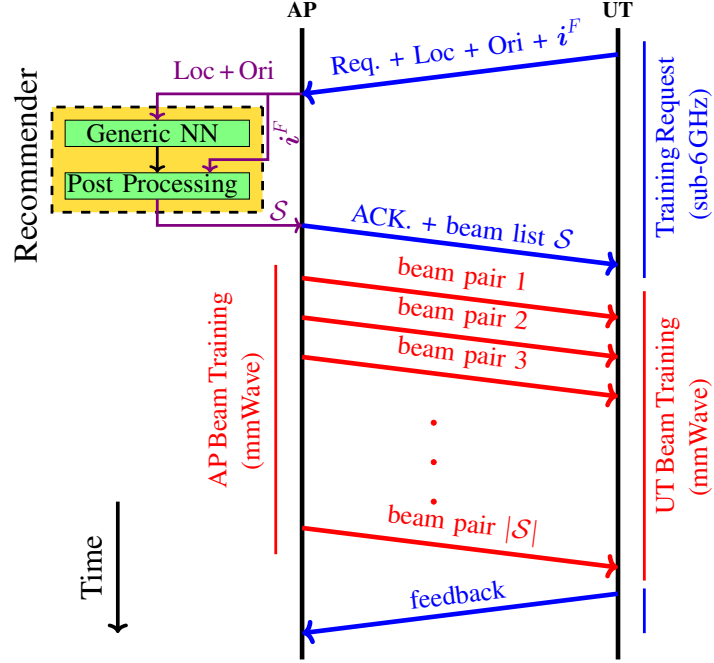
\begin{figure}
    \centering
    \scalebox{0.35}{
    \begin{tikzpicture}
        \draw[line width=5pt] (0, 20) node[above = 5pt ]{\Huge \textbf{AP}} -- (0, -4);
        \draw[line width=5pt] (12, 20) node[above = 5pt ]{\Huge \textbf{UT}} -- (12, -4);
        \draw[->, line width=5pt, color=blue] (12, 19) -- (0, 17.5) node[midway, sloped, above = 4pt]{\scalebox{1.2}{\Huge Req. + Loc + Ori + $\boldsymbol{i}^{F}$}};
        
        \draw[->, line width=5pt, color=blue] (0, 12.5) -- (12, 11) node[midway, sloped, above = 4pt]{\scalebox{1.2}{\Huge ACK. + beam list $\mathcal{S}$}};
        
        \draw[->, line width=5pt, color=red] (0, 10.5) -- (12, 9) node[midway, sloped, above = 4pt]{\scalebox{1.2}{\Huge beam pair 1}};
        
        \draw[->, line width=5pt, color=red] (0, 9) -- (12, 7.5) node[midway, sloped, above = 4pt]{\scalebox{1.2}{\Huge beam pair 2}};
        
        \draw[->, line width=5pt, color=red] (0, 7.5) -- (12, 6) node[midway, sloped, above = 4pt]{\scalebox{1.2}{\Huge beam pair 3}};
        
        \draw[red,fill=red] (5,5) circle (.5ex);
        
        \draw[red,fill=red] (5,3.5) circle (.5ex);
        
        \draw[red,fill=red] (5,2) circle (.5ex);

        \draw[->, line width=5pt, color=red] (0, 1) -- (12, -0.5) node[midway, sloped, above = 4pt]{\scalebox{1.2}{\Huge beam pair $|\mathcal{S}|$}};
        
        \draw[->, line width=5pt, color=blue] (12, -1.5) -- (0, -3) node[midway, sloped, above = 4pt]{\scalebox{1.2}{\Huge feedback}};
        
        \draw [decorate,xshift=-1cm,yshift=0pt, line width=3pt, color=red]
        (0, 0) -- (0, 11) node [red,rotate=90, midway,yshift=1.5cm, text width=7cm] 
        {\scalebox{1.2}{\Huge \hspace{0.3cm} AP Beam Training} \\[10pt] \scalebox{1.2}{\Huge \hspace{1cm}(mmWave)}};
        
        \draw [decorate,xshift=1cm,yshift=0pt, line width=3pt, color=red]
        (12, -1) -- (12, 10) node [red,rotate=90, midway,yshift=-1.5cm, text width=7cm] 
        {\scalebox{1.2}{\Huge \hspace{0.3cm} UT Beam Training} \\[10pt] \scalebox{1.2}{\Huge \hspace{1cm}(mmWave)}};
        
        \draw [decorate,xshift=1cm,yshift=0pt, line width=3pt, color=blue]
        (12, 19.5) -- (12, 10.5) node [blue,rotate=90, midway,yshift=-1.5cm, text width=7cm] 
        {\scalebox{1.2}{\Huge \hspace{0.3cm} Training Request} \\[10pt] \scalebox{1.2}{\Huge \hspace{1cm}(sub-6 GHz)}};
        
        \draw [decorate,xshift=1cm,yshift=0pt, line width=3pt, color=blue]
        (12, -1.3) -- (12, -3);
        
        \draw[draw=black, dash pattern=on 10pt off 10pt, line width=3pt, fill=Goldenrod] (-1.5,17) rectangle ++(-8,-4) node[rotate=90, midway,yshift=5cm] {\scalebox{1.5}{\Huge Recommender }};
        
        \draw[draw=black, fill=green!50] (-2,16.5) rectangle ++(-7,-1) node[pos=.5] {\scalebox{1.2}{\Huge Generic NN }};
        \draw[draw=black, fill=green!50] (-2,14.5) rectangle ++(-7,-1) node[pos=.5] {\scalebox{1.2}{\Huge Post Processing }};
        \draw[draw=black, ->, line width=3pt] (-5.5, 15.5) -- (-5.5, 14.5);
        \draw[violet, ->, line width=3pt] (0, 17.5) node [above left =10pt and -80pt, text width=7.5cm] {\scalebox{1.2}{\Huge Loc + Ori}} -- (-5.5, 17.5) -- (-5.5, 16.5);
        \draw[decorate, violet, ->, line width=3pt] (-1.3, 17.5) -- (-1.3, 15) node [rotate=90, midway,xshift=3cm,yshift=-0.7cm, text width=7.5cm] {\scalebox{1.2}{\Huge $\boldsymbol{i}^{F}$}} -- (-3.5, 15) -- (-3.5, 14.5);
        \draw[violet, ->, line width=3pt] (-5.5, 13.5) -- (-5.5, 12.5) node [above right=0pt and 70pt, text width=-3.5cm] {\scalebox{1.2}{\Huge $\mathcal{S}$}} -- (0, 12.5);
        
        \draw[->, line width=4pt] (-7, 2) -- (-7, -3) node [rotate=90, midway,yshift=1cm]{\scalebox{1.4}{\Huge Time}};
    \end{tikzpicture}
    }
    \caption{The device agnostic DL-based mmWave beam training procedure using location and orientation information.}
    \label{fig:Sensing}
\end{figure}
Taking advantage of the device-agnostic framework, we propose a generic context-aware beam selection procedure that uses the location and orientation of the UT to recommend a beam candidate list. Fig.~\ref{fig:Sensing} shows the proposed device/codebook-agnostic beam selection procedure where the connection request and feedback are communicated in sub-6 GHz links. A generic model processes the location and orientation information to predict the optimal AP beam - UT direction for establishing a mmWave link. A post processing framework is needed to map the directions in the UT LCS to the UT beams. 

\begin{figure}[t]
	\centering
	\subfloat[ $\mathit{NET}_{\mathrm{I}}$\label{I}]{%
    \centering
	\definecolor{mycolor1}{RGB}{224,9,2} 
	\definecolor{mycolor2}{RGB}{255,174,66} 
	\definecolor{mycolor3}{RGB}{0,120,210} 
	\definecolor{mycolor4}{RGB}{0, 128, 0} 
	\definecolor{mycolor5}{RGB}{0,191,255} 
	\definecolor{mycolor6}{RGB}{0,0,139} 
	\definecolor{mycolor7}{RGB}{255,0,255} 
	\definecolor{mycolor8}{RGB}{238,130,238} 
	\definecolor{mycolor9}{RGB}{128,0,128} 
	\scalebox{.78}{
		\begin{tikzpicture}[cross/.style={path picture={ 
				\draw[red, line width=3pt]
				(path picture bounding box.south east) -- (path picture bounding box.north west) (path picture bounding box.south west) -- (path picture bounding box.north east);
		}}]
		\foreach \i in {-1,...,1} {
			\foreach \h in {0,...,-1} {
				\draw[black, ->-=0.95, line width=1pt](0.15 , 0.5*\i) -- (1.75 ,0.6 * \h - 1.7);
				\draw[black, ->-=0.95, line width=1pt](0.15 , 0.5*\i) -- (1.75 ,-0.6 * \h + 1.7);
			}
		}
		
		\foreach \i in {0,...,-1} {
			\foreach \h in {-1,...,1} {
				\draw[black, line width=1pt](2.2 ,0.6 * \i - 1.7) -- (3.8-0.5 , 1.3 * \h  - 0.5/1.6*1.3*\h +0.5/1.6*1.3*\i);
				\draw[black, line width=1pt](2.2 ,-0.6 * \i + 1.7) -- (3.8-0.5 , 1.3 * \h  - 0.5/1.6*1.3*\h -0.5/1.6*1.3*\i);
			}
		}

		\foreach \i in {-1,...,1} {
			\foreach \h in {0,...,-1} {
				\draw[black, ->-=0.95, line width=1pt](4.2 +0.5,1.3 * \i + 0.5*1.3*\h/1.6 - 0.5*1.3*\i/1.6) -- (5.75 ,0.6 * \h - 1.7);
				\draw[black, ->-=0.95, line width=1pt](4.2 + 0.5,1.3 * \i - 0.5*1.3*\h/1.6 - 0.5*1.3*\i/1.6) -- (5.75 ,-0.6 * \h + 1.7);
			}
		}

		\foreach \i in {0,...,-1} {
			\foreach \h in {-1,...,-1} {
                \draw[black, ->-=0.95, line width=1pt](6.25 ,0.6 * \i - 1.7) -- (7.85 ,1.3 * \h- 0.35);
                \draw[black, ->-=0.95, line width=1pt](6.25 ,-0.6 * \i + 1.7) -- (7.85 ,1.3 * \h- 0.35);
			}
		}
		
		\foreach \i in {0,...,-1} {
			\foreach \h in {0,...,1} {
				\draw[black, ->-=0.95, line width=1pt](6.25 ,0.6 * \i - 1.7) -- (7.85 ,0.7 * \h+ 0.95);
				\draw[black, ->-=0.95, line width=1pt](6.25 ,-0.6 * \i + 1.7) -- (7.85 ,0.7 * \h+ 0.95);
			}
		}
		
		\path (3.8,0) -- (4.2,0) node [black, font=\Huge, midway, sloped] {\textbf{$...$}};
		
		\foreach \i in {-1,...,1} {
    		\draw[fill=mycolor1] (0,0.5*\i) circle (0.2);  
    	}
		
		\draw[fill=mycolor2] (2,-2.3) circle (0.2);  
		\draw[fill=mycolor2] (2,-1.7) circle (0.2);  
		\path (2,-1.7) -- (2,0) node [black, font=\Huge, midway, sloped] {$...$};
		\path (2,0) -- (2,1.7) node [black, font=\Huge, midway, sloped] {$...$};
		\draw[fill=mycolor2] (2,1.7) circle (0.2);  
		\draw[fill=mycolor2] (2,2.3) circle (0.2);  
		
		\coordinate (dm1) at (2,-2.4);
        \coordinate (dm2) at (2,2.4);
        \node[rectangle, draw, mycolor5, very thick, minimum width=0.5cm] [fit = (dm1) (dm2)] (bx4) {};
		\node[align=center,font=\large,rotate=90] at (bx4.center) {};
		
		\draw[fill=mycolor2] (6,-2.3) circle (0.2);  
		\draw[fill=mycolor2] (6,-1.7) circle (0.2);  
		\path (6,-1.7) -- (6,0) node [black, font=\Huge, midway, sloped] {$...$};
		\path (6,0) -- (6,1.7) node [black, font=\Huge, midway, sloped] {$...$};
		\draw[fill=mycolor2] (6,1.7) circle (0.2);  
		\draw[fill=mycolor2] (6,2.3) circle (0.2);  
		
		\coordinate (dm1) at (6,-2.4);
        \coordinate (dm2) at (6,2.4);
        \node[rectangle, draw, mycolor5, very thick, minimum width=0.5cm] [fit = (dm1) (dm2)] (bx4) {};
		\node[align=center,font=\large,rotate=90] at (bx4.center) {};
		
		\draw[fill=mycolor4] (8,-1.65) circle (0.2);  
		\path (8,-1.65) -- (8,0.95) node [black, font=\Huge, midway, sloped] {$...$};
		\draw[fill=mycolor4] (8,0.95) circle (0.2);  
		\draw[fill=mycolor4] (8,1.65) circle (0.2);  
		
		\draw[->, line width=1pt] (-1,-0.5*1) node [font=\Large, left] {$z_\mathrm{UT}$} -- (-0.2,-0.5*1);
		\draw[->, line width=1pt] (-1,0) node [font=\Large, left] {$y_\mathrm{UT}$} -- (-0.2,0);
		\draw[->, line width=1pt] (-1,0.5*1) node [font=\Large, left] {$x_\mathrm{UT}$} -- (-0.2,0.5*1);

		\draw[->, line width=1pt, mycolor4] (8.15,-1.65) -- (9,-1.65) node [font=\Large, right] {$P_{N_\mathrm{AP}}$};
		\draw[->, line width=1pt, mycolor4] (8.15,0.95) -- (9,0.95) node [font=\Large, right] {$P_{2}$};
		\draw[->, line width=1pt, mycolor4] (8.15,1.65) -- (9,1.65) node [font=\Large, right] {$P_{1}$};
		
		\coordinate (dm11) at (1,2);
        \coordinate (dm12) at (1,-2);
		\node[rectangle, fill=gray!50, minimum width=0.7cm, opacity=.8] [fit = (dm11) (dm12)] (bx4) {};
		\node[align=center,font=\large,rotate=90] at (bx4.center) {Dense};
		\coordinate (dm11) at (2.8,2);
        \coordinate (dm12) at (2.8,-2);
		\node[rectangle, fill=gray!50, minimum width=0.7cm, opacity=.8] [fit = (dm11) (dm12)] (bx4) {};
		\node[align=center,font=\large,rotate=90] at (bx4.center) {Dense};
		\coordinate (dm11) at (5.2,2);
        \coordinate (dm12) at (5.2,-2);
		\node[rectangle, fill=gray!50, minimum width=0.7cm, opacity=.8] [fit = (dm11) (dm12)] (bx4) {};
		\node[align=center,font=\large,rotate=90] at (bx4.center) {Dense};
		\coordinate (dm11) at (7,2);
        \coordinate (dm12) at (7,-2);
		\node[rectangle, fill=gray!50, minimum width=0.7cm, opacity=.8] [fit = (dm11) (dm12)] (bx4) {};
		\node[align=center,font=\large,rotate=90] at (bx4.center) {Dense};
		\end{tikzpicture}}}
		\\[0.05cm]
	\hspace{-1cm}
  \subfloat[$\mathit{NET}_{\mathrm{II}}$\label{II}]{%
    \centering
	\definecolor{mycolor1}{RGB}{224,9,2} 
	\definecolor{mycolor2}{RGB}{255,174,66} 
	\definecolor{mycolor3}{RGB}{0,120,210} 
	\definecolor{mycolor4}{RGB}{0, 128, 0} 
	\definecolor{mycolor5}{RGB}{0,191,255} 
	\definecolor{mycolor6}{RGB}{0,0,139} 
	\definecolor{mycolor7}{RGB}{255,0,255} 
	\definecolor{mycolor8}{RGB}{238,130,238} 
	\definecolor{mycolor9}{RGB}{128,0,128} 
	\scalebox{.78}{
		\begin{tikzpicture}[cross/.style={path picture={ 
				\draw[red, line width=3pt]
				(path picture bounding box.south east) -- (path picture bounding box.north west) (path picture bounding box.south west) -- (path picture bounding box.north east);
		}}]
		\foreach \i in {-1,...,4} {
			\foreach \h in {0,...,-1} {
				\draw[black, ->-=0.95, line width=1pt](0.15 , 0.5*\i+0.7) -- (1.75 ,-0.6 * \h + 1.7);
			}
		}
		\foreach \i in {-1,...,4} {
				\draw[black, ->-=0.95, line width=1pt](0.15 , 0.5*\i+0.7) -- (1.75 , 0.4);
		}
		\draw[black, ->-=0.95, line width=1pt](0.15, -0.5*4+0.7) -- (1.75, -0.5*4+0.7);

		\foreach \i in {0,...,-1} {
			\foreach \h in {-1,...,1} {
				\draw[black, line width=1pt](2.2+2 ,0.6 * \i - 1.55) -- (3.8-0.5+2 , 1.3 * \h  - 0.5/1.6*1.3*\h +0.5/1.6*1.3*\i);
				\draw[black, line width=1pt](2.2+2 ,-0.6 * \i + 1.55) -- (3.8-0.5+2 , 1.3 * \h  - 0.5/1.6*1.3*\h -0.5/1.6*1.3*\i);
			}
		}
		
		\foreach \h in {-1,...,1} {
			\draw[black, line width=1pt](2.2+2 ,0.3) -- (3.8-0.5+2 , 1.3 * \h  - 0.5/1.6*1.3*\h +0.5/1.6*1.3*0);
			\draw[black, line width=1pt](2.2+2 ,-0.3) -- (3.8-0.5+2 , 1.3 * \h  - 0.5/1.6*1.3*\h -0.5/1.6*1.3*0);
		}

		\foreach \i in {-1,...,1} {
			\foreach \h in {0,...,-1} {
				\draw[black, ->-=0.95, line width=1pt](4.2+2 +0.5,1.3 * \i + 0.5*1.3*\h/1.6 - 0.5*1.3*\i/1.6) -- (5.75+2 ,0.6 * \h - 1.55);
				\draw[black, ->-=0.95, line width=1pt](4.2 + 0.5+2,1.3 * \i - 0.5*1.3*\h/1.6 - 0.5*1.3*\i/1.6) -- (5.75+2 ,-0.6 * \h + 1.55);
			}
		}

		\foreach \i in {0,...,-1} {
			\foreach \h in {-1,...,-1} {
                \draw[black, ->-=0.95, line width=1pt](6.25+2 ,0.6 * \i - 1.55) -- (7.85+2 ,1 * \h);
                \draw[black, ->-=0.95, line width=1pt](6.25+2 ,-0.6 * \i + 1.55) -- (7.85+2 ,1 * \h);
			}
		}
		
		\foreach \i in {0,...,-1} {
			\foreach \h in {0,...,1} {
				\draw[black, ->-=0.95, line width=1pt](6.25+2 ,0.6 * \i - 1.55) -- (7.85+2 ,0.7 * \h+ 0.3);
				\draw[black, ->-=0.95, line width=1pt](6.25+2 ,-0.6 * \i + 1.55) -- (7.85+2 ,0.7 * \h+ 0.3);
			}
		}
		
		\path (3.8+2,0) -- (4.2+2,0) node [black, font=\Huge, midway, sloped] {\textbf{$...$}};
		
		\foreach \i in {-1,...,4} {
    		\draw[fill=mycolor1] (0,0.5*\i+0.7) circle (0.2);  
    	}
    	\draw[fill=mycolor1] (0,-0.5*4+0.7) circle (0.2);  
		
		\draw[fill=mycolor7] (2,-2.3) circle (0.2);  
		\draw[fill=mycolor7] (2,-1.7) circle (0.2);  
		\path (2,-1.7) -- (2,-0.4) node [black, font=\Huge, midway, sloped] {$...$};
		\draw[fill=mycolor7] (2,-0.4) circle (0.2);  
		\draw[fill=mycolor2] (2,0.4) circle (0.2);  
		\path (2,0.4) -- (2,1.7) node [black, font=\Huge, midway, sloped] {$...$};
		\draw[fill=mycolor2] (2,1.7) circle (0.2);  
		\draw[fill=mycolor2] (2,2.3) circle (0.2);  
		
		\coordinate (dm1) at (2,-.25);
        \coordinate (dm2) at (2,-2.4);
        \node[rectangle, draw, mycolor3, very thick, minimum width=0.5cm] [fit = (dm1) (dm2)] (bx4) {};
		\node[align=center,font=\large,rotate=90] at (bx4.center) {};
		
		\coordinate (dm1) at (2,0.25);
        \coordinate (dm2) at (2,2.4);
        \node[rectangle, draw, mycolor5, very thick, minimum width=0.5cm] [fit = (dm1) (dm2)] (bx4) {};
		\node[align=center,font=\large,rotate=90] at (bx4.center) {};
		
		\draw[fill=mycolor7] (2+2,-2.15) circle (0.2);  
		\draw[fill=mycolor7] (2+2,-1.55) circle (0.2);  
		\path (2+2,-1.55) -- (2+2,-0.25) node [black, font=\Huge, midway, sloped] {$...$};
		\draw[fill=mycolor7] (2+2,-0.25) circle (0.2);  
		\draw[fill=mycolor2] (2+2,0.25) circle (0.2);  
		\path (2+2,0.25) -- (2+2,1.55) node [black, font=\Huge, midway, sloped] {$...$};
		\draw[fill=mycolor2] (2+2,1.55) circle (0.2);  
		\draw[fill=mycolor2] (2+2,2.15) circle (0.2);  
		
		\coordinate (dm1) at (2+2,-0.1);
        \coordinate (dm2) at (2+2,-2.25);
        \node[rectangle, draw, mycolor3, very thick, minimum width=0.5cm] [fit = (dm1) (dm2)] (bx4) {};
		\node[align=center,font=\large,rotate=90] at (bx4.center) {};
		
		\coordinate (dm1) at (2+2,0.1);
        \coordinate (dm2) at (2+2,2.25);
        \node[rectangle, draw, mycolor5, very thick, minimum width=0.5cm] [fit = (dm1) (dm2)] (bx4) {};
		\node[align=center,font=\large,rotate=90] at (bx4.center) {};
		
		\coordinate (A) at (2.25,1.2);
		\coordinate (B) at (2.25,-1.2);
        \coordinate (C) at (3.75,0);
		\draw[line width=0.5mm, dotted] (A) to[out=0,in=180] (C);
		\draw[line width=0.5mm, dotted] (B) to[out=0,in=180] (C);
		
		\draw[fill=mycolor2] (6+2,-2.15) circle (0.2);  
		\draw[fill=mycolor2] (6+2,-1.55) circle (0.2);  
		\path (6+2,-1.55) -- (6+2,0) node [black, font=\Huge, midway, sloped] {$...$};
		\path (6+2,0) -- (6+2,1.55) node [black, font=\Huge, midway, sloped] {$...$};
		\draw[fill=mycolor2] (6+2,1.55) circle (0.2);  
		\draw[fill=mycolor2] (6+2,2.15) circle (0.2);  
		
		\coordinate (dm1) at (6+2,-2.25);
        \coordinate (dm2) at (6+2,2.25);
        \node[rectangle, draw, mycolor5, very thick, minimum width=0.5cm] [fit = (dm1) (dm2)] (bx4) {};
		\node[align=center,font=\large,rotate=90] at (bx4.center) {};
		
		\draw[fill=mycolor4] (8+2,-1) circle (0.2);  
		\path (8+2,-1.65) -- (8+2,0.95) node [black, font=\Huge, midway, sloped] {$...$};
		\draw[fill=mycolor4] (8+2,0.3) circle (0.2);  
		\draw[fill=mycolor4] (8+2,1) circle (0.2);  

		\draw[->, line width=1pt] (-1,-0.5*4+0.7) node [font=\Large, above left = 0pt and 0pt] {AP beam} -- (-0.2,-0.5*4+0.7);
		\draw[->, line width=1pt] (-1,-0.5*4+0.7) node [font=\Large, below left = 0pt and 0pt] {index $i$} -- (-0.2,-0.5*4+0.7);
		\draw[->, line width=1pt] (-1,-0.5*1+0.7) node [font=\Large, left] {$\gamma_\mathrm{UT}$} -- (-0.2,-0.5*1+0.7);
		\draw[->, line width=1pt] (-1,-0.5*0+0.7) node [font=\Large, left] {$\beta_\mathrm{UT}$} -- (-0.2,-0.5*0+0.7);
		\draw[->, line width=1pt] (-1,0.5*1+0.7) node [font=\Large, left] {$\alpha_\mathrm{UT}$} -- (-0.2,0.5*1+0.7);
		\draw[->, line width=1pt] (-1,0.5*2+0.7) node [font=\Large, left] {$z_\mathrm{UT}$} -- (-0.2,0.5*2+0.7);
		\draw[->, line width=1pt] (-1,0.5*3+0.7) node [font=\Large, left] {$y_\mathrm{UT}$} -- (-0.2,0.5*3+0.7);
		\draw[->, line width=1pt] (-1,0.5*4+0.7) node [font=\Large, left] {$x_\mathrm{UT}$} -- (-0.2,0.5*4+0.7);

		\draw[->, line width=1pt, mycolor4] (8.15+2,-1) -- (9+2,-1) node [font=\Large, right] {$P_{N_{G}|i}$};
		\draw[->, line width=1pt, mycolor4] (8.15+2,0.3) -- (9+2,0.3) node [font=\Large,right] {$P_{2|i}$};
		\draw[->, line width=1pt, mycolor4] (8.15+2,1) -- (9+2,1) node [font=\Large, right] {$P_{1|i}$};
		
		\coordinate (dm11) at (1,-2.4);
        \coordinate (dm12) at (1,-0.3);
		\node[rectangle, fill=gray!50, minimum width=0.7cm, opacity=.8] [fit = (dm11) (dm12)] (bx4) {};
		\node[align=center,font=\large,rotate=90] at (bx4.center) {Embedding};
		\coordinate (dm11) at (1,2.5);
        \coordinate (dm12) at (1,0.3);
		\node[rectangle, fill=gray!50, minimum width=0.7cm, opacity=.8] [fit = (dm11) (dm12)] (bx4) {};
		\node[align=center,font=\large,rotate=90] at (bx4.center) {Dense};
		\coordinate (dm11) at (3,1.5);
        \coordinate (dm12) at (3,-1.5);
		\node[rectangle, fill=gray!50, minimum width=0.7cm, opacity=.8] [fit = (dm11) (dm12)] (bx4) {};
		\node[align=center,font=\large,rotate=90] at (bx4.center) {Concatenate};
		\coordinate (dm11) at (4.8,2);
        \coordinate (dm12) at (4.8,-2);
		\node[rectangle, fill=gray!50, minimum width=0.7cm, opacity=.8] [fit = (dm11) (dm12)] (bx4) {};
		\node[align=center,font=\large,rotate=90] at (bx4.center) {Dense};
		\coordinate (dm11) at (7.2,2);
        \coordinate (dm12) at (7.2,-2);
		\node[rectangle, fill=gray!50, minimum width=0.7cm, opacity=.8] [fit = (dm11) (dm12)] (bx4) {};
		\node[align=center,font=\large,rotate=90] at (bx4.center) {Dense};
		\coordinate (dm11) at (9,2);
        \coordinate (dm12) at (9,-2);
		\node[rectangle, fill=gray!50, minimum width=0.7cm, opacity=.8] [fit = (dm11) (dm12)] (bx4) {};
		\node[align=center,font=\large,rotate=90] at (bx4.center) {Dense};
		\end{tikzpicture}}}
	\caption{Multi-network structure of the proposed generic neural network predicting the optimality probability for AP beam-UT directions, including (a) $\mathit{NET}_{\mathrm{I}}$ used to select AP beam based on UE location, (b) $\mathit{NET}_{\mathrm{II}}$ used to select UE beamforming direction based on AP beam and UE location and orientation.}
	\label{Fig:GenericNN}
\end{figure}
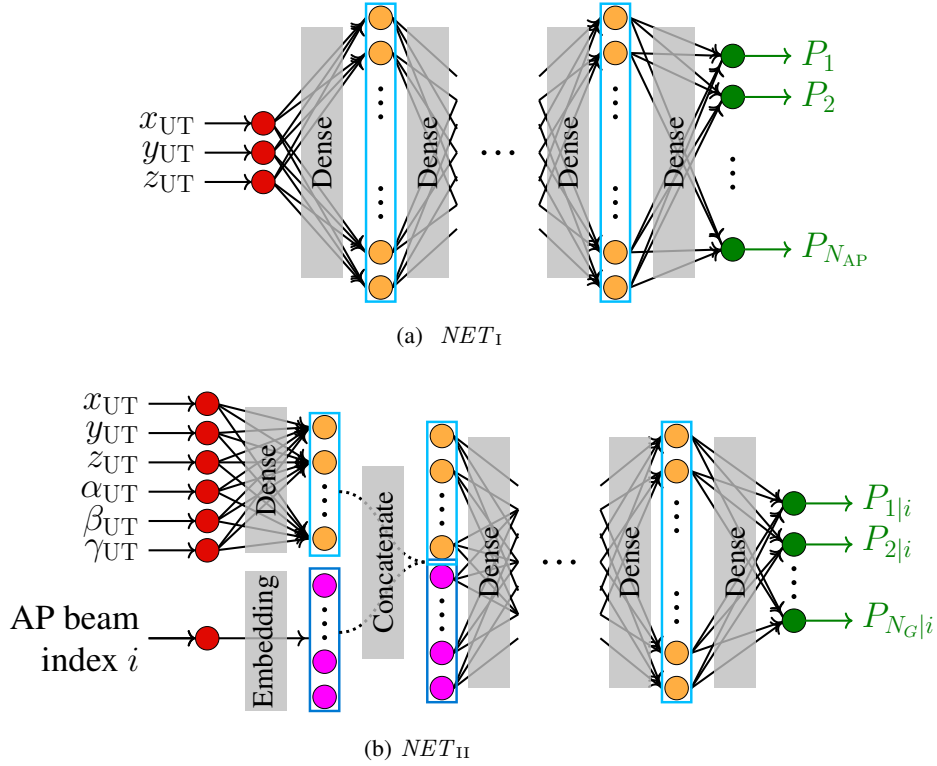
Inspired by our previous work in \cite{rezaie_location-_2022}, we propose a multi-network structure that can predict proper directions of transmission covering the LOS or strong NLOS paths. Fig.~\ref{Fig:GenericNN} shows our proposed design for the generic NN, which is able to point directions instead of the UT beams. The first network, $\mathit{NET}_{\mathrm{I}}$, predicts the optimality probabilities of the AP beams, i.e.,
\begin{equation}
    P_{i} = \mathbb{P} \left[ i = i^\star \right]
\end{equation}
Subsequently, $\mathit{NET}_{\mathrm{II}}$ predicts the conditional probabilities of optimality given the AP beam index for the directions given by the FG in the UT LCS as
\begin{equation}
    P_{k|i}^{D} = \mathbb{P} \left[ k = k^\star | i = i^\star \right]
\end{equation}
where $k^\star$ denotes the optimal direction of transmission in the UT LCS. Thus, $\mathit{NET}_{\mathrm{II}}$ needs to be executed $N_\mathrm{AP}$ times, once for each AP beam. Both the  $\mathit{NET}_{\mathrm{I}}$ and $\mathit{NET}_{\mathrm{II}}$ have $N_h$ hidden layers where each hidden layer is made of $n_h$ neurons. $\mathit{NET}_{\mathrm{II}}$ includes an embedding layer that maps the AP beam index as an integer to a point in hyperspace $\mathbb{R}^{n_h/2}$. The weights of the embedding layer are trained as part of the learning process \cite{rezaie_location-_2022}. Thus, the conditional probabilities of UT beam optimality given the AP beam $i$ can be written as
\begin{equation}
    P_{j|i}^{B} = \sum_{k \in \mathcal{I}^{j}} P_{k|i}^{D}.
\end{equation}
The joint optimality probability of beam pair $(i, j)$ reads
\begin{equation}
    P_{i, j} = P_{j|i}^{B} P_{i}.
\end{equation}
The beam pair candidate list $\mathcal{S}$ includes the index of beam pairs with the highest joint probabilities.

\subsection{Dataset Generation and Labeling}
To construct the training and evaluation datasets, the AP is kept fixed in the environment, and the UT is placed randomly in different positions with different orientations. All the possible transceivers beam pairs are sensed for each UT location and orientation. Considering beam pair $(i^\star, j^\star)$ provides the highest RSS, the labeling vector of $\mathit{NET}_{\mathrm{I}}$ is
\begin{equation} \label{eq:NetI}
    L^{\mathrm{I}}_{i} = 
    \begin{cases}
    	1, & \text{if $i = i^\star$,} \\
    	0, & \text{otherwise}.
    \end{cases}
\end{equation}
For training $\mathit{NET}_{\mathrm{II}}$, the AP beam index used as input to the network is set to $i^\star$, and we mark all the directions where the UT beam $j^\star$ provides the highest gain in the labeling vector, i.e.,
\begin{equation} \label{eq:NetII}
    L^{\mathrm{II}}_{k} = 
    \begin{cases}
    	1, & \text{if $k \in \mathcal{I}^{j^\star}$,} \\
    	0, & \text{otherwise}.
    \end{cases}
\end{equation}
Thus, each dataset sample is made of UT location and orientation, the RSS measurements, and the labeling vectors.

\subsection{Simulation Results}
\begin{figure}[t]
	\centering
	\scalebox{0.2}{
		\includegraphics[trim={0cm 0cm 0cm 0cm},clip]{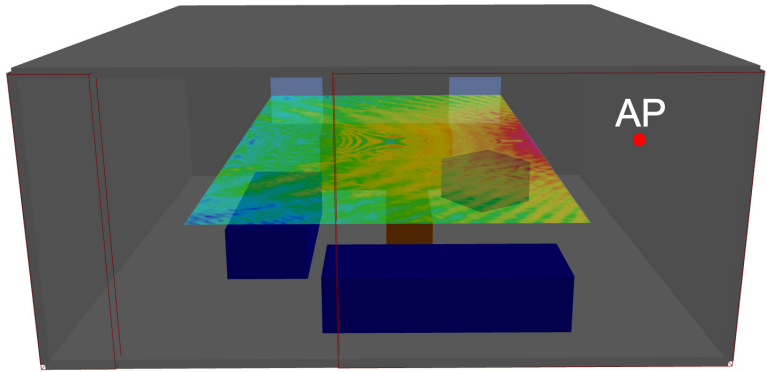}
	}
	\caption{The IEEE standard indoor scenario. In this environment, a rectangle of $4\times7$ meters at $1.5$m height above the floor is considered as the user grid. The LOS power is shown for the user grid.}
	\label{Fig:LR}
\end{figure}
In the simulation setup, we consider the living room presented in the IEEE 802.11ad task group as a standard indoor environment, shown in Fig. \ref{Fig:LR}. The room has $7\times7\times3$ (m) dimensions, and the propagation properties of the environment are defined precisely in \cite{maltsev_channel_2010}. A UPA panel with $\{1, 8, 8\}$ antenna elements is used for the AP. For all the antenna placement shown in Fig. \ref{Fig:Designs}, we consider ULAs with $4$ antenna elements at the edge panels. Also, UPAs with $\{2, 2, 1\}$ antenna elements are used for the face and back panels. To model the antenna radiation pattern of a patch antenna, the model proposed in 
\cite{3gpp_study_2020} is used. Also, we set $P_t = 24~$dBm and $\sigma_n^2 = -84~$dBm in the simulations. we consider $N_h = 5$ hidden layers with $n_h = 128$ neurons for the proposed generic NN, and we stick to the training policy described in \cite{rezaie_location-_2020}.

The AP is placed close to the center of one of the side walls, and the LOS power is shown in Fig. \ref{Fig:LR} for the considered user grid. The UT is placed uniformly in the user grid, and the UT orientation is uniformly chosen between portrait and landscape modes. In the portrait and landscape mode, an orientation is uniformly drawn form the angle ranges $\{ \alpha_\mathrm{UT} \in [-\pi, \pi), \beta_\mathrm{UT} = 0, \gamma_\mathrm{UT} \in [0, \pi/2]\}$ and $\{ \alpha_\mathrm{UT} \in [-\pi, \pi), \beta_\mathrm{UT} \in [-\pi/2, 0], \gamma_\mathrm{UT} = 0 \}$, respectively. In addition, $50\%$ of realizations are generated in the LOS condition and the other $50\%$ in NLOS condition. In this study, we use training and evaluation datasets, respectively, with $56,000$ and $14,000$ samples. In obtaining the results, three random initialization of NNs were trained and averaged to de-emphasize the effects of initial weights on the model performance. $\mathbb{D}^{E}$, $\mathbb{D}^{F}$, and $\mathbb{D}^{EF}$ denote the datasets with the E, F, and EF antenna placement designs, respectively.

\begin{figure}[t]
	\centering
	\subfloat[\label{Fig:BeamRegion_a}]{%
       \scalebox{1}{\includegraphics[trim={2cm 0.2cm 4.5cm 1cm},clip,width=0.4\textwidth, height=0.2\textwidth]{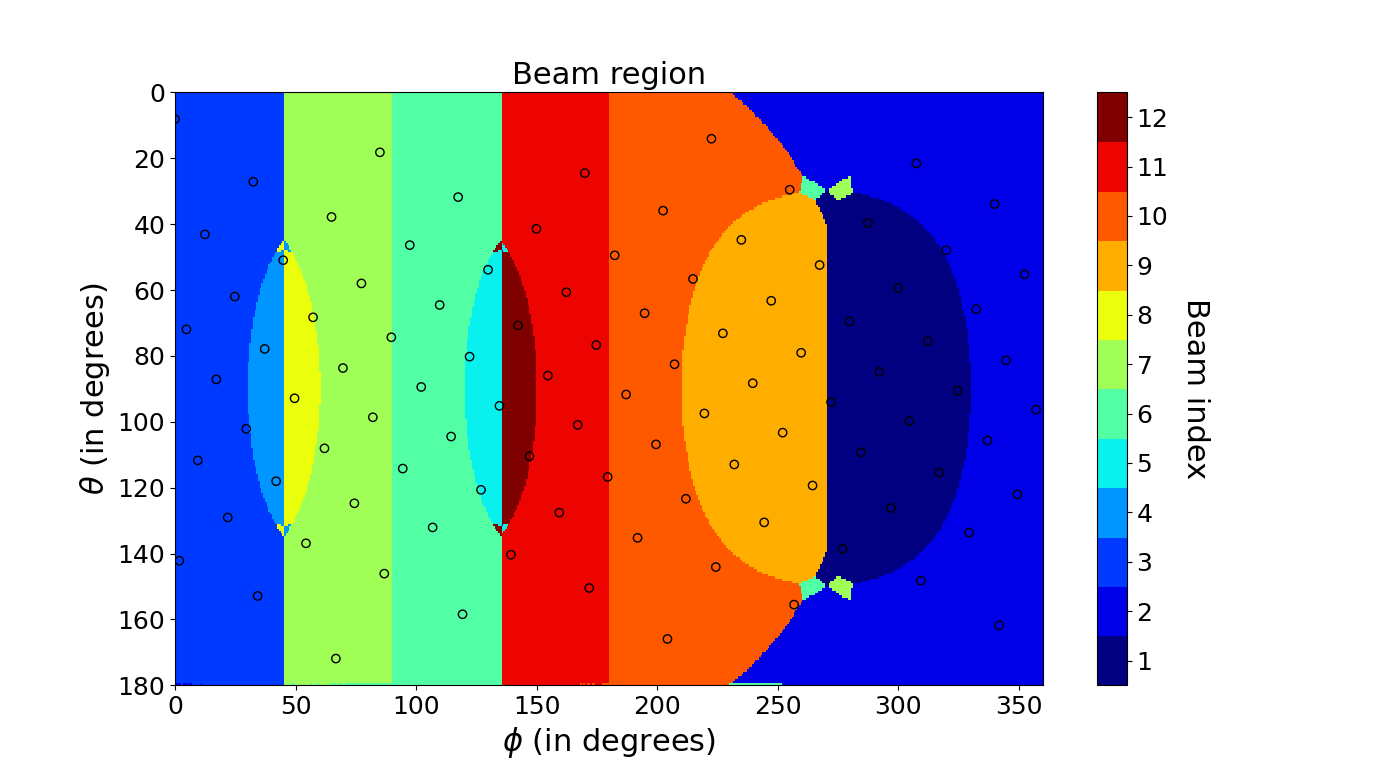}}}
    \\[0.05cm]
    \subfloat[\label{Fig:BeamRegion_b}]{%
        \scalebox{1}{\includegraphics[trim={2cm 0.2cm 4.5cm 1cm},clip,width=0.4\textwidth, height=0.2\textwidth]{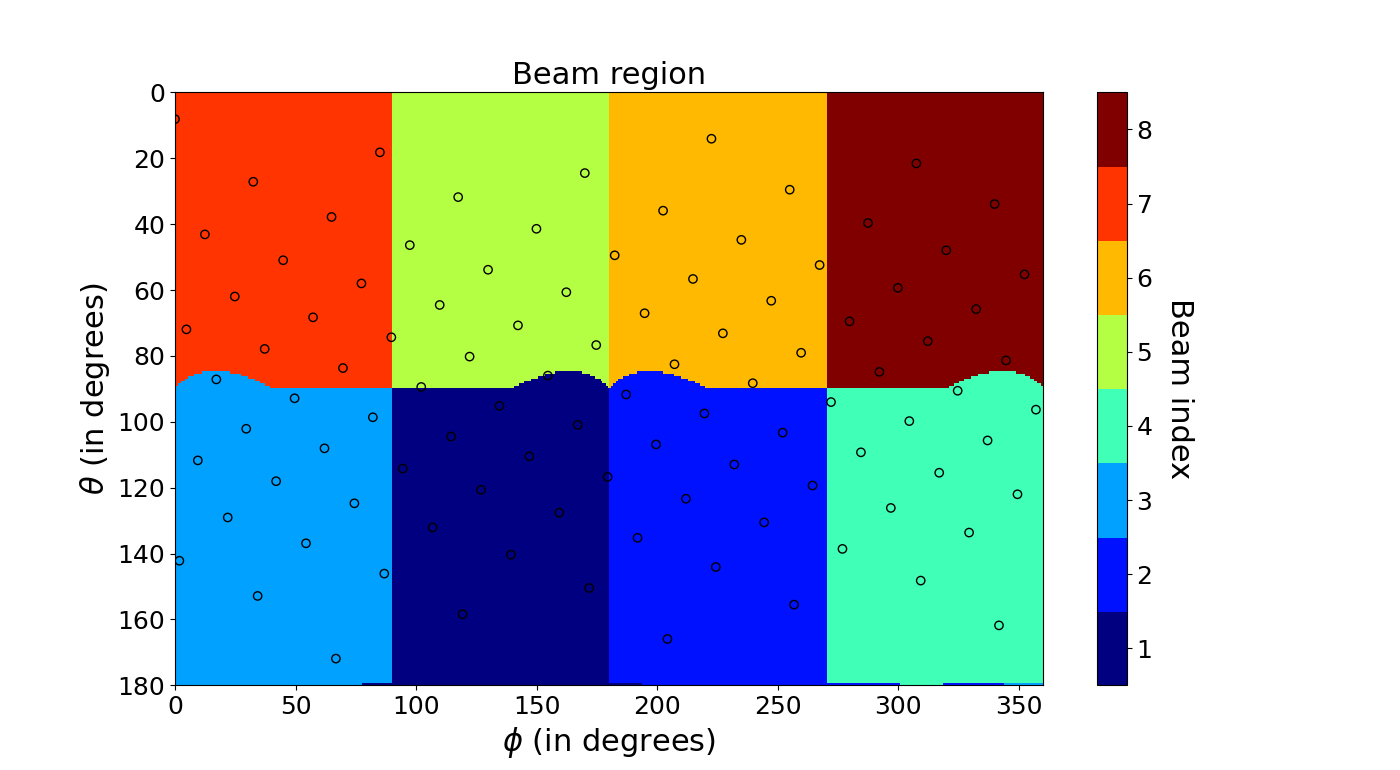}}}
    \\[0.05cm]
    \subfloat[\label{Fig:BeamRegion_c}]{%
        \scalebox{1}{\includegraphics[trim={2cm 0.2cm 4.5cm 1cm},clip,width=0.4\textwidth, height=0.2\textwidth]{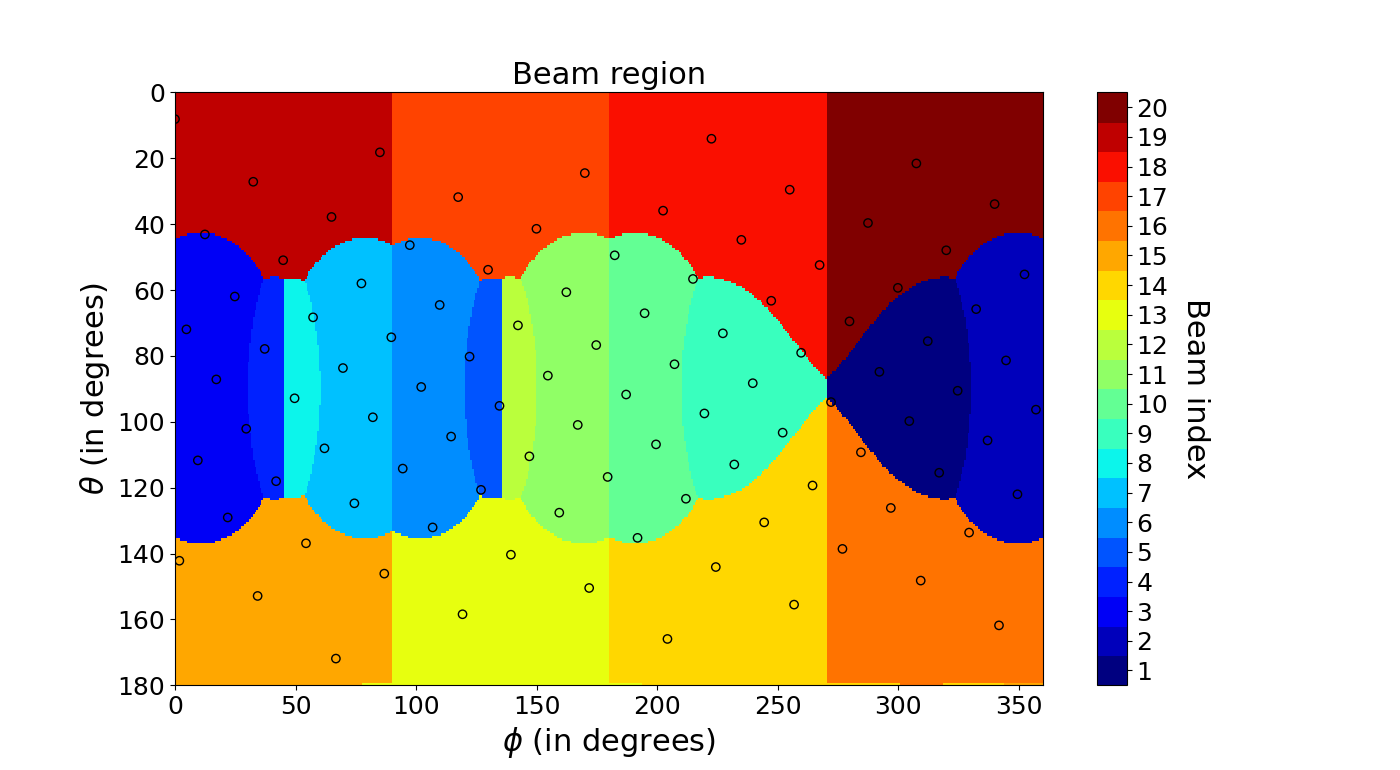}}}
	\caption{Beam decision regions (a) edge design (b) face design (c) edge-face design.}
	\label{Fig:BeamRegion}
\end{figure}
Fig.~\ref{Fig:BeamRegion} shows the beam decision region in azimuth-elevation space for all the three edge, face, and edge-face antenna placements. In addition, the Fibonacci points for a grid of $100$ points are marked to illustrate the distribution of the points for each antenna placement/codebook. The mapping $\boldsymbol{i}^{F}$ for each antenna placement can be constructed by using the beam indices of the beam regions in which the Fibonacci points are located. The considered baseline to evaluate the performance of the proposed generic NN is the multi-network beam selection (MN-BS) device-specific method presented in \cite{rezaie_location-_2022}. 

We use the misalignment probability as a performance measure of missing the beam pair with the highest RSS, i.e.,
\begin{equation} \label{eq:MisAMetric}
P_{m}(\mathcal{S}) = \mathbb{P} \left[R_{\hat{i},\hat{j}} < \max\limits_{(p, q) \in \mathcal{B}} R_{p,q}\right]
\end{equation}
where $(\hat{i},\hat{j})$ denote the selected beam pair after sensing the environment with all the beam pairs in the beam list $\mathcal{S}$. Also, the effective spectral efficiency (ESE) of the beam selection procedure is: 
\begin{equation}
    \text{SE}_{\text{eff}} = \frac{T_{fr} - N_b T_s}{T_{fr}} \log_2 (1+ \text{SNR}_{\hat{i},\hat{j}}), \hspace{10pt} N_b T_s \leq T_{fr}
\end{equation}
where $N_b = | \mathcal{S} |$ is the number of required time slots for sensing the environment in the beam alignment phase. Also, $\text{SNR}_{i, j}$ denotes the SNR of $i$th and $j$th beamforming vectors at the transceivers, i.e.
\begin{equation}
    \text{SNR}_{i, j} = \frac{\Big \lVert \sqrt{P_\mathrm{AP}} \boldsymbol{v}_{j}^H \boldsymbol{H}^{(p_j)} \boldsymbol{u}_{i} s \Big \rVert ^2}{\sigma_n^2},
\end{equation}
where $p_j$ is the corresponding panel of combiner $\boldsymbol{v}_j$. In this work, we use a frame duration of $T_{fr}=20$ms and beam-sensing time of $T_{s}=0.1$ms \cite{hussain_second-best_2019}. The perfect alignment method, as a genie-aided beam alignment approach, selects the beam pair with highest SNR out of all the possible beam pairs in $\mathcal{B}$. As a result, the perfect alignment is an upper bound for ESE of the device-specific and device-agnostic beam selection methods.

\subsubsection{Numerical Evaluation}
\begin{figure}[t]
	\centering
	\subfloat[ \label{Fig:EdgeFace_a}]{%
       \scalebox{1}{\includegraphics[trim={1.5cm 0.1cm 1cm 2cm},clip,width=0.6\textwidth]{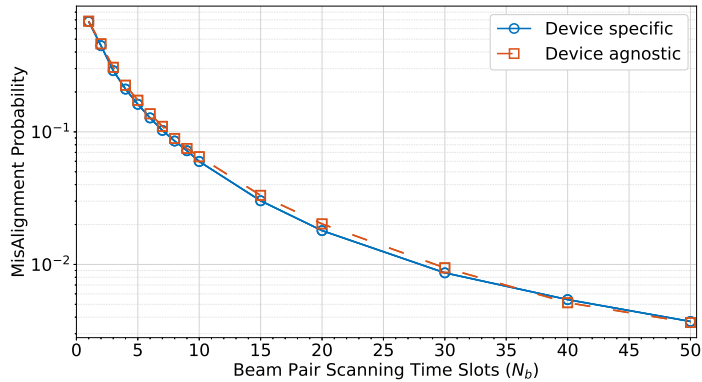}}}
    \\[0.05cm]
   \subfloat[\label{Fig:EdgeFace_b}]{%
        \scalebox{1}{\includegraphics[trim={1.5cm 0.1cm 1cm 2cm},clip,width=0.6\textwidth]{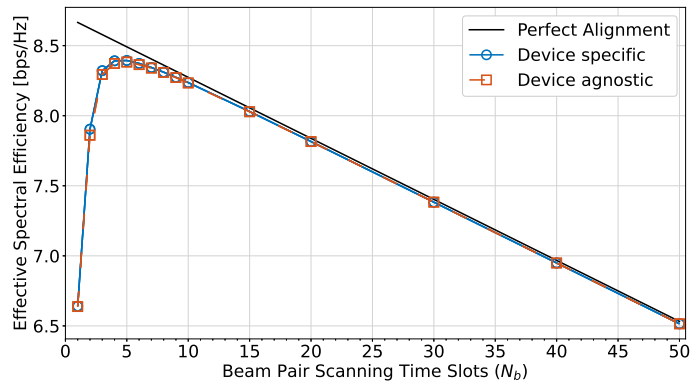}}}
    \caption{Performance of the device-specific and device-agnostic BA methods for the edge-face design.}
	\label{Fig:EdgeFace}
\end{figure}
Fig.~\ref{Fig:EdgeFace} shows the performance in terms of misalignment probability and spectral efficiency as a function of the candidate beam list size $N_b$ of different BA methods for the EF antenna placement. 
The performance of the generic NN is as good as the MN-BS method, which shows that the degradation by pointing the directions instead of UT beams is negligible. 

\begin{figure}[t]
	\centering
        \scalebox{1}{\includegraphics[trim={1.5cm 0.1cm 1cm 2cm},clip,width=0.6\textwidth]{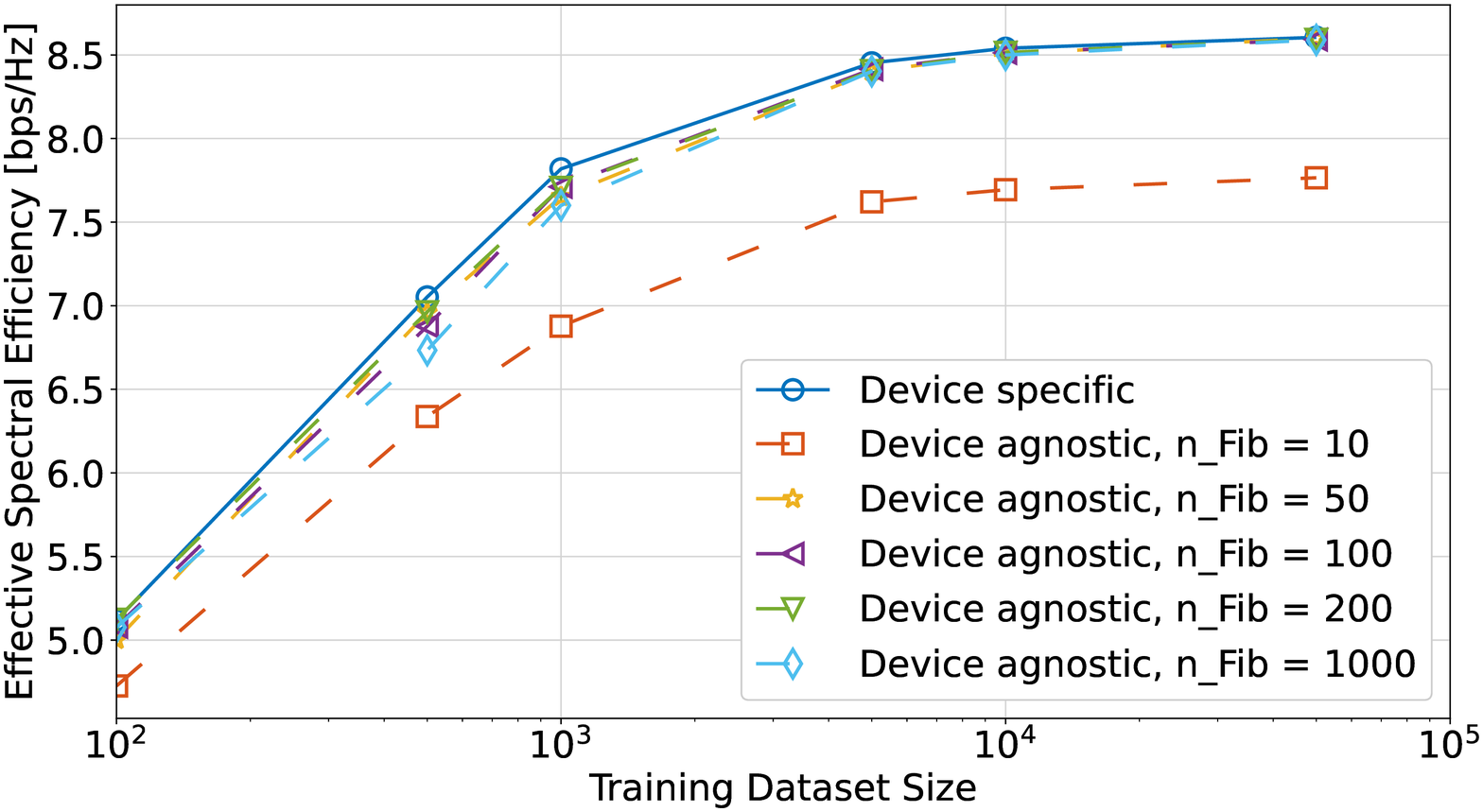}}
    \caption{Performance of the generic NN method with different Fibonacci grid size for the edge-face design with a beam list including $5$ beam pairs ($N_b = 5$).}
	\label{Fig:FibPts}
\end{figure}
Since the number of Fibonacci points, $n_{Fib}$, is a hyperparameter of the proposed generic NN, the performance of this method for different $n_{Fib}$ is shown in Fig.~\ref{Fig:FibPts}. In this experiment, we consider that the beam candidate list, $\mathcal{S}$, includes $5$ beam pairs, and evaluate the effective spectral efficiency as a function of the training dataset sizec. The generic NN performance is almost as good as the MN-BS method as an upper bound for $n_{Fib} > 50$. Although the NN is quite robust to this hyperparameter, a tiny number of Fibonacci points like $n_{Fib} = 10$ causes performance degradation because some of the beams may not include any Fibonacci points in their beam regions. It is worth pointing out that the proposed device-agnostic framework is quite robust to the FG size.  

\begin{figure}[t]
	\centering
        \scalebox{1}{\includegraphics[trim={1.5cm 0.1cm 1cm 2cm},clip,width=0.6\textwidth]{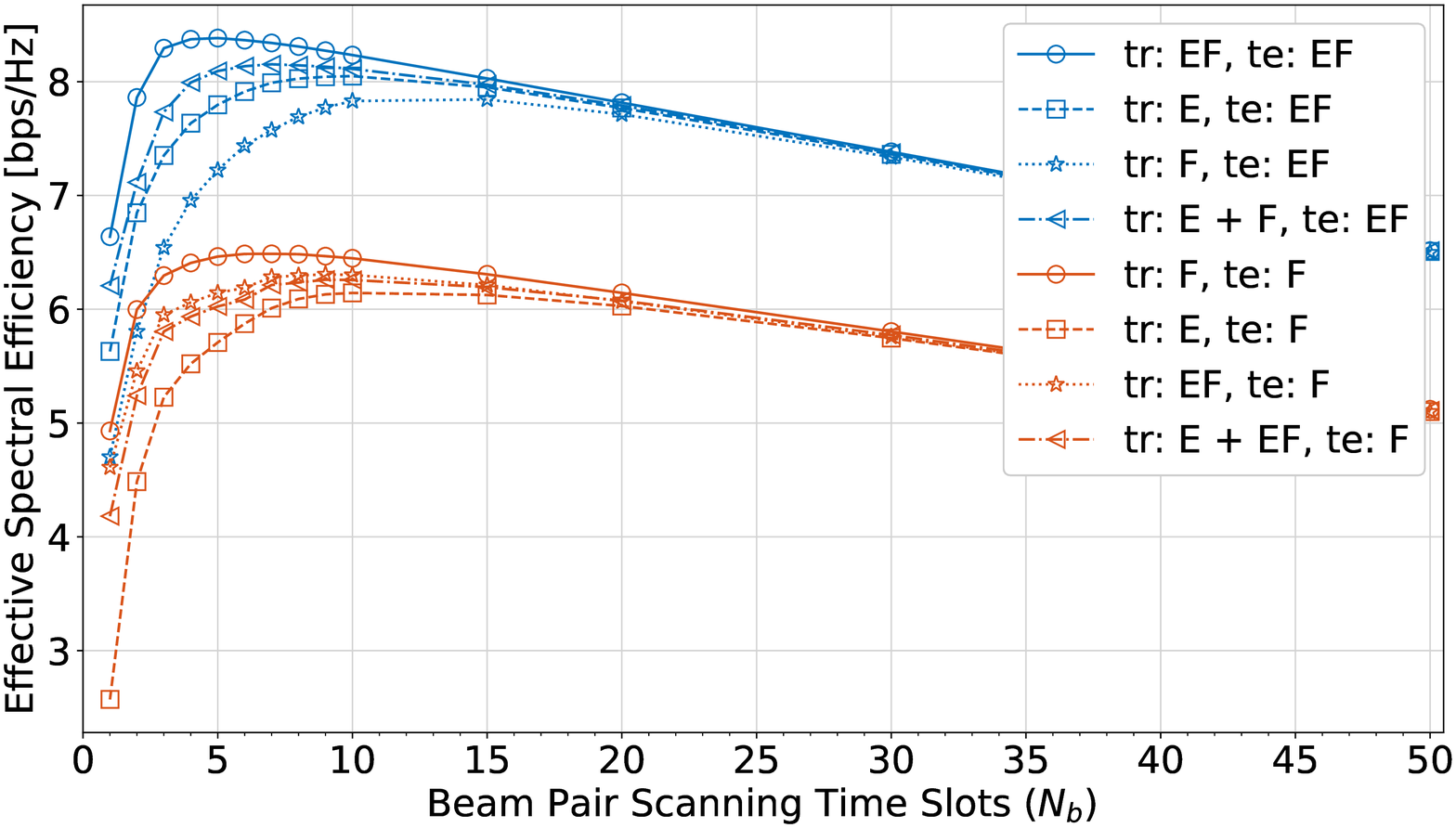}}
    \caption{Generalization of the proposed device/codebook-agnostic method for unseen configuration in the training datasets. In the legend, 'tr' and 'te' terms, respectively, are used to determine the datasets used for training and testing of the proposed method.}
	\label{Fig:Generalization}
\end{figure}
Fig.~\ref{Fig:Generalization} shows the performance of the proposed generic NN with a mismatch in the antenna placement/codebook in the training and test samples. To see the degradation because of mismatched datasets, the performance of the matched cases is also shown. As the EF antenna placement design has more beams, resulting in smaller beam regions, it is likely that a beam region in the EF design can be entirely covered by only one beam in the F design. Thus, the degradation of training with the $\mathbb{D}^{EF}$ and test on the $\mathbb{D}^{F}$ is not significant. We see more degradation by training with $\mathbb{D}^{E}$ and testing on $\mathbb{D}^{F}$, as the beam regions of these two antenna configurations are quite different. Note that even in this case, the generic NN can provide an acceptable performance with limited degradation by sensing larger beam lists like $N_b = 10$. 

On the other hand, we see a degradation when the generic NN is trained with $\mathbb{D}^{F}$ and is tested with $\mathbb{D}^{EF}$. The same conclusions can be drawn from training with $\mathbb{D}^{E}$ while evaluating with $\mathbb{D}^{EF}$. The performance degradation is due to the fact that the beam regions in the F or E design are larger than in the EF design, so the generic NN cannot pinpoint the directions precisely enough. Using $\mathbb{D}^{E+F}$, including samples mixed of both E and F antenna designs, helps to reduce the performance degradation over test samples $\mathbb{D}^{EF}$, even though the antenna configuration of test samples is not observed in the training samples. Overall, the results show the ability of the proposed generic NN in generalization for unseen antenna placement/codebook configurations without the need for any additional training samples.

\begin{figure}[t]
	\centering
        \scalebox{1}{\includegraphics[trim={1.5cm 0.1cm 1cm 2cm},clip,width=0.6\textwidth]{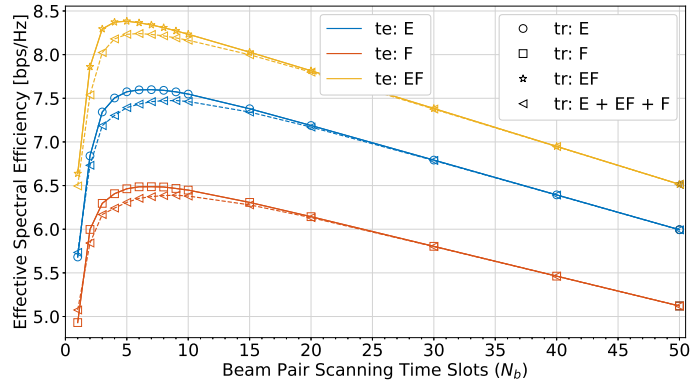}}
    \caption{Performance of the proposed generic NN using a dataset mixed of all antenna configurations compared to the matched training datasets. For example, for evaluating the EF dataset (yellow curves), the curve with tr:EF shows the performance of the matched condition since the same dataset is used in the training and evaluation processes. However, when the mixed dataset is used for training the device-agnostic framework (tr: E + EF + F), the curve shows the performance in a mismatched condition.}
	\label{Fig:Mix}
\end{figure}
Fig.~\ref{Fig:Mix} illustrates the performance of the generic NN with a training dataset mixed of all the antenna configurations, $\mathbb{D}^{E+F+EF}$. It is promising that the trained network with the mixed dataset can achieve almost the same performance as the matched training datasets for all three antenna configurations.

\section{Beam Selection using sub-6 GHz channel}\label{Sec:Sub6}
As another example of integrating the proposed device-agnostic framework in beam selection methods proposed in literature, in this section expand the method using sub-6 GHz channel state information (CSI) proposed in \cite{alrabeiah_deep_2020}. 
\subsection{System Model}\label{Sec:SysModelSub6}
Consider a point-to-point OFDM system where the base station (BS) and mobile user are equipped with sub-6 GHz and mmWave antenna elements. The BS has two ULAs made of $M_{\text{sub-6}}$ and $M_{\text{mmW}}$ antenna elements at the sub-6 GHz and mmWave bands, respectively. The user employs a single antenna at the sub-6 GHz band. However, different from the assumption in \cite{alrabeiah_deep_2020}, we assume a ULA of size $M_{\text{mmW}}^{U}$ is used by the mobile user. The sub-6 GHz and mmWave bands are used for the uplink signaling and the downlink data transmission, respectively. The uplink received signal at the $k$th subcarrier,  $\boldsymbol{y}_{\text{sub-6}}[k]$, may be written as:
\begin{equation}
    \boldsymbol{y}_{\text{sub-6}}[k] = \boldsymbol{h}_{\text{sub-6}}[k] s_{\text{p}}[k] + \boldsymbol{n}_{\text{sub-6}}[k], k = 1, \dots, K
\end{equation}
where $\boldsymbol{h}_{\text{sub-6}}[k] \in \mathbb{C}^{M_{\text{sub-6}} \times 1}$, $s_{\text{p}}[k]$, and $\boldsymbol{n}_{\text{sub-6}}[k] \in \mathbb{C}^{M_{\text{sub-6}} \times 1}$ respectively are the uplink channel vector, uplink pilot signal and received noise at the BS sub-6 GHz array at the $k$th subcarrier. $\mathbb{E}[s_{\text{p}}[k]] = P_{\text{sub-6}}/K$, where $P_{\text{sub-6}}$ is the transmit power at the mobile user and $K$ is the number of sub-6GHz subcarriers. Also, $\boldsymbol{n}_{\text{sub-6}}[k] \in \mathbb{C}^{M_{\text{sub-6}} \times 1}$ is a zero-mean complex Gaussian vector with variance $\sigma_{\text{sub-6}}^2$. The uplink channel estimation at the baseband processor is assumed to be available using digital beamforming at the BS sub-6 GHz array.

For the mmWave downlink transmission, we consider analog beamforming at the BS and user. Using $\boldsymbol{f} \in \mathbb{C}^{M_{\text{mmW}} \times 1}$ and $\boldsymbol{w} \in \mathbb{C}^{M_{\text{mmW}}^{U} \times 1}$ respectively for beamforming at the BS and user, the received signal by the user at the $\bar{k}$th subcarrier, $\boldsymbol{y}_{\text{sub-6}}[\bar{k}]$, is
\begin{equation}
    \boldsymbol{y}_{\text{mmW}}[\bar{k}] = \boldsymbol{w}^T \boldsymbol{H}_{\text{mmW}}^T[\bar{k}] \boldsymbol{f} s_{\text{d}}[\bar{k}] + \boldsymbol{n}_{\text{mmW}}[\bar{k}], \bar{k} = 1, \dots, \bar{K}
\end{equation}
where $\boldsymbol{H}_{\text{mmW}}[\bar{k}] \in \mathbb{C}^{M_{\text{mmW}} \times M_{\text{mmW}}^{U}}$ denotes the MIMO downlink channel at the $\bar{k}$th subcarrier. $s_{\text{d}}[\bar{k}]$ is the known downlink signal with $\mathbb{E}[s_{\text{d}}[\bar{k}]] = P_{\text{mmW}}/\bar{K}$, where $P_{\text{mmW}}$ and $\bar{K}$ are the transmission power at the BS and the number of mmWave subcarriers. Also, $\boldsymbol{n}_{\text{mmW}}[\bar{k}] \sim \mathcal{CN}(\boldsymbol{0}_{M_{\text{mmW}}}, \sigma_{\text{mmW}}^2\boldsymbol{I}_{M_{\text{mmW}}})$ is a complex Gaussian vector. 
$\mathcal{F}$ and $\mathcal{W}$ respectively are the beamforming codebooks at the BS and user with cardinality $|\mathcal{F}| = N_{\text{CB}}^{\text{BS}}$ and $|\mathcal{W}| = N_{\text{CB}}^{\text{U}}$. A geometric channel model is considered for the sub-6 GHz and mmWave channels. The uplink sub-6 GHz channel may be written as
\begin{equation}
    \boldsymbol{h}_{\text{sub-6}}[k] = \sum_{d=0}^{D_c-1} \sum_{l=1}^{L} \alpha_l \hspace{2pt} e^{-j \frac{2 \pi k}{K} d} p(d \hspace{2pt} T_s - \tau_l) \boldsymbol{a}_{\text{BS-sub-6}}(\phi_{l},\theta_{l}).
\end{equation}
Also, the downlink mmWave channel can be written as
\begin{equation}
    \boldsymbol{H}_{\text{mmW}}[\bar{k}] = \sum_{d=0}^{D_c-1} \sum_{l=1}^{L} \alpha_l \hspace{2pt} e^{-j \frac{2 \pi \bar{k} d}{\bar{K}}} p(d \hspace{2pt} T_s - \tau_l) \boldsymbol{a}_{\text{U-mmW}}(\phi_{l},\theta_{l}) \boldsymbol{a}_{\text{BS-mmW}}(\psi_{l},\omega_{l}).
\end{equation}
$L$, $D_c$, and $T_s$ denote number of channel components, the cyclic prefix length, and the the sampling time, respectively. $\alpha_l$ and $\tau_l$ are the complex path gain and delay of the $l$th path, and $p(\cdot)$ denotes the pulse shaping filter. In addition, $\phi_{l}$, $\theta_{l}$, $\psi_{l}$, and $\omega_{l}$ are the azimuth AoA, elevation AoA, azimuth AoD, elevation AoD, respectively.
$\boldsymbol{a}_{\text{BS-sub-6}}$ and $\boldsymbol{a}_{\text{BS-mmW}}$ are the sub-6 GHz and mmWave antenna array responses at the BS. Also, $\boldsymbol{a}_{\text{U-mmW}}$ is the mmWave antenna array response at the mobile user. Note that $L$, $D_c$, and $T_s$ may be different for the sub-6 GHz and mmWave channels. 

\subsection{Beam prediction using sub-6 GHz channel}
The optimal pair of mmWave beamforming vectors at the transceivers provides the highest spectral efficiency, i.e.,
\begin{equation}
    (\boldsymbol{f}^\star, \boldsymbol{w}^\star) = \underset{\boldsymbol{f} \in \mathcal{F}, \boldsymbol{w} \in \mathcal{W}}{\mathrm{arg\hspace{2pt}max}} \hspace{2pt} R(\boldsymbol{H}_{\text{mmW}}, \boldsymbol{f}, \boldsymbol{w})
\end{equation}
where the achievable spectral efficiency can be expressed as
\begin{equation}\label{eq:ASE}
    R(\boldsymbol{H}_{\text{mmW}}, \boldsymbol{f}, \boldsymbol{w}) = \sum_{\bar{k}=1}^{\bar{K}} \log_2 (1+ \text{SNR} |\boldsymbol{w} \boldsymbol{H}_{\text{mmW}}^T[\bar{k}] \boldsymbol{f}|^2).
\end{equation}
However, exhaustive search over all the beam pairs leads to large training overhead and latency. In this study, the sub-6 GHz channel $\boldsymbol{h}_{\text{sub-6}}$ is used to reduce the overhead in mmWave link. The objective of this work is to maximize the success probability in predicting the optimal beam pair $(\boldsymbol{f}^\star, \boldsymbol{w}^\star)$, i.e.,
\begin{equation}
    \kappa_1 = \mathbb{P}\left((\hat{\boldsymbol{f}}, \hat{\boldsymbol{w}}) = (\boldsymbol{f}^\star, \boldsymbol{w}^\star) | \boldsymbol{h}_{\text{sub-6}}\right)
\end{equation}
where $\hat{\boldsymbol{f}} \in \mathcal{F}$ and $\hat{\boldsymbol{w} \in \mathcal{W}}$ are, respectively, the predicted beamforming vectors at the BS and user using the sub-6 GHz channel. The mapping function from the sub-6 GHz channel to the mmWave channel can be interpreted as a fingerprinting approach that uses the propagation properties to propose the probable proper beam pairs. Because of the high non-linearity of the mapping, deep learning is an appropriate choice for this application.

\begin{figure}[t]
    \centering
	\definecolor{mycolor1}{RGB}{224,9,2} 
	\definecolor{mycolor2}{RGB}{255,174,66} 
	\definecolor{mycolor3}{RGB}{0,120,210} 
	\definecolor{mycolor4}{RGB}{0, 128, 0} 
	\definecolor{mycolor5}{RGB}{0,191,255} 
	\definecolor{mycolor6}{RGB}{0,0,139} 
	\definecolor{mycolor7}{RGB}{255,0,255} 
	\definecolor{mycolor8}{RGB}{238,130,238} 
	\definecolor{mycolor9}{RGB}{128,0,128} 
	\scalebox{.58}{
		\begin{tikzpicture}[cross/.style={path picture={ 
				\draw[red, line width=3pt]
				(path picture bounding box.south east) -- (path picture bounding box.north west) (path picture bounding box.south west) -- (path picture bounding box.north east);
		}}]
		
		\draw[->, line width=1pt] (-1.5,0) node [font=\Large, above, rotate around={90:(0,0)}] {Input Sub-6GHz Vectorized channel} -- (-0.5,0);
		\foreach \h in {0,...,4} {
		    \foreach \i in {-2,...,-1} {
    		    \draw[fill=mycolor1] (\h*3,\i) circle (0.4);
    		    \path (\h*3,-1) -- (\h*3,1) node [black, font=\Huge, midway, sloped] {\textbf{$...$}};
    		    \draw[fill=mycolor1] (\h*3,-\i) circle (0.4);
    		}
	        \node at (\h*3,0) [rectangle,draw,line width=1pt,minimum width=1cm, minimum height=5cm,rounded corners=1mm] (v1) {};
	        \node at (\h*3+1.1,0) [rectangle,draw,line width=1pt,minimum width=5cm, minimum height=0.7cm,rounded corners=1mm, rotate=90] (v2) {Drop-out Layer};
	        \draw[->, line width=1pt] (\h*3+1.6,0) -- (\h*3+2.4,0);
	        \tikzmath{\hh=int(\h + 1);}
	        \node[align=left] at (\h*3+0.5,5.7) {\large \textbf{Stack} $\boldsymbol{\hh}$};
	        \node[align=left] at (\h*3+0.5,5.2) {\small Neurons + ReLU+};
	        \node[align=left] at (\h*3+0.5,4.8) {\small dropout};
    	}
    	
    	\draw[fill=mycolor1] (5*3,3) circle (0.4);
    	\draw[fill=mycolor1] (5*3,2) circle (0.4);
	    \path (5*3,1.5) -- (5*3,0.5) node [black, font=\Huge, midway, sloped] {\textbf{$...$}};
	    \draw[fill=mycolor1] (5*3,0) circle (0.4);
	    \draw[fill=mycolor1] (5*3,-1) circle (0.4);
	    \path (5*3,-1) -- (5*3,-3) node [black, font=\Huge, midway, sloped] {\textbf{$...$}};
	    \draw[fill=mycolor1] (5*3,-3) circle (0.4);
	    
	    \node at (5*3,0) [rectangle,draw,line width=1pt,minimum width=1cm, minimum height=7cm,rounded corners=1mm] (v1) {};
        \node at (5*3+1.1,0) [rectangle,draw,line width=1pt,minimum width=7cm, minimum height=0.7cm,rounded corners=1mm, rotate=90] (v2) {Soft-max};

        \node[align=left] at (5*3+0.5,5.7) {\large \textbf{Stack} $\boldsymbol{6}$};
        \node[align=left] at (5*3+0.5,5.2) {\small Neurons + Softmax};

	    \draw[->, line width=1pt, mycolor4] (5*3+1.45,3) -- (6*3+1.3,3) node [font=\Large, above left = -2pt and 0pt] {$P_{1, 1}$};
		\draw[->, line width=1pt, mycolor4] (5*3+1.45,2) -- (6*3+1.3,2) node [font=\Large, above left = -2pt and 0pt] {$P_{1, 2}$};
		\path (5*3+2,2) -- (5*3+2,0.5) node [mycolor4, font=\Huge, midway, sloped] {\textbf{$...$}};
		\draw[->, line width=1pt, mycolor4] (5*3+1.45,0) -- (6*3+1.3,0) node [font=\Large, above left = -2pt and 0pt] {$P_{1, n_{Fib}}$};
		\draw[->, line width=1pt, mycolor4] (5*3+1.45,-1) -- (6*3+1.3,-1) node [font=\Large, above left = -2pt and 0pt] {$P_{2, 1}$};
		\path (5*3+2,-1) -- (5*3+2,-2.5) node [mycolor4, font=\Huge, midway, sloped] {\textbf{$...$}};
		\draw[->, line width=1pt, mycolor4] (5*3+1.45,-3) -- (6*3+1.3,-3) node [font=\Large, above left = -2pt and 0pt] {$P_{|\mathcal{F}|, n_{Fib}}$};
		
		\node at (6*3+2,0) [rectangle,draw,line width=1pt,minimum width=7cm, minimum height=1.4cm,rounded corners=1mm, rotate=90] (v3) {Post Processing};
		
		\draw[->, line width=1pt, mycolor5] (6*3+2.7,3) -- (7*3+2,3) node [font=\Large, above left = -2pt and 0pt] {$P_{1, 1}$};
		\draw[->, line width=1pt, mycolor5] (6*3+2.7,2) -- (7*3+2,2) node [font=\Large, above left = -2pt and 0pt] {$P_{1, 2}$};
		\path (7*3,2) -- (7*3,0.5) node [mycolor5, font=\Huge, midway, sloped] {\textbf{$...$}};
		\draw[->, line width=1pt, mycolor5] (6*3+2.7,0) -- (7*3+2,0) node [font=\Large, above left = -2pt and 0pt] {$P_{1, |\mathcal{W}|}$};
		\draw[->, line width=1pt, mycolor5] (6*3+2.7,-1) -- (7*3+2,-1) node [font=\Large, above left = -2pt and 0pt] {$P_{2, 1}$};
		\path (7*3,-1) -- (7*3,-2.5) node [mycolor5, font=\Huge, midway, sloped] {\textbf{$...$}};
		\draw[->, line width=1pt, mycolor5] (6*3+2.7,-3) -- (7*3+2,-3) node [font=\Large, above left = -2pt and 0pt] {$P_{|\mathcal{F}|, |\mathcal{W}|}$};
		
		\draw[->, line width=1pt] (6*3+2,4.5) -- (6*3+2,3.5) node [sloped, font=\Large, above = 31pt] {$\boldsymbol{i}^{F}$};

		\end{tikzpicture}}
	\caption{The device agnostic design for predicting the optimal mmWave beam pair using the sub-6 GHz channel. This design is a modified version of the device specific network proposed in \cite{alrabeiah_deep_2020}. The output layer includes $|\mathcal{F}| \times n_{Fib}$ neurons. The post processing unit provides the prediction probability of optimality for all the beam pairs with $|\mathcal{F}| \times |\mathcal{W}|$ outputs.}
	\label{Fig:Sub6NN}
\end{figure}
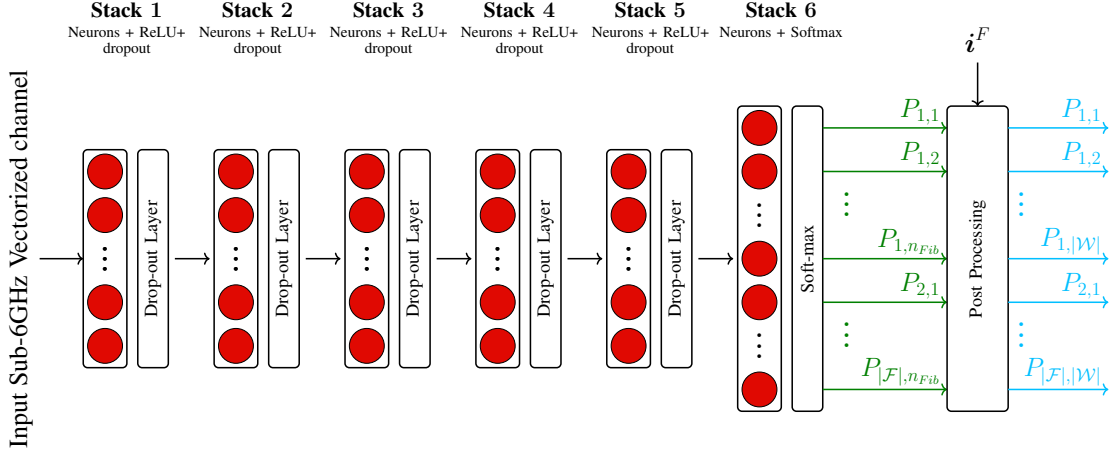
Fig.~\ref{Fig:Sub6NN} shows the beam selection network in the device-agnostic framework. This network uses several hidden layers to extract the information in the sub-6 GHz channel and map it to the beams at the BS and AoAs at the user. Thus, the generic network has $|\mathcal{F}| \times n_{Fib}$ neurons for the optimality probability of all the possible BS beams and user AoAs at the transceivers. The post processing unit uses the mapping information between the user terminal beams and AoAs, $\boldsymbol{i}^{F}$, to provide the optimality probability of the beam pairs at the transceivers based on the generic network's outputs.

\subsection{DL Training and Deployment Phases}
\textbf{DL Training Phase:} The DL training phase needs dual-band communication in sub-6 GHz and mmWave bands. In a coherence time, all the beam pairs of the mmWave codebooks at the transceivers, $\mathcal{F}$ and $\mathcal{W}$, are exhaustively sensed, and the achievable spectral efficiency by each beam pair is stored in $R$. We store index of the optimal beam pair $(\boldsymbol{f}^\star, \boldsymbol{w}^\star)$ in $(i^\star, j^\star)$. Also, the mapping from the beam indices to the Fibonacci points, $\boldsymbol{i}^{F}$, is needed for each dataset sample. Furthermore, the sub-6 GHz channel is sounded and stored in $\boldsymbol{h}_{\text{sub-6}}$. Thus, each training sample is made of $\{ \boldsymbol{h}_{\text{sub-6}}, \boldsymbol{i}^{F}, i^\star, j^\star \}$. We train the generic NN using the training samples and corresponding labels as ground truth. We use categorical cross entropy as the loss function of the generic NN, i.e.,
\begin{align}
    \mathcal{L} = - \sum_{i=1}^{|\mathcal{F}|} \sum_{j=1}^{n_{Fib}} L_{i,j} \log(P_{i,j})
\end{align}
where $L_{i,j}$ and $P_{i,j}$ are the labels and outputs of the generic NN for the $i$th and $j$th beamforming vectors at the BS and user, respectively. In a similar way that is explained in \eqref{eq:LabelGen}, the label can be defined as
\begin{equation}
    L_{i,j} = 
    \begin{cases}
    	1, & \text{if $i = i^\star \hspace{2pt} \text{and} \hspace{2pt} k \in \mathcal{I}^{j^\star}$,} \\
    	0, & \text{otherwise}
    \end{cases}
\end{equation}
where $\mathcal{I}^{j^\star}$ includes indices of the FG with $\boldsymbol{i}^{F}_k = j^\star$.

\textbf{DL Deployment Phase:} Dual-band communication is operated in the deployment phase. The trained NN has learned the mapping from the sub-6 GHz channel to the optimal beam pair. The user only sends an uplink pilot at sub-6 GHz for estimation of $\boldsymbol{h}_{\text{sub-6}}$ at the BS. The device-agnostic framework uses the estimated channel $\boldsymbol{h}_{\text{sub-6}}$ for predicting the optimal mmWave beam pairs.

\begin{figure}[t]
	\centering
	\subfloat[\label{Fig:scenario_a} 'O$1$' scenario]{%
       \scalebox{1}{\includegraphics[trim={0cm 0cm 0cm 0cm},clip,width=0.4\textwidth]{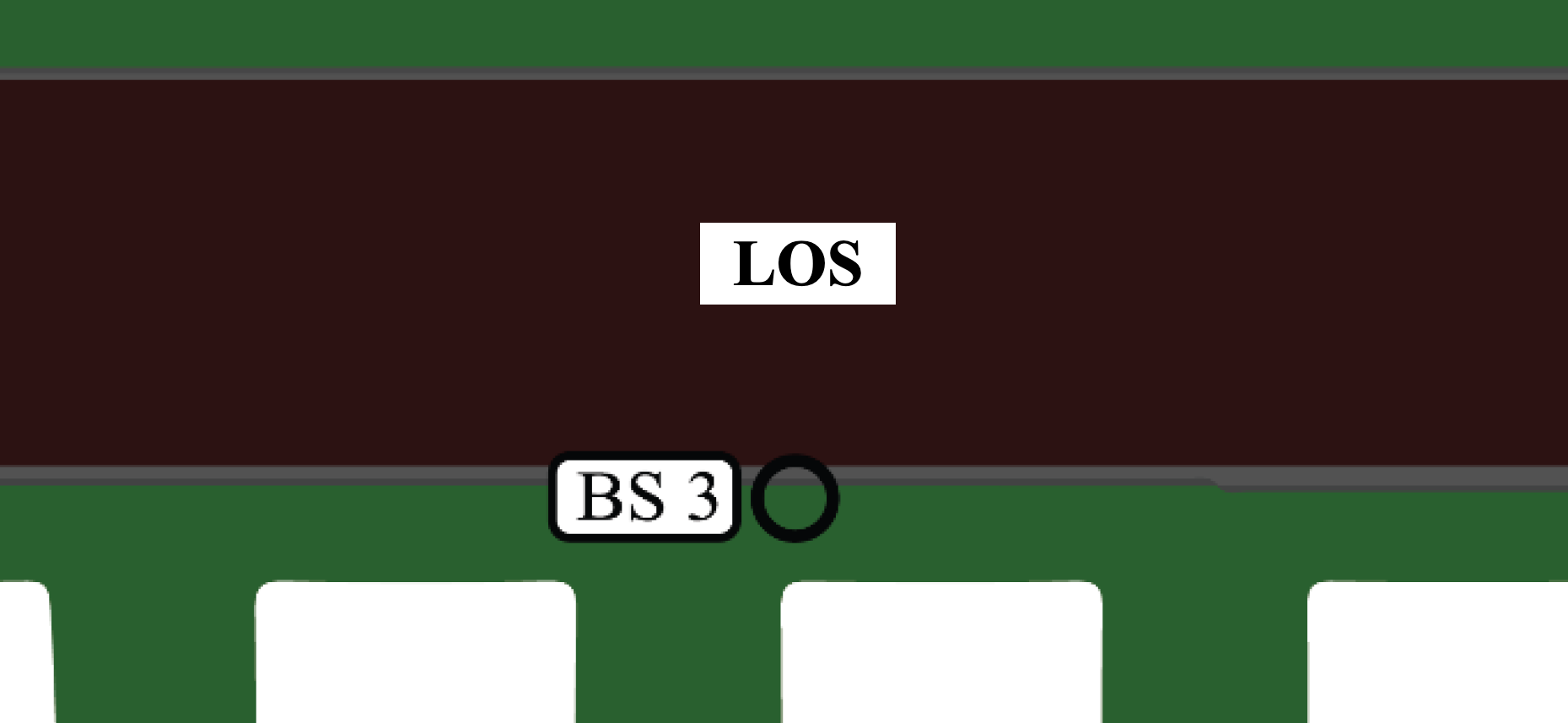}}}
    \hspace{20pt}%
    \subfloat[\label{Fig:scenario_b} 'O$1$ Blockage' scenario]{%
        \scalebox{1}{\includegraphics[trim={0cm 0cm 0cm 0cm},clip,width=0.4\textwidth]{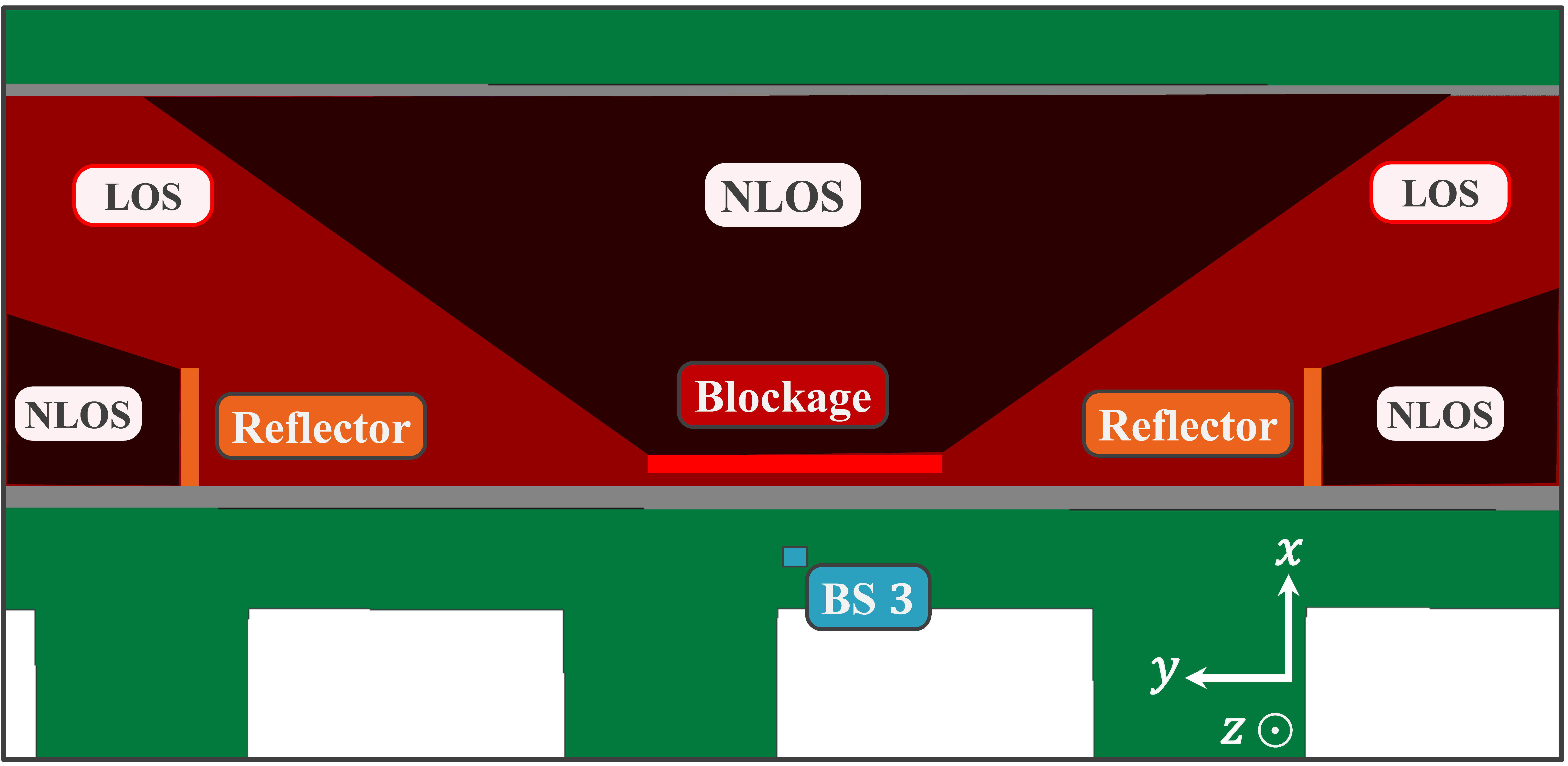}}}
	\caption{Top view of the considered 'O$1$' and 'O$1$ Blockage' outdoor scenarios \cite{alkhateeb_deepmimo:_2019}.}
	\label{Fig:scenario}
\end{figure}
\begin{table}
    \caption{DeepMIMO dataset  \cite{alkhateeb_deepmimo:_2019} parameters for the 'O$1$' and 'O$1$ Blockage' scenarios}
    \label{T1}
    \centering
    \scalebox{0.8}{
    \begin{tabular}{ c | c  c | c  c}
        \hline \hline
        Parameters & \multicolumn{2}{c|}{'O$1$'} & \multicolumn{2}{c}{'O$1$' Blockage'}\\
          & mmW & sub-6 & mmW & sub-6\\
        \hline
        scenario name & O$1\_28$ & O$1\_3\text{p}5$ & O$1\_28$B & O$1\_3\text{p}5$B\\
        \hline
        Active BS & $3$ & $3$ & $3$ & $3$\\
        \hline
        Active users & $700-1300$ & $700-1300$ & $700-1300$ & $700-1300$\\
        \hline
        Number of BS antennas & $64$ & $4$ & $64$ & $4$\\
        \hline
        Number of user antennas & $4$ or $8$ & $1$ & $4$ or $8$ & $1$\\
        \hline
        Antenna spacing (wave-length) & $0.5$ & $0.5$ & $0.5$ & $0.5$\\
        \hline
        Bandwidth (GHz) & $0.5$ & $0.02$ & $0.5$ & $0.02$\\
        \hline
        Number of OFDM subcarriers & $512$ & $32$ & $512$ & $32$\\
        \hline
        Number of paths & $5$ & $15$ & $5$ & $15$\\
        \hline \hline
    \end{tabular}}
\end{table}
\subsection{Simulation Results}
We consider two outdoor scenarios to evaluate performance of the proposed device-agnostic framework from the DeepMIMO dataset \cite{alkhateeb_deepmimo:_2019}. As shown in Fig.~\ref{Fig:scenario}, the 'O$1$' scenario considers LOS condition for all the points in the user region. In the 'O$1$ Blockage' scenario, a blocker and two reflectors are placed in the environment. The LOS and NLOS user regions are illustrated in Fig.~\ref{Fig:scenario}. In the 'O$1$' and 'O$1$ Blockage' scenarios, dual-band communication with $3.5$ GHz and $28$ GHz links are considered. We consider co-located ULAs with $M_{\text{sub-6}} = 4$ and $M_{\text{mmW}} = 64$ antenna elements at the BS3 that enables dual-band communication. The user is equipped with a single antenna at the sub-6 GHz and a ULA including $4$ or $8$ mmWave antenna elements. An ideal pulse shaping filter is assumed in the generation of channel responses. Table \ref{T1} provides parameters used in the dataset generation of the 'O$1$' and 'O$1$ Blockage' scenarios. For each scenario and configuration, a dataset with $27391$ samples is collected and split randomly $70\%-30\%$ into training and test data, respectively. Thus, the training dataset includes $19173$ samples while the test data is made of $8218$ samples. For training the device-specific and device-agnostic ML models, we consider $40$ epochs with a batch size of $64$ samples. Each network is trained and averaged out with three random weight initialization, which leads to playing down the initialization effects.

We consider Top-$n$ accuracy metric, denoted $A_{\text{Top-}n}$, to evaluate the performance of beam prediction methods as classifiers. Top-$n$ accuracy can be defined as
\begin{equation}
    A_{\text{Top-}n} = \frac{1}{N_{\text{test}}} \sum_{k=1}^{N_{test}} \mathbb{1}_{(\boldsymbol{f}^\star_k, \boldsymbol{w}^\star_k) \in \mathcal{S}^n_k}
\end{equation}
where $\mathcal{S}^n_k$ is a candidate list of beam pairs that provide the $n$ highest optimality likelihood, $P$, for the $k$th test sample. $\mathbb{1}_{(\boldsymbol{f}^\star_k, \boldsymbol{w}^\star_k) \in \mathcal{S}^n_k}$ results in $1$ when the $n$-long beam list includes the optimal beam pair, otherwise $0$. Moreover, we use the Top-$n$ spectral efficiency as the average spectral efficiency provided by the best beam pair in the $n$-long beam pair candidate list, $\mathcal{S}^n$. For each sample in the test dataset, spectral efficiency of the best beam pair in the candidate list is calculated using \eqref{eq:ASE}. 

\begin{figure}[t]
	\centering
	\subfloat[\label{Fig:comp_Fib100_a} ]{%
       \scalebox{1}{\includegraphics[trim={0cm 0cm 0cm 0cm},clip,width=0.4\textwidth]{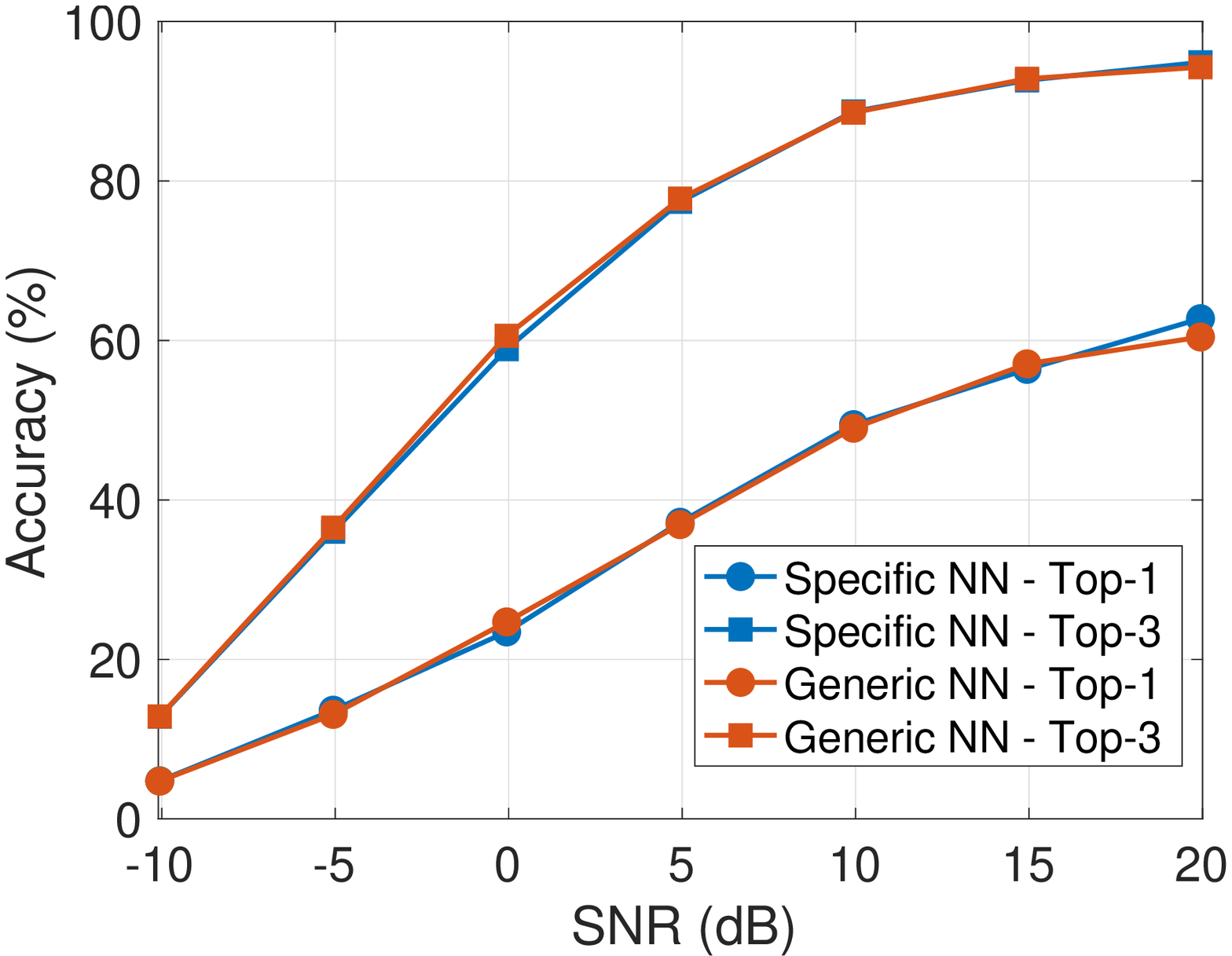}}}
    \hspace{20pt}%
    \subfloat[\label{Fig:comp_Fib100_b} ]{%
       \scalebox{1}{\includegraphics[trim={0cm 0cm 0cm 0cm},clip,width=0.4\textwidth]{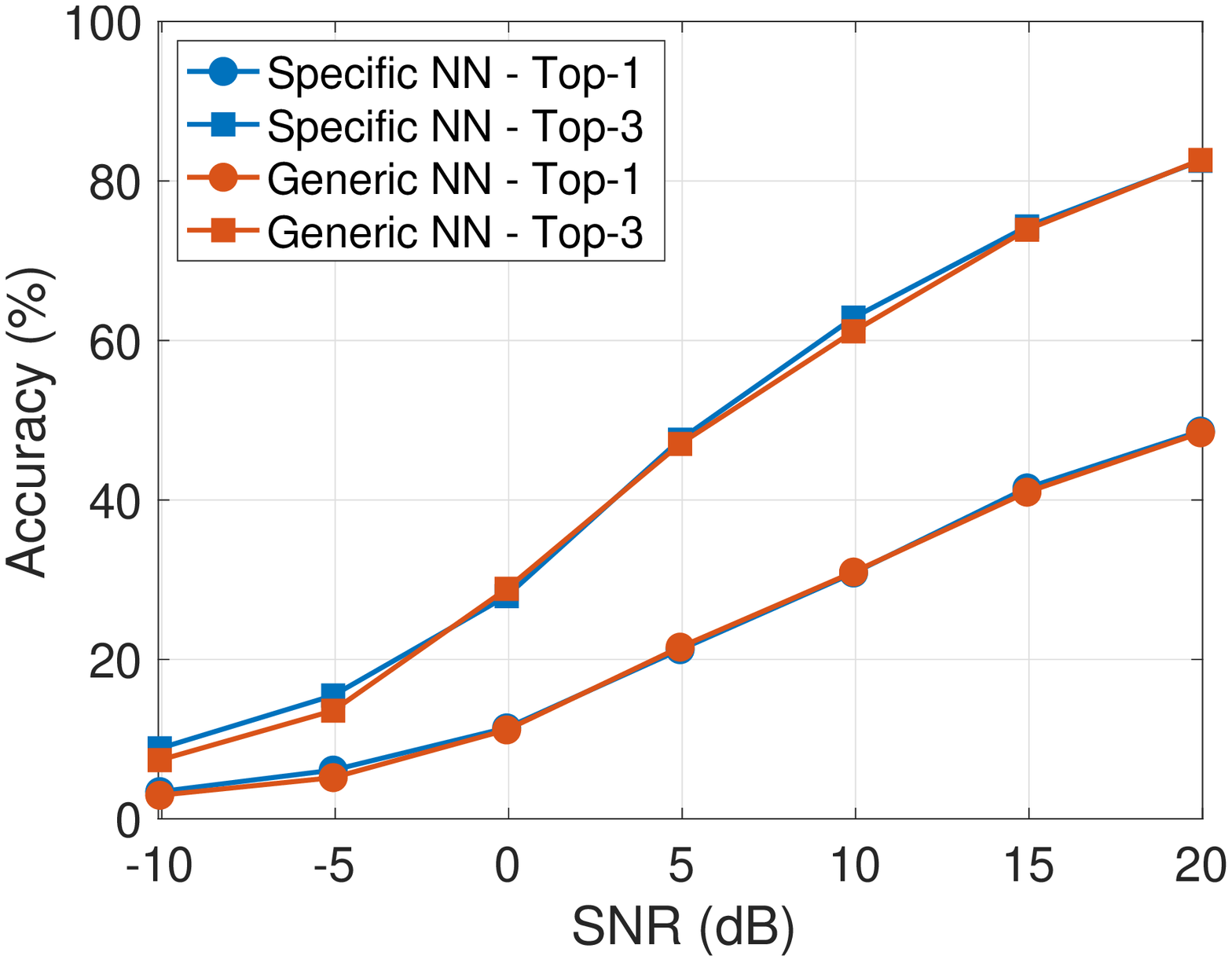}}}
    \\[0.05cm]
    \subfloat[\label{Fig:comp_Fib100_c} ]{%
       \scalebox{1}{\includegraphics[trim={0cm 0cm 0cm 0cm},clip,width=0.4\textwidth]{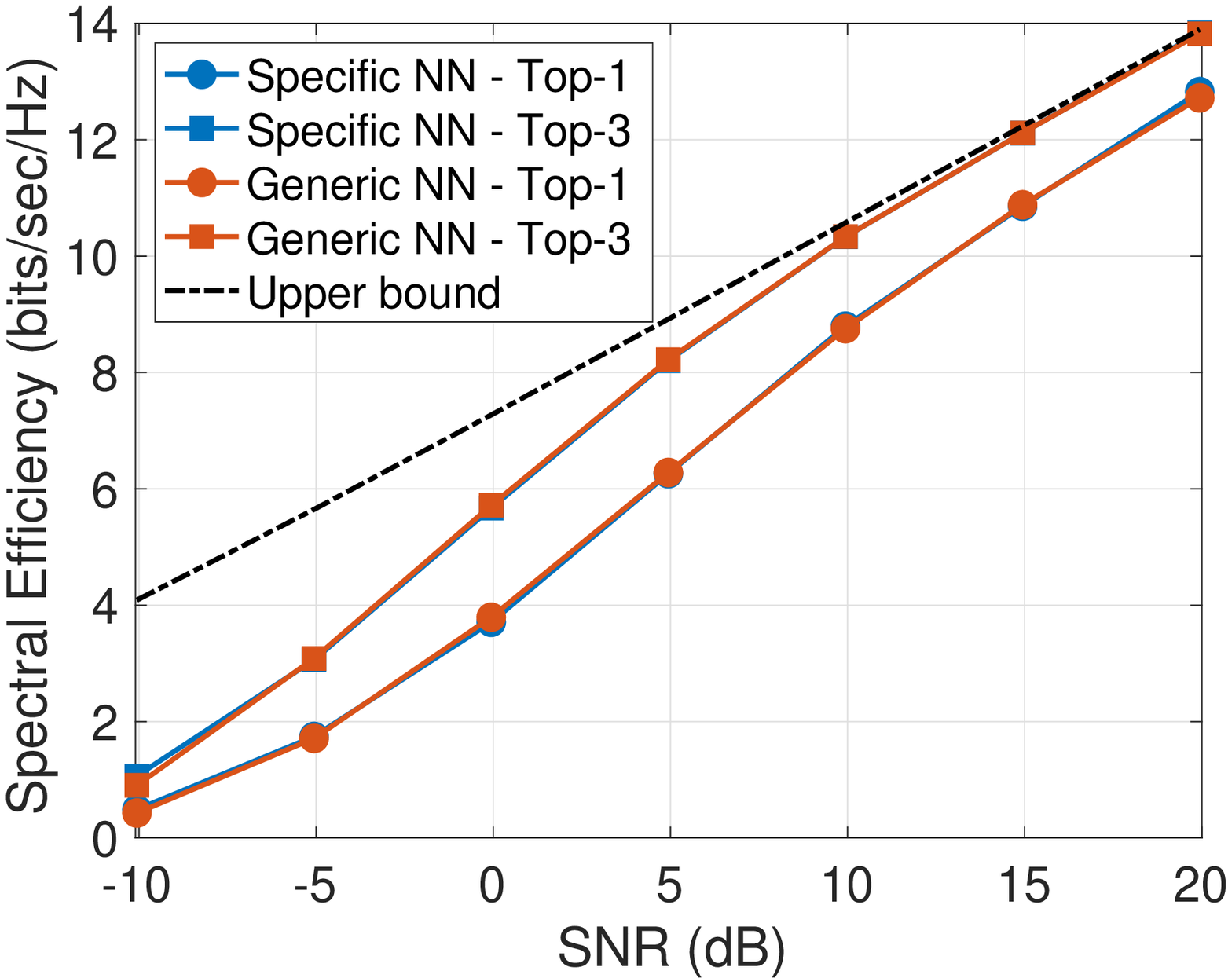}}}
    \hspace{20pt}%
    \subfloat[\label{Fig:comp_Fib100_d} ]{%
       \scalebox{1}{\includegraphics[trim={0cm 0cm 0cm 0cm},clip,width=0.4\textwidth]{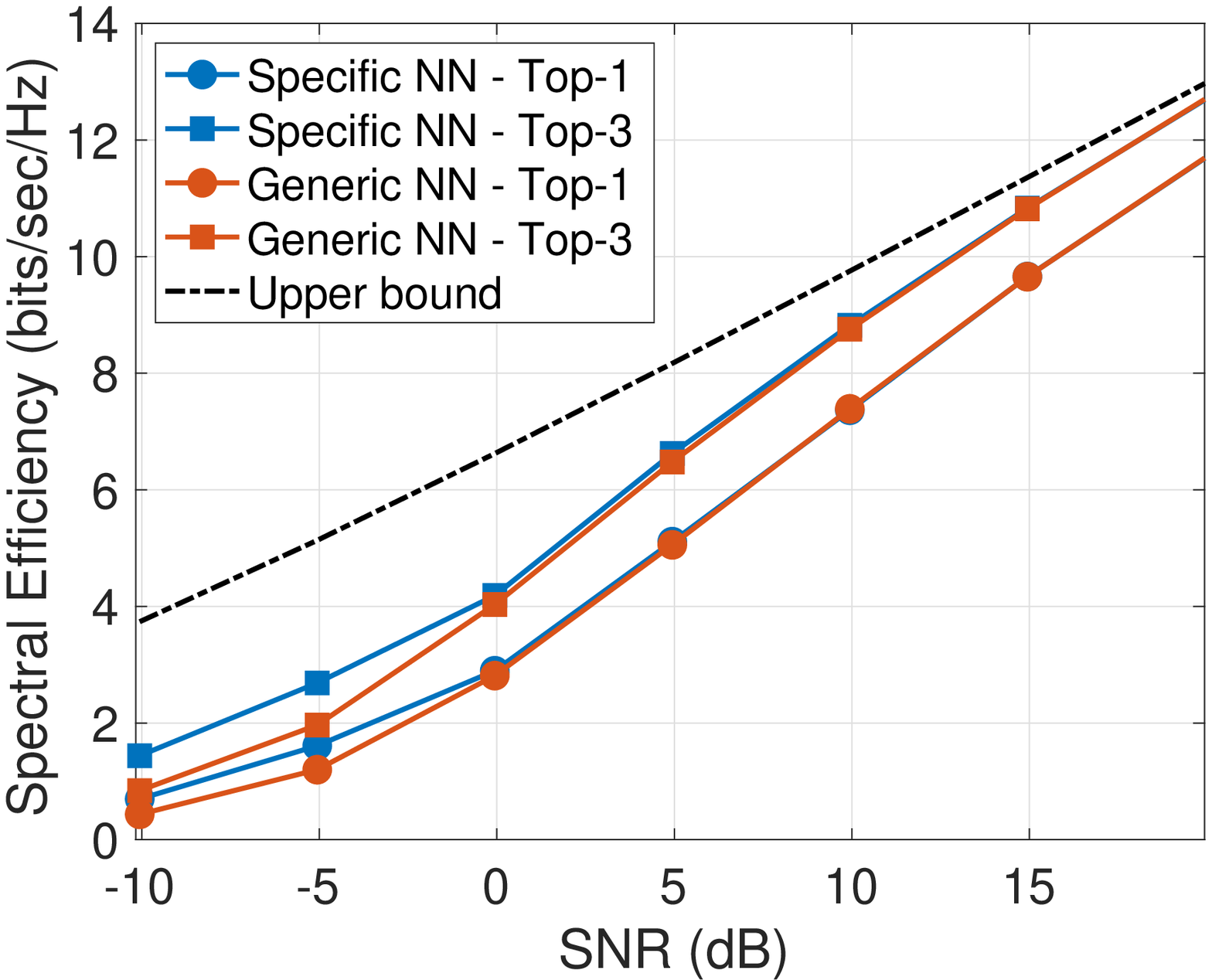}}}
	\caption{Performance of the device-specific \cite{alrabeiah_deep_2020} compared to the proposed device-agnostic framework with $M_{\text{mmW}}^{U} = 8$ and $n_{Fib} = 100$ in the 'O$1$' (a, c) and 'O$1$ Blockage' (b, d) scenarios.}
	\label{Fig:comp_Fib100}
\end{figure}
Fig.~\ref{Fig:comp_Fib100} shows the Top-1 and Top-3 accuracy and Top-1 and Top-3 spectral efficiency of the device-specific and device-agnostic methods for the 'O$1$' and 'O$1$ Blockage' scenarios. A ULA array made of $8$ antenna elements is considered for the mmWave communication at the user. We consider a FG with $n_{Fib} = 100$ points for the device-agnostic framework. There is almost no performance degradation by using the proposed device-agnostic framework compared to the device-specific one. The results show that the performance of the generic network providing the optimality probability for the FG is as good as the device-specific network pointing the beam pairs.

\begin{figure}[t]
	\centering
	\subfloat[\label{Fig:comp_UE_4_8_Fib25_a} $M_{\text{mmW}}^{U} = 8$, 'O$1$' scenario]{%
       \scalebox{1}{\includegraphics[trim={0cm 0cm 0cm 0cm},clip,width=0.4\textwidth]{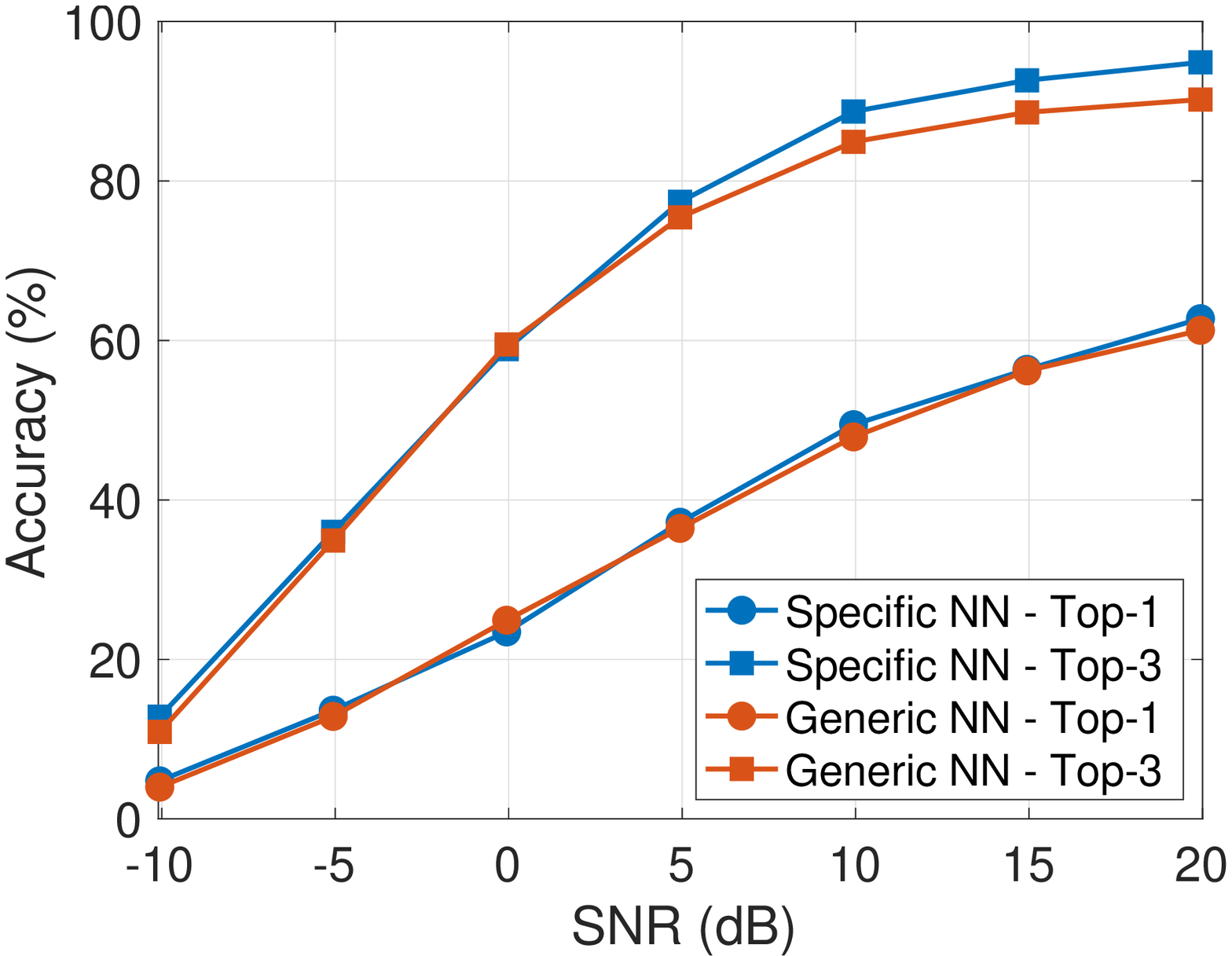}}}
    \hspace{20pt}%
    \subfloat[\label{Fig:comp_UE_4_8_Fib25_b} $M_{\text{mmW}}^{U} = 8$, 'O$1$ Blockage' scenario]{%
       \scalebox{1}{\includegraphics[trim={0cm 0cm 0cm 0cm},clip,width=0.4\textwidth]{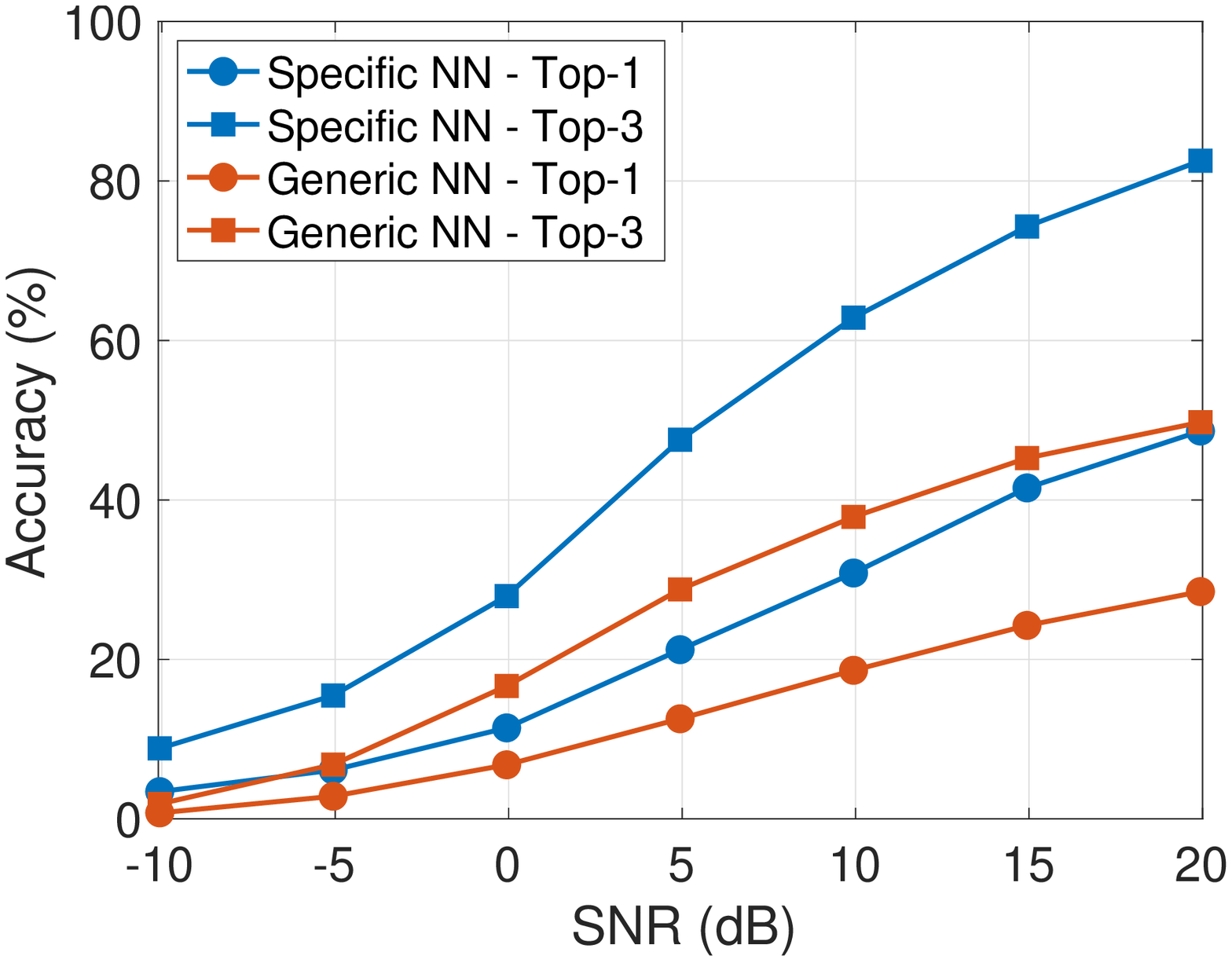}}}
    \\[0.05cm]
    \subfloat[\label{Fig:comp_UE_4_8_Fib25_c} $M_{\text{mmW}}^{U} = 4$, 'O$1$' scenario]{%
       \scalebox{1}{\includegraphics[trim={0cm 0cm 0cm 0cm},clip,width=0.4\textwidth]{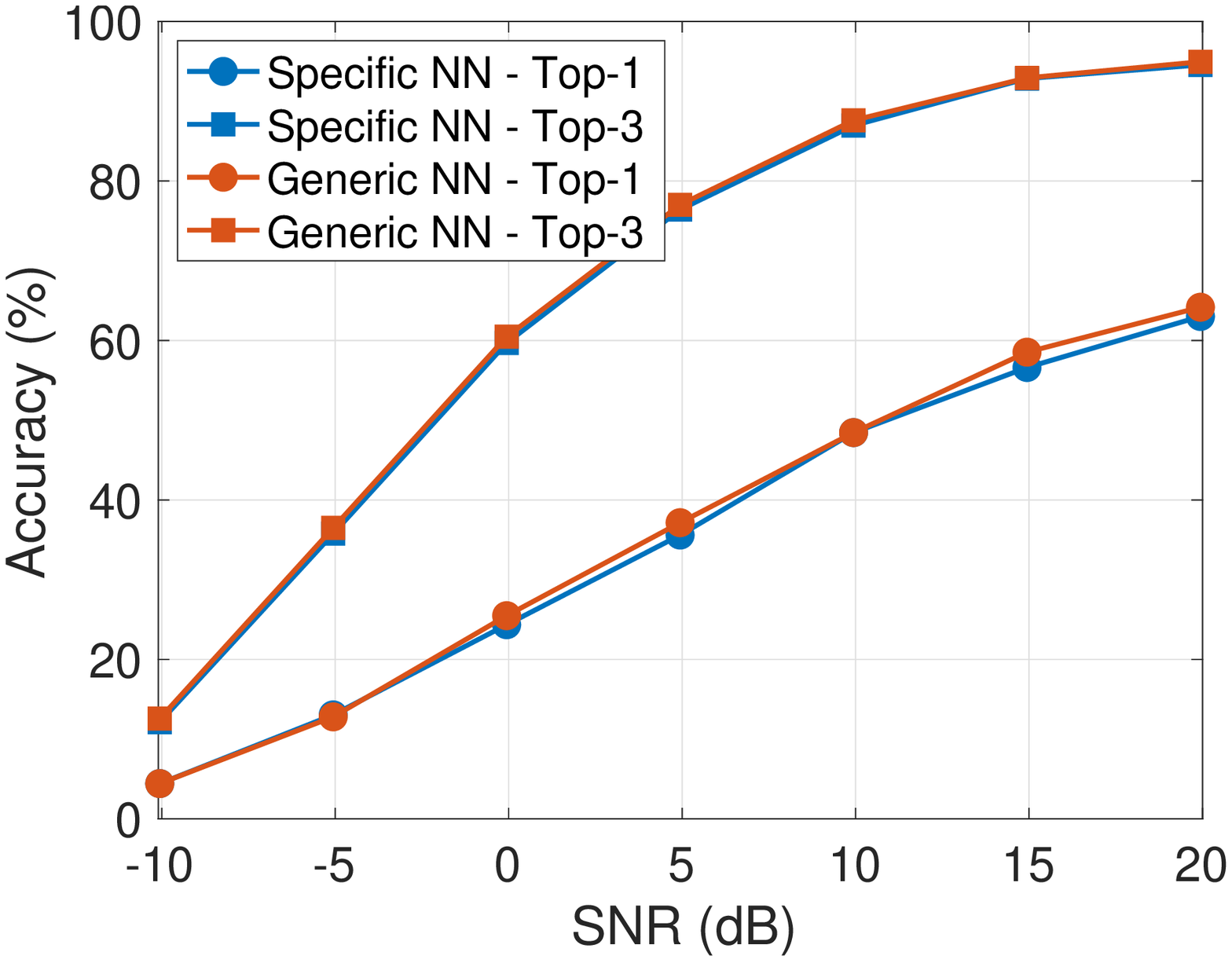}}}
    \hspace{20pt}%
    \subfloat[\label{Fig:comp_UE_4_8_Fib25_d} $M_{\text{mmW}}^{U} = 4$, 'O$1$ Blockage' scenario]{%
       \scalebox{1}{\includegraphics[trim={0cm 0cm 0cm 0cm},clip,width=0.4\textwidth]{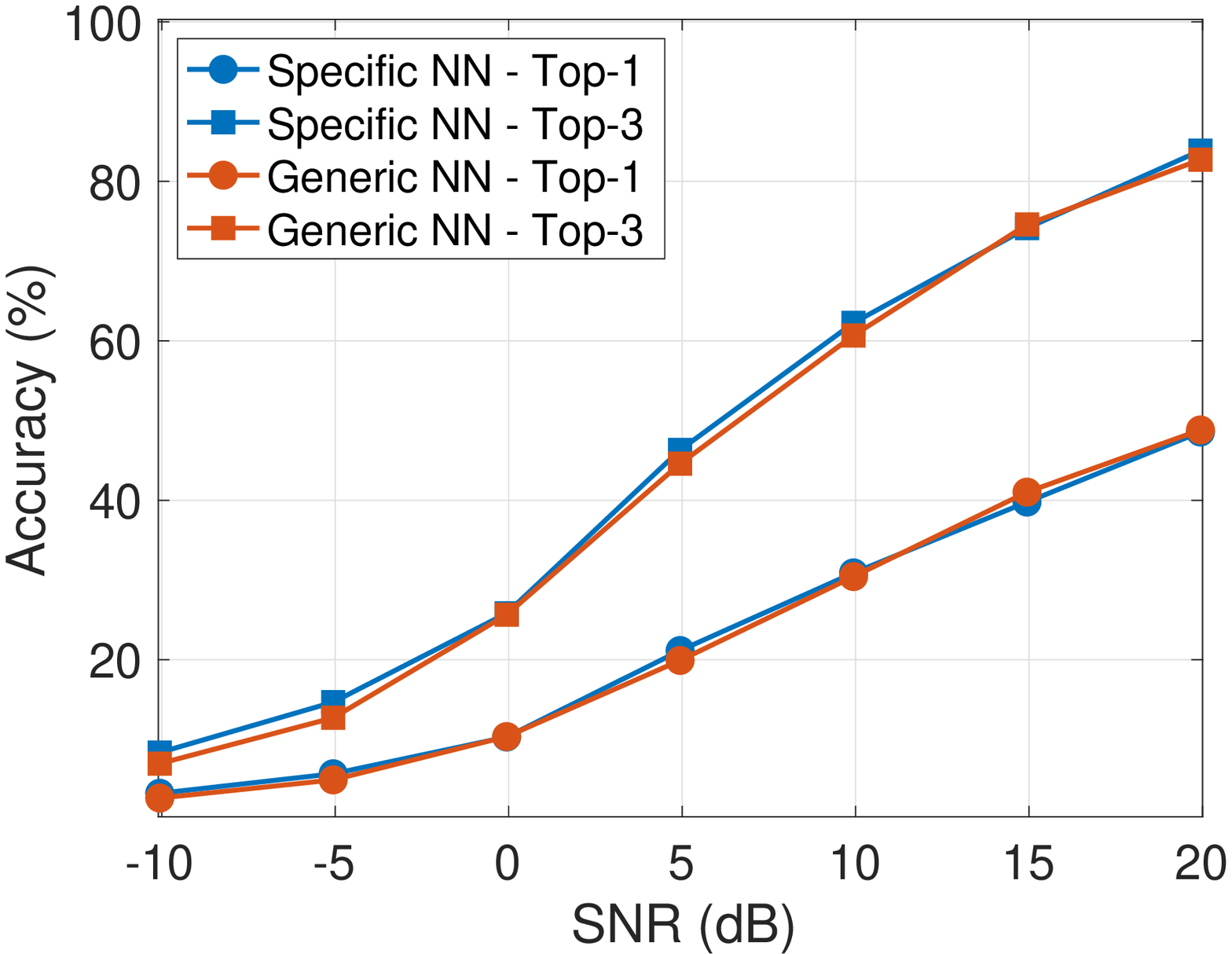}}}
	\caption{Accuracy of the device-agnostic framework where the Fibonacci grid is made of $25$ points ($n_{Fib} = 25$).}
	\label{Fig:comp_UE_4_8_Fib25}
\end{figure}

\begin{figure}[t]
	\centering
	\subfloat[\label{Fig:BeamReg_4_8_a} $M_{\text{mmW}}^{U} = 8$]{%
       \scalebox{1}{\includegraphics[trim={0cm 0cm 0cm 0cm},clip,width=0.55\textwidth]{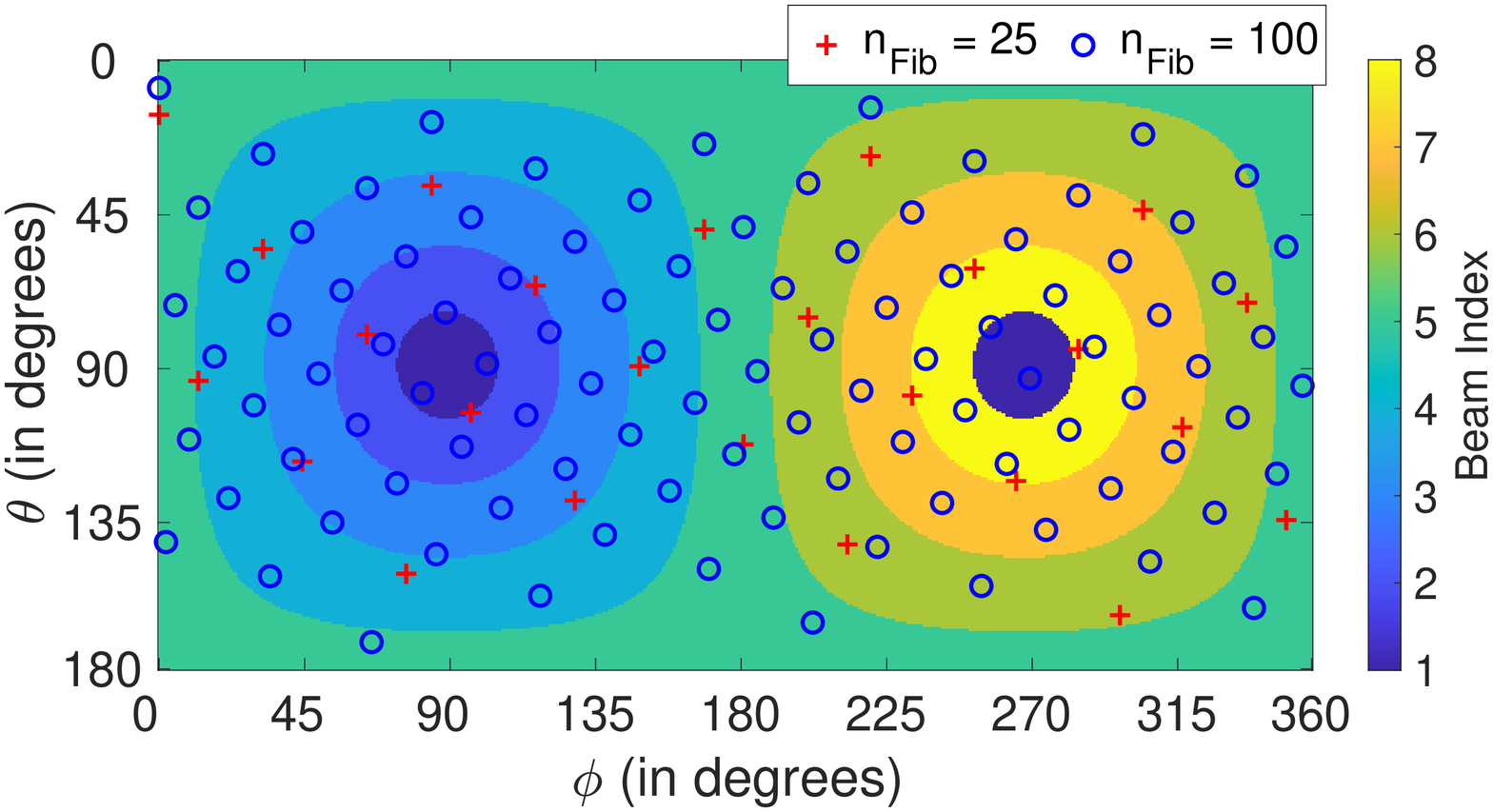}}}
    \hspace{20pt}%
    \subfloat[\label{Fig:BeamReg_4_8_b} $M_{\text{mmW}}^{U} = 8$]{%
       \scalebox{1}{\includegraphics[trim={0cm 0cm 0cm 0cm},clip,width=0.4\textwidth]{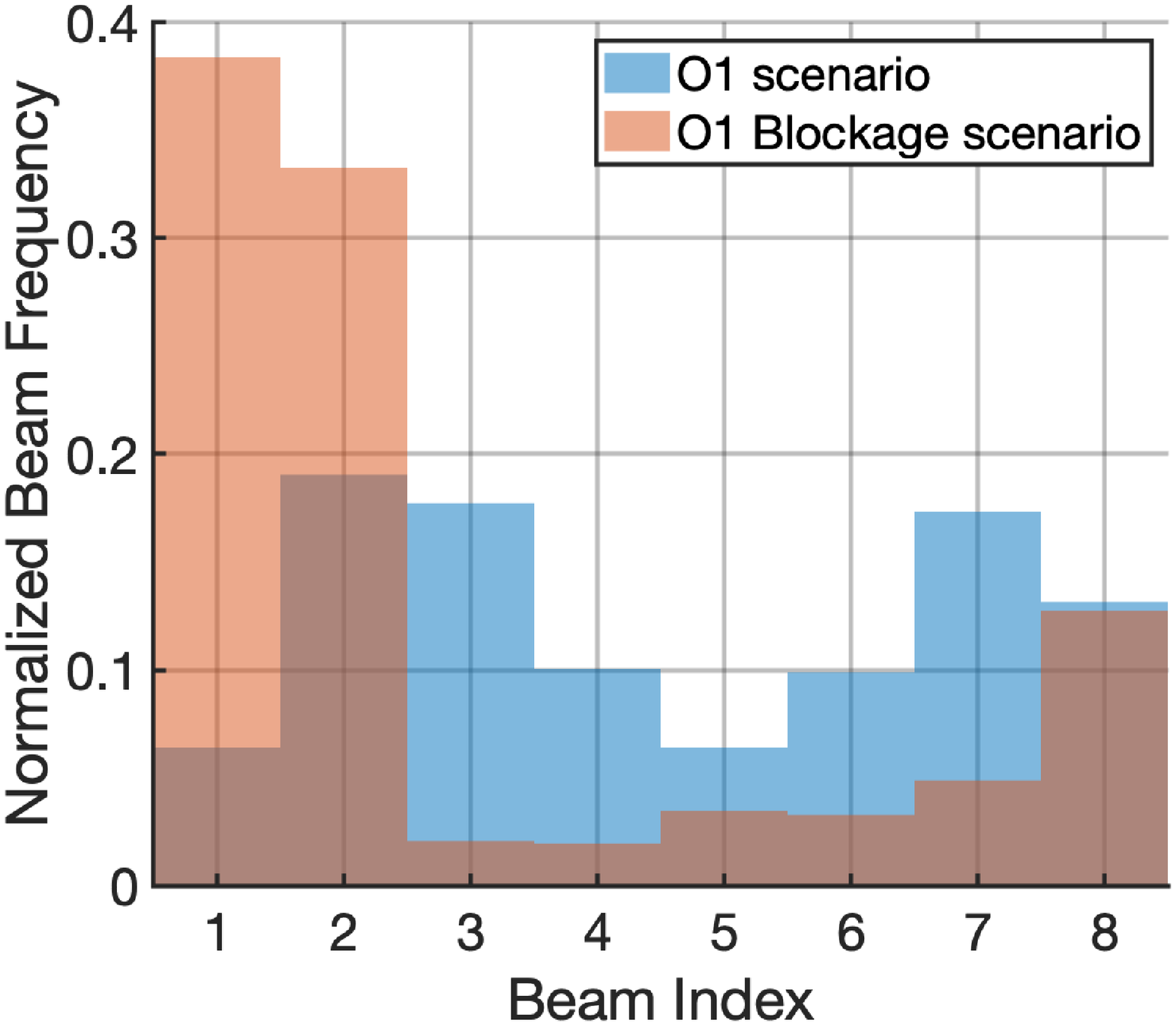}}}
    \\[0.05cm]
    \subfloat[\label{Fig:BeamReg_4_8_c} $M_{\text{mmW}}^{U} = 4$]{%
       \scalebox{1}{\includegraphics[trim={0cm 0cm 0cm 0cm},clip,width=0.55\textwidth]{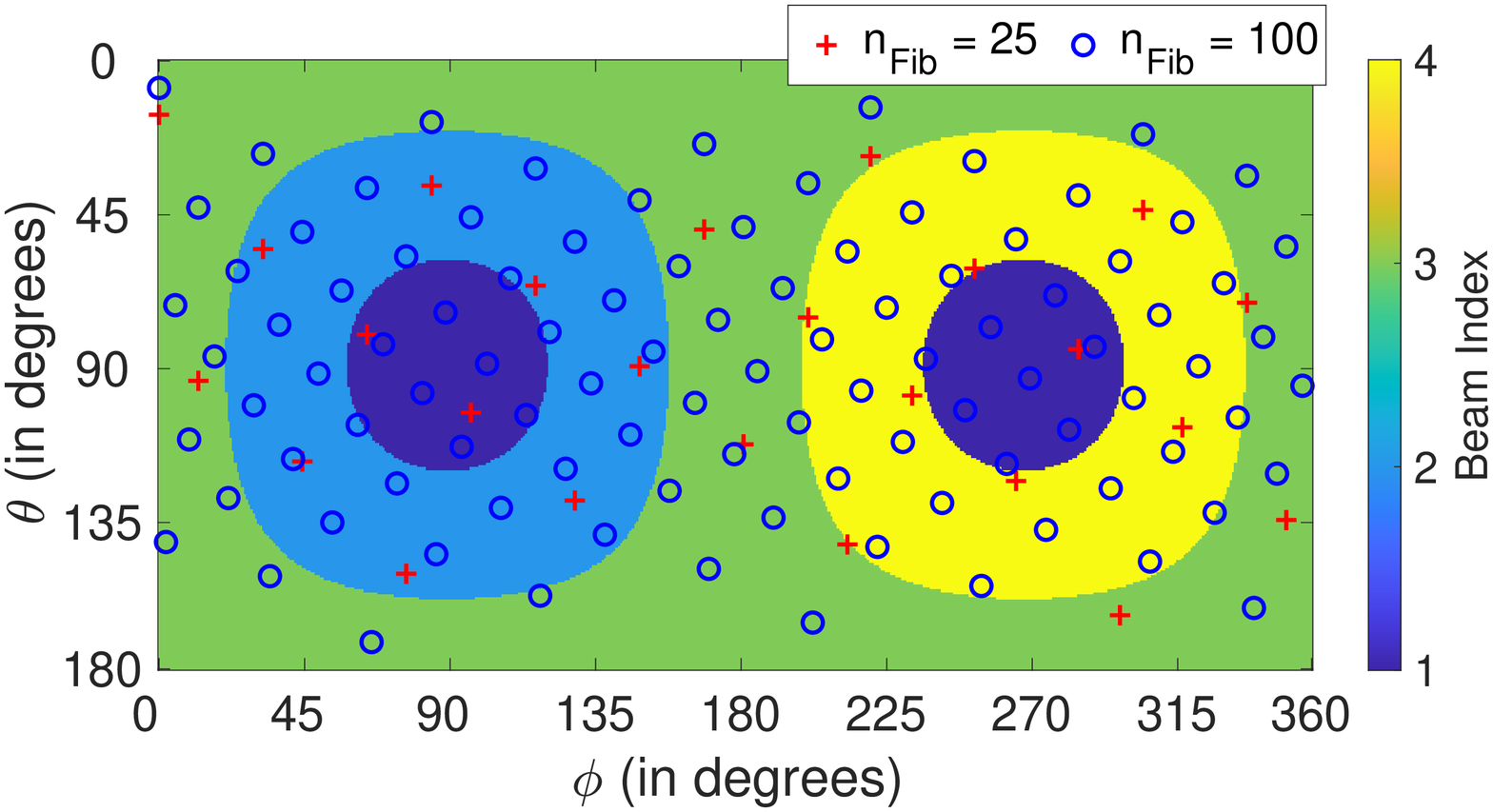}}}
    \hspace{20pt}%
    \subfloat[\label{Fig:BeamReg_4_8_d} $M_{\text{mmW}}^{U} = 4$]{%
       \scalebox{1}{\includegraphics[trim={0cm 0cm 0cm 0cm},clip,width=0.4\textwidth]{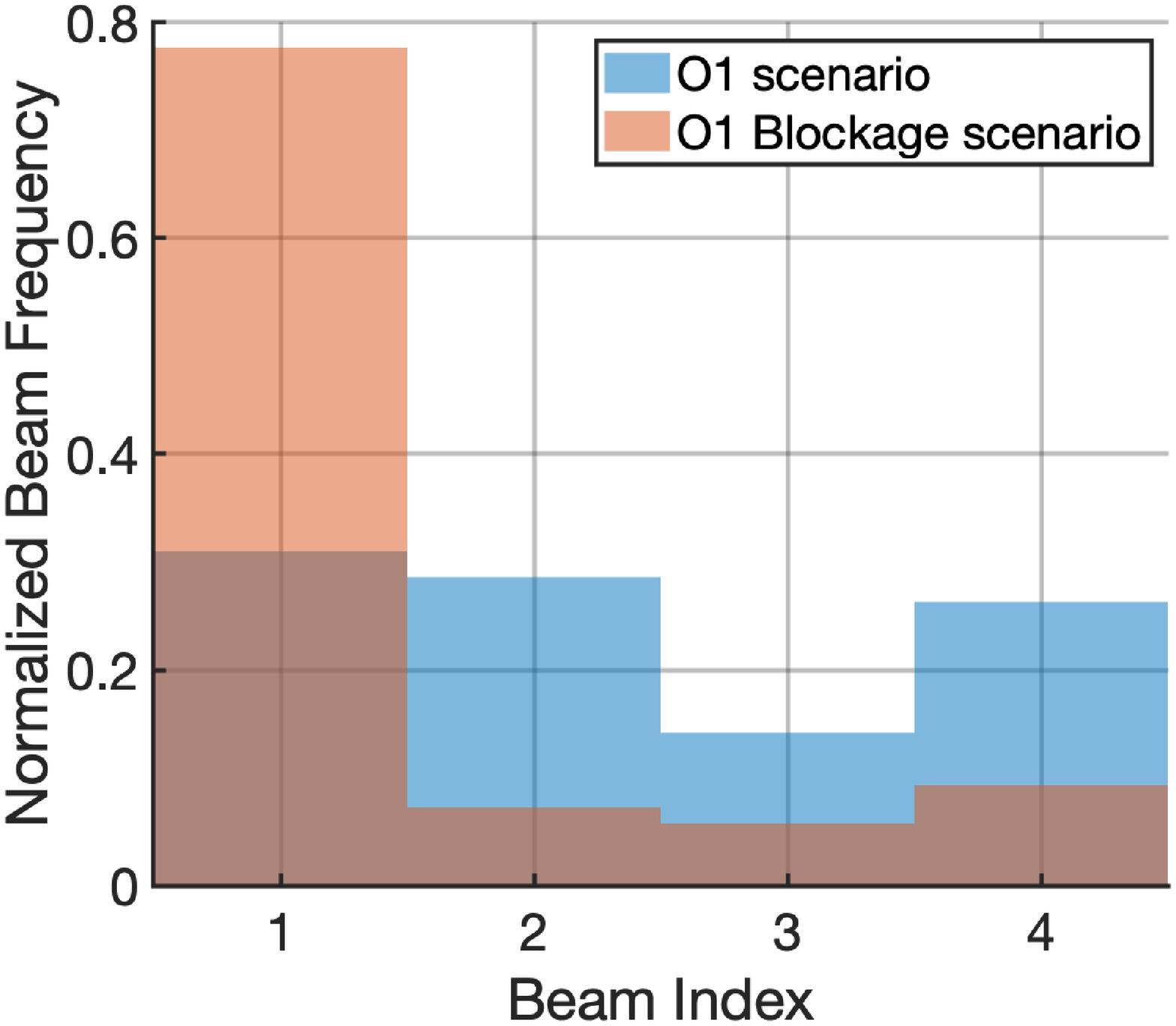}}}
	\caption{Beam region and normalized histogram of the user beam index with $4$ or $8$ antenna elements. In the case $M_{\text{mmW}}^{U} = 8$, a FG with $25$ points has no point inside the beam region of first beam.}
	\label{Fig:BeamReg_4_8}
\end{figure}
The effects of considering a Fibonacci Grid with $25$ points are shown in Fig.~\ref{Fig:comp_UE_4_8_Fib25}. In the case $M_{\text{mmW}}^{U} = 8$, the device-agnostic framework cannot achieve the same performance as the device-specific method. Moreover, a FG with $n_{Fib} = 25$ incurs more severe performance degradation in the 'O$1$ Blockage'. However, for the 'O$1$' and 'O$1$ Blockage' scenarios, there is no performance loss when $M_{\text{mmW}}^{U} = 4$. These results show that selecting the number of points in the FG may depend on the number of antenna elements codebook size at the user side. Figs.~\ref{Fig:BeamReg_4_8}\subref{Fig:BeamReg_4_8_a}, \subref{Fig:BeamReg_4_8_c} depict the beam regions for devices with $8$ and $4$ antenna elements. We plot the Fibonacci points for two cases with $25$ and $100$ points. As it is illustrated in  Fig.~\ref{Fig:BeamReg_4_8}\subref{Fig:BeamReg_4_8_a}, for the FG with $n_{Fib} = 25$, no points are placed in the beam region for beam $1$. This explains the performance loss in the case of $M_{\text{mmW}}^{U} = 8$ and a FG with $n_{Fib} = 25$. Figs.~\ref{Fig:BeamReg_4_8}\subref{Fig:BeamReg_4_8_b}, \subref{Fig:BeamReg_4_8_d} show the normalized frequency of best user beam for devices with $M_{\text{mmW}}^{U} = 8$ and $M_{\text{mmW}}^{U} = 4$. The frequency of beam $1$ in the 'O$1$ Blockage' scenario is several times more than its frequency in the 'O$1$' scenario, which justifies the more severe performance degradation in the 'O$1$ Blockage' scenario.

\begin{figure}[t]
	\centering
    \subfloat[\label{Fig:TrainSize_a}]{%
       \scalebox{1}{\includegraphics[trim={0cm 0cm 0cm 0cm},clip,width=0.45\textwidth]{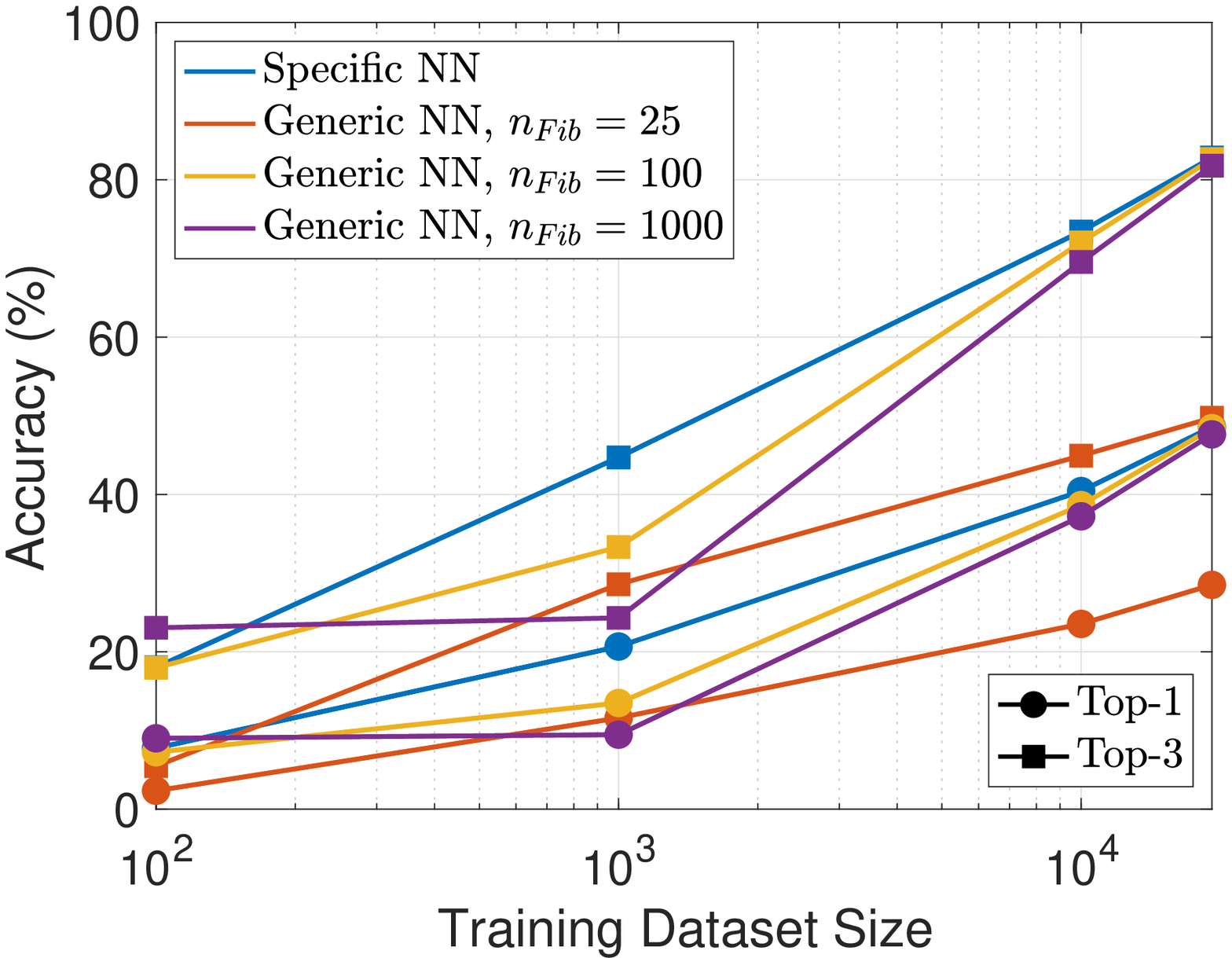}}}
    \hspace{20pt}%
    \subfloat[\label{Fig:TrainSize_b}]{%
       \scalebox{1}{\includegraphics[trim={0cm 0cm 0cm 0cm},clip,width=0.45\textwidth]{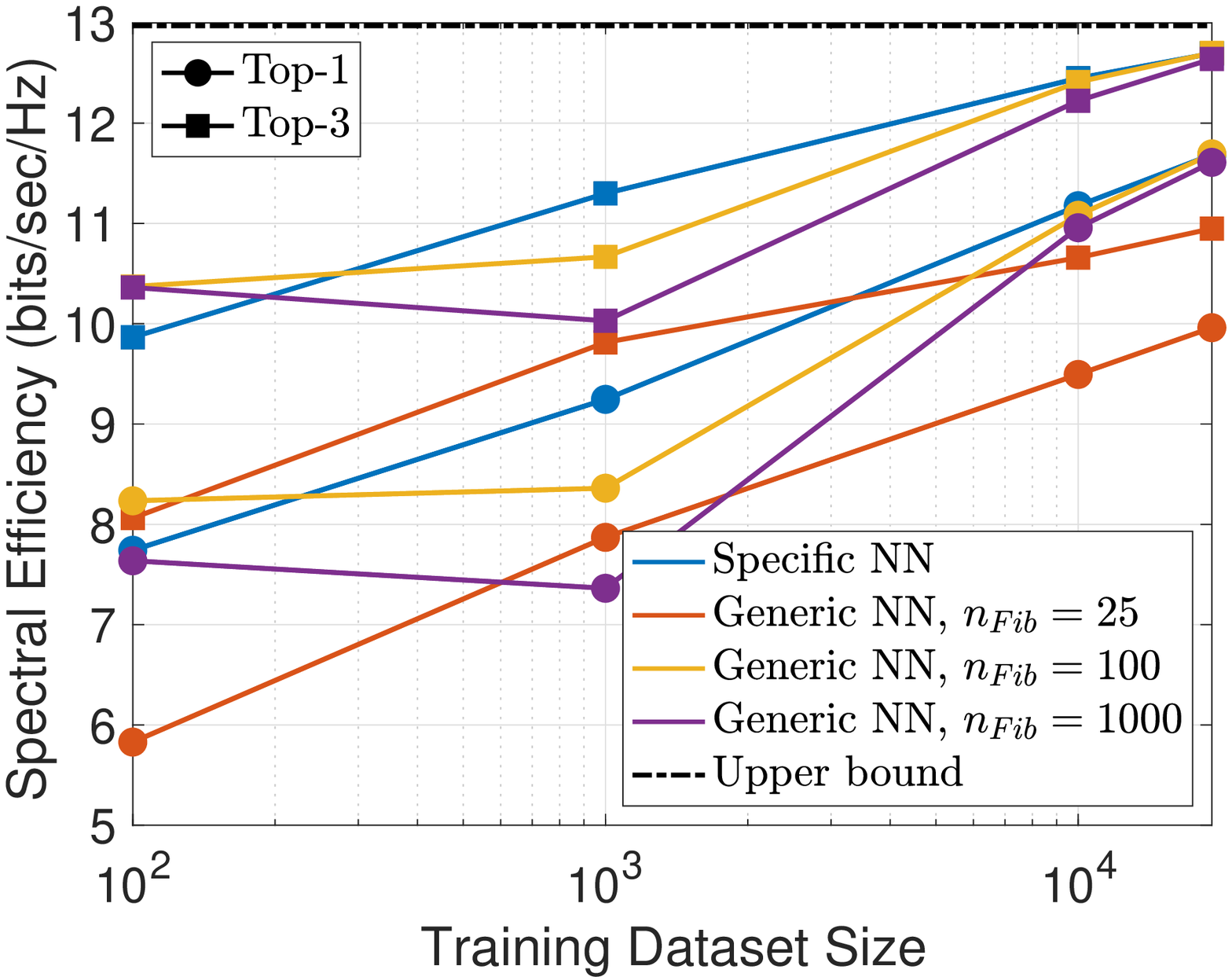}}}
	\caption{Effect of training dataset size on the performance of the device-agnostic framework.}
	\label{Fig:TrainSize}
\end{figure}
Fig.~\ref{Fig:TrainSize} shows the Top-3 accuracy and achievable spectral efficiency for different training dataset sizes with $25$, $100$, and $1000$ points in the FG. Considering insufficient points in the FG may not cover some beams in the user codebook. For example, a significant performance degradation for the device-agnostic framework with $n_{Fib}=25$ can be seen in different training dataset sizes. On the other hand, a FG with too many points results in a huge number of training parameters at the output layer. In the case with $n_{Fib} = 1000$, the accuracy of the device-agnostic framework with limited training samples is dropped. However, with a large training dataset, the device-agnostic framework with a FG with a moderate or large number of points provides the same performance as the device-specific method.

\begin{figure}[t]
	\centering
    \subfloat[\label{Fig:MisMatch_a}]{%
       \scalebox{1}{\includegraphics[trim={0cm 0cm 0cm 0cm},clip,width=0.45\textwidth]{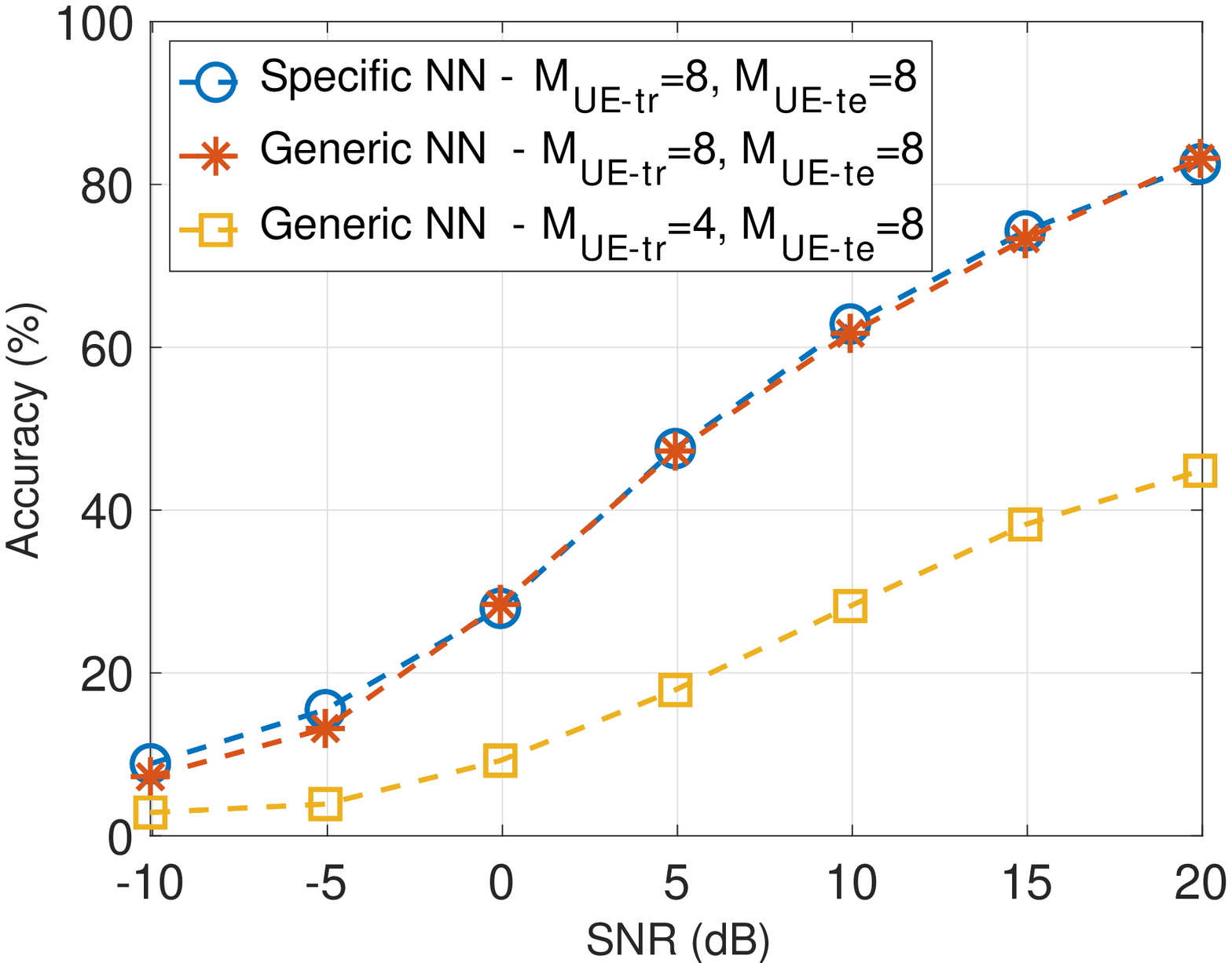}}}
    \hspace{20pt}%
    \subfloat[\label{Fig:MisMatch_b}]{%
       \scalebox{1}{\includegraphics[trim={0cm 0cm 0cm 0cm},clip,width=0.45\textwidth]{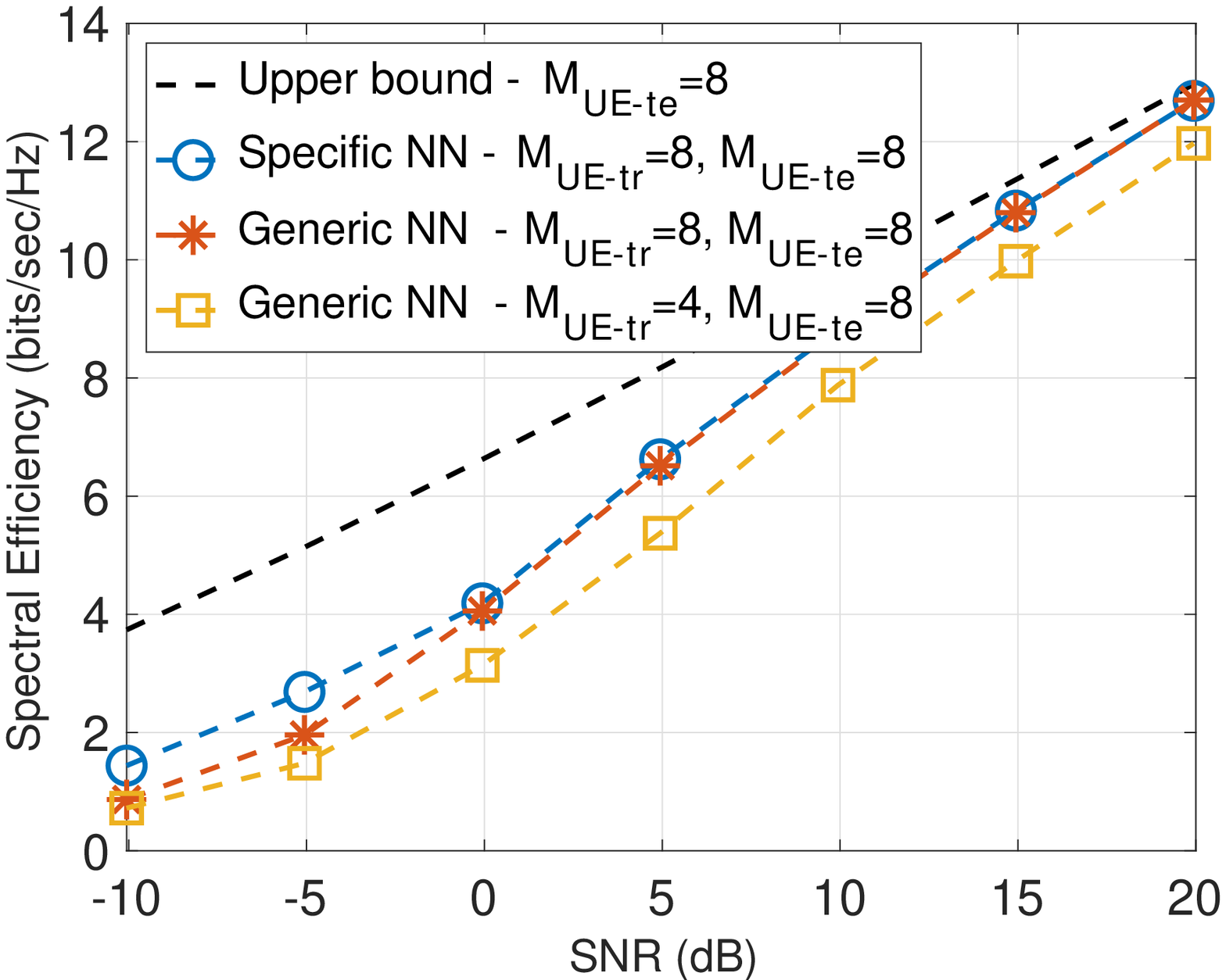}}}
    \\[0.05cm]
    \subfloat[\label{Fig:MisMatch_c}]{%
       \scalebox{1}{\includegraphics[trim={0cm 0cm 0cm 0cm},clip,width=0.45\textwidth]{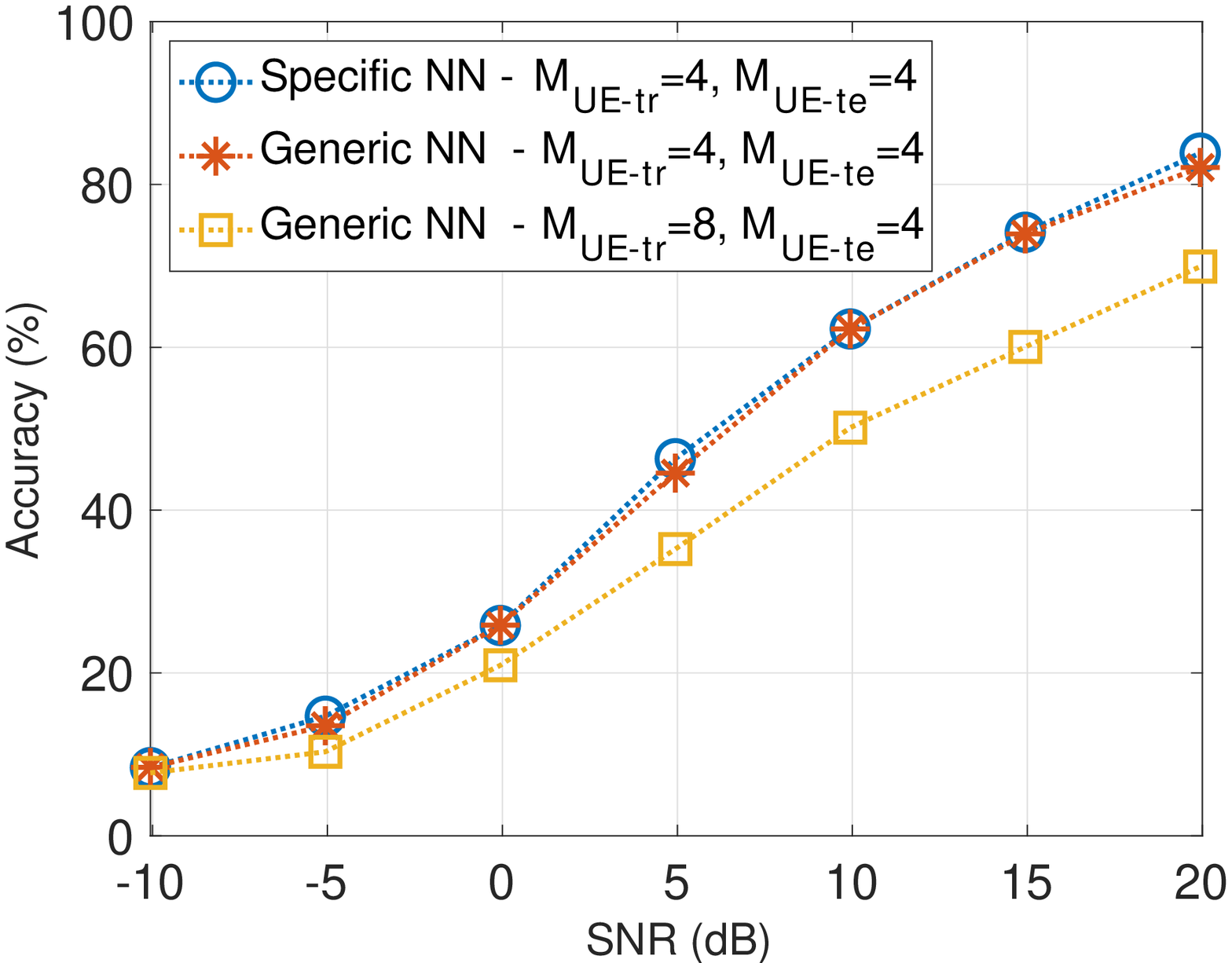}}}
    \hspace{20pt}%
    \subfloat[\label{Fig:MisMatch_d}]{%
       \scalebox{1}{\includegraphics[trim={0cm 0cm 0cm 0cm},clip,width=0.45\textwidth]{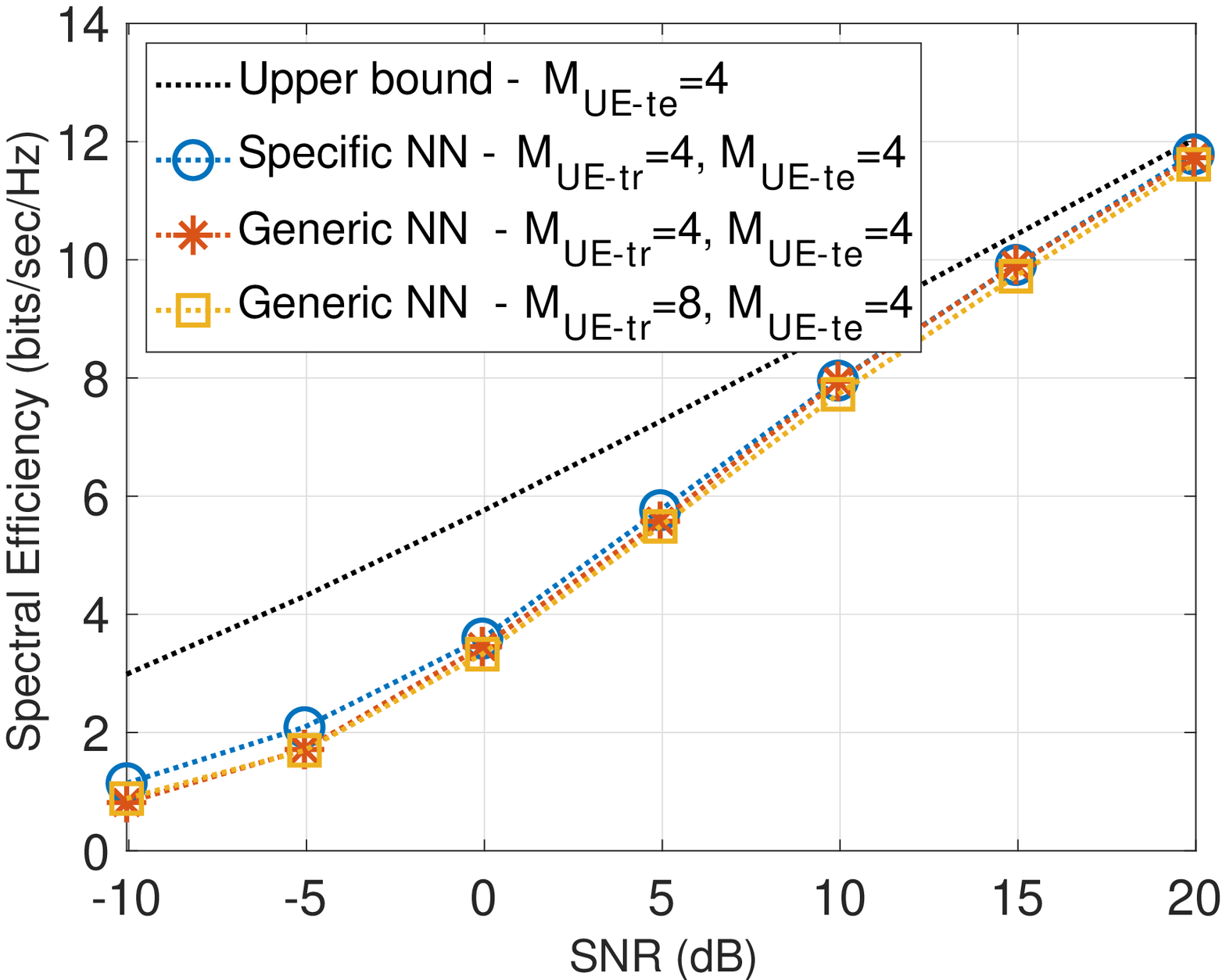}}}
	\caption{The Top-3 accuracy and spectral efficiency of the proposed device-agnostic framework when there is a mismatch in the training and test datasets. The 'O$1$ Blockage' scenario and a FG with $n_{Fib} = 100$ are considered.}
	\label{Fig:MisMatch}
\end{figure}
The proposed device-agnostic framework can be trained with a specific device and be used for other devices with different antenna configurations in the deployment phase. Fig.~\ref{Fig:MisMatch} shows the Top-3 accuracy of the device-agnostic framework in the 'O$1$ Blockage' scenario when there is a mismatch in the antenna configuration of the devices in the training and deployment phases. The performance degradation is more noticeable when the training samples are collected with devices with $4$ antenna elements, but the framework is evaluated for devices with $8$ antenna elements. However, in the 
opposite case, when the device is trained with an antenna of $8$ elements and generalized to operate with an antenna of $4$ elements, there is less loss in performance. Although there is a performance loss due to the mismatch effects, the results are promising as the device-agnostic framework achieves acceptable accuracy without additional training samples. 

\section{Conclusions}\label{Sec:Conc}
In this paper, we have shown that a generic neural network followed by a post processing unit can be used to recommend an accurate beam candidate list for the initial beam selection process of devices having diverse beam codebook and hardware configurations. We use the Fibonacci grid to point the directions in the unit sphere so that each point covers an equal area from the sphere. The generic network leverages the information from inputs (e.g., location, orientation, sub-6 GHz CSI) to pinpoint appropriate beamforming directions in the Fibonacci grid. The proposed framework works well with unseen antenna and hardware configurations in the training dataset. Also, it can be trained with data collected with different device antenna configurations/codebooks. 

In this study, two ML-based beam alignment methods are considered that receive position and orientation information and out-of-band channel response as the models' inputs. These two beam alignment problems are intended as mere illustrations of the applicability of our proposed approach, which is much more general. The proposed framework can be used not only with other types of context information (CI) but also in non-CI-based beam management and other codebook-specific applications. As an example, future research can investigate the feasibility of the proposed technique in beam refinement and tracking problems.

\bibliographystyle{IEEEtran} 
\bibliography{BeamAlignment_Ref_2} 

\end{document}